\documentclass[prd,onecolumn,preprintnumbers,aps]{revtex4}

\usepackage{graphicx}
\usepackage{subfigure}

\newcommand{\vect}[1]{\mathbf{#1}}  
\newcommand{\state}[1]{\left| #1 \right>}
\newcommand{\text}{\mathrm}
\newcommand{\degrees}{{^{\circ}}}

\newcommand{\mvp}[1][]{|\vect{p_{#1}}|}
\newcommand{\MeV}{\text{MeV}}
\newcommand{\rr}{\tilde{\Gamma}}

\begin{document}

\title{Composition of the Pseudoscalar $\eta$ and $\eta'$ Mesons}
\author{C. E. Thomas}
\date{1 October 2007}
\email[E-mail: ]{c.thomas1@physics.ox.ac.uk}
\affiliation{Rudolf Peierls Centre for Theoretical Physics, University of Oxford,\\ 1 Keble Road, Oxford, OX1 3NP}
\preprint{OUTP-07-06P}
\preprint{arXiv:0705.1500 [hep-ph]}

\begin{abstract}
The composition of the $\eta$ and $\eta'$ mesons has long been a source of discussion and is of current interest with new experimental results appearing.  We investigate what can be learnt from a number of different processes: $V \rightarrow P \gamma$ and $P \rightarrow V \gamma$ ($V$ and $P$ are light vector and pseudoscalar mesons respectively), $P \rightarrow \gamma \gamma$, $J/\psi,\psi' \rightarrow P \gamma$, $J/\psi,\psi' \rightarrow P V$, and $\chi_{c0,2} \rightarrow PP$.  These constrain the $\eta-\eta'$ mixing angle to a consistent value, $\phi \approx 42\degrees$; we find that the $c\bar{c}$ components are $\lesssim 5\%$ in amplitude.  We also find that, while the data hint at a small gluonic component in the $\eta'$, the conclusions depend sensitively on unknown form factors associated with exclusive dynamics.  In addition, we predict $BR(\psi' \rightarrow \eta' \gamma) \approx 1 \times 10^{-5}$ and $BR(\chi_{c0} \rightarrow \eta \eta') \approx 2\times 10^{-5} - 1\times 10^{-4}$.  We provide a method to test the mixing using $\chi_{c2} \rightarrow \eta \eta,\  \eta' \eta'\ \text{and}\ \eta \eta'$ modes and make some general observations on $\chi_{c0,2}$ decays.  We also survey the semileptonic and hadronic decays of bottom and charmed mesons and find some modes where the mixing angle can be extracted cleanly with the current experimental data, some where more data will allow this, and some where a more detailed knowledge of the different amplitudes is required.
\end{abstract}

\maketitle

\section{Introduction}
\label{sec:Introduction}

Although the vector mesons are almost ideally mixed, this is not the case for the light pseudoscalar mesons $\eta$ and $\eta'$ and there has long been uncertainty in their composition.  Motivated by the recent papers on the glue content of the $\eta'$ by KLOE\cite{Ambrosino:2006gk} and Li et.\ al.\cite{Li:2007ky}, we set out to determine the gluonic composition and mixing angle from a range of different processes.  While this work was under completion, an additional paper appeared by Escribano and Nadal\cite{Escribano:2007cd} which highlights the current interest in this topic.  We will compare our results with these papers in later sections.

In this paper we will investigate what can be learnt phenomenologically about the $\eta$ and $\eta'$ mesons from a range of processes.  We will test whether these can be explained consistently using one mixing angle or whether any glue content is required.  We use experimental data from the PDG Review 2006\cite{PDG06} as a basis and consider more recent data where it is available.  We will discuss why we think that the analysis of the glue content of the $\eta'$ by KLOE\cite{Ambrosino:2006gk} is inconsistent and independently reach a similar conclusion to Escribano and Nadal\cite{Escribano:2007cd}.  We highlight how the conclusions depend on the data and theoretical assumptions and clarify the logic to show where more theoretical control and experimental data are needed.

The physical $\eta$ and $\eta'$ wavefunction can be written in general as: 
\begin{eqnarray}
\left|\eta\right> &=& X_{\eta} \left|n\bar{n}\right> + Y_{\eta} \left|s\bar{s}\right> + Z_{\eta} \left|G\right>\\
\left|\eta'\right> &=& X_{\eta'} \left|n\bar{n}\right> + Y_{\eta'} \left|s\bar{s}\right> + Z_{\eta'} \left|G\right>
\end{eqnarray}
where $\state{G}$ represents any intrinsic glue component in the wavefunctions, radially excited states and $\state{c\bar{c}}$ components, and $\state{n\bar{n}} \equiv \frac{1}{\sqrt{2}} (\state{u\bar{u}} + \state{d\bar{d}})$.

If we assume that the $\eta$ and $\eta'$ mesons consist only of the lowest level $n\bar{n}$ and $s\bar{s}$ states we can parameterise the mixing in terms of one angle $\phi$:
\begin{eqnarray}
X_{\eta} = \cos \phi && Y_{\eta} = -\sin \phi \\
X_{\eta'} = \sin \phi && Y_{\eta'} = \cos \phi
\end{eqnarray}

Alternatively, allowing for some glue component in the $\eta$ and $\eta'$, we can parameterise the mixing in terms of $\phi$, $\phi_{G1}$ and $\phi_{G2}$:
\begin{eqnarray}
X_{\eta} &=& \cos \phi \cos \phi_{G1} - \sin \phi \sin \phi_{G1} \sin \phi_{G2} \nonumber \\
Y_{\eta} &=& -\sin \phi \cos \phi_{G1} - \cos \phi \sin \phi_{G1} \sin \phi_{G2} \nonumber \\
Z_{\eta} &=& \sin \phi_{G1} \cos \phi_{G2} \\
X_{\eta'} &=& \sin \phi \cos \phi_{G2} \nonumber \\ 
Y_{\eta'} &=& \cos \phi \cos \phi_{G2} \nonumber \\ 
Z_{\eta'} &=& \sin \phi_{G2} 
\end{eqnarray}
In this case we would expect another pseudoscalar meson with $n\bar{n}$, $s\bar{s}$ and gluonic components.  Other pseudoscalars are known, for example the $\eta(1295)$ and the $\eta(1405/1475)$.  There has been much discussion in the literature as to whether these are conventional radially excited mesons, glueballs or a mixture (see for example Refs.\ \cite{rosner:1982ey,Haber:1985cv,Seiden:1988rr,Klempt:2004xg,Close:1996yc,Li:2007ky}).  Processes involving the heavier $\eta^*$ such as $\eta^* \rightarrow \gamma \rho, \gamma \omega, \gamma \phi$ possible in $\psi \rightarrow \gamma \gamma V$ would give more information\cite{Close:2004ip}.  The purpose of this paper is to focus on the $\eta$ and $\eta'$ where there are data.

The mixing reduces to a simpler form when we only allow glue in the $\eta'$:
\begin{eqnarray}
X_{\eta} = \cos \phi & Y_{\eta} = -\sin \phi & Z_{\eta} = 0 \\
X_{\eta'} = \sin \phi \cos \phi_{G2} & Y_{\eta'} = \cos \phi \cos \phi_{G2} & Z_{\eta'} = \sin \phi_{G2} 
\end{eqnarray}

In contrast, the vector mesons are almost ideally mixed, i.e. the $\omega$ is mostly $n\bar{n}$ and the $\phi$ is mostly $s\bar{s}$.  We parametrise the mixing by an angle $\theta_V$:
\begin{eqnarray}
\left|\omega\right> &=& \cos \theta_V \left|n\bar{n}\right> - \sin \theta_V \left|s\bar{s}\right> \\
\left|\phi\right> &=& \sin \theta_V \left|n\bar{n}\right> + \cos \theta_V \left|s\bar{s}\right>
\end{eqnarray}
Ideal mixing corresponds to $\theta_V = 0$ and so we expect $\theta_V$ to be small.  We do not expect this mixing to significantly change the results but will discuss this below.

Above we have used the quark basis to describe the mixing.  Another basis used in the literature is the singlet-octet basis:
\begin{eqnarray}
\eta_8 &\equiv& \frac{1}{\sqrt{6}}\left(\state{u\bar{u}} + \state{d\bar{d}} - 2\state{s\bar{s}} \right) \\
\eta_1 &\equiv& \frac{1}{\sqrt{3}}\left(\state{u\bar{u}} + \state{d\bar{d}} + \state{s\bar{s}} \right)
\end{eqnarray}
 The $\eta$ and $\eta'$ mixing is then parameterised by angle $\theta_p = \phi - \arctan{\sqrt{2}}$:
\begin{eqnarray}
\state{\eta} &=& \cos \theta_P \state{\eta_8} - \sin \theta_P \state{\eta_1} \\
\state{\eta'} &=& \sin \theta_P \state{\eta_8} + \cos \theta_P \state{\eta_1}
\end{eqnarray}

In this work we will consistently use the quark flavour mixing basis rather than the singlet-octet mixing basis.  It has been shown in the literature (see for example Refs.\ \cite{Feldmann:1998vh,Feldmann:1998sh,escribano:1999nh,escribano:2005qq,escribano:2005ci}) that for some processes two mixing angles are required in singlet-octet basis but experimentally only one is required in the quark basis (in particular $\pi,\eta,\eta' \rightarrow \gamma \gamma$).  This is expected if the quark basis is the relevant basis with different decay constants for strange and non-strange quarks.  If this is transformed to the singlet-octet basis there is not a simple parametrisation in terms of singlet and octet decay constants and one angle: then two mixing angles are required to describe the data.  In addition, the determination of mixing angle is more consistent in the quark basis and the quark basis provides a clear physical interpretation and extension to excited states.  In the framework of large $N_C$ (number of colours) chiral perturbation theory, the difference in mixing angles in the singlet-octet basis is due to SU(3) breaking effects and so is reasonably significant.  In the quark flavour basis the mixing angles differences are due to OZI violating effects and so expected to be smaller\cite{Kaiser:1998ds,Kaiser:2000gs,Email:RafelEscribano}.

Throughout this work we will assume that there is no isospin violation ($m_d = m_u$ and there is no isosinglet - isotriplet mixing in either the vector or pseusdoscalar mesons).  Isospin symmetry breaking was found to be small by Kroll \cite{kroll:2005sd,kroll:2004rs}.

The composition of the $\eta$ and $\eta'$ mesons has been studied in many places in the literature.  We give specific references in the following sections, for now we note that Rosner\cite{rosner:1982ey}, Gilman and Kauffman\cite{Gilman:1987ax}, Ball et.\ al.\cite{ball:1995zv}, Bramon et.\ al.\ \cite{Bramon:1997va} and Feldmann et.\ al.\cite{Feldmann:1998vh,Feldmann:1998sh} have reviewed and studied this puzzle with a number of different processes.

The mixing angle $\phi$ has been determined in many places in the literature using different modes and under different assumptions.  These range from $\approx33\degrees$ to $\approx42\degrees$.  For example (in chronological order): $34.7\degrees$ to $37.7\degrees$\cite{ball:1995zv}, $39.2\pm1.3\degrees$\cite{Bramon:1997va}, $37.8\pm1.7\degrees$\cite{Bramon:1997mf}, $39.3\pm1.0\degrees$\cite{Feldmann:1998vh}, $39.8\pm1.8\degrees$\cite{Cao:1999fs}, $37.7\pm2.4\degrees$\cite{bramon:2000fr}, $32.7\pm1\degrees$\cite{Gerard:2004gx}, $39.0\pm1.7\degrees$ and $40.8\pm0.9\degrees$\cite{escribano:2005qq}, $41.8\pm1.2\degrees$ and $41.5\pm1.2\degrees$\cite{escribano:2005ci},  $41.5\pm2.2\degrees$ and $41.2\pm1.1\degrees$\cite{kroll:2005sd}, $38.6\degrees$\cite{Xiao:2005af}, and $38.0\pm1.0\pm2.0\degrees$\cite{Huang:2006as}.  These tend to favour a mixing angle of $\approx 40\degrees$ while leaving uncertainty on precision and error.

Our approach is first to make the conservative assumption that $\eta$ and $\eta'$ only consist of ground state light quarks and test whether this is consistent with the data.  We will then investigate whether any glue is allowed or required by the data.

We start in Section \ref{sec:RadDecays} with the radiative decays of light mesons and continue with the decays of light pseudoscalar mesons to two photons in Section \ref{sec:TwoPhoton}.  We then discuss decays of the $J/\psi$ and $\psi'$: the radiative decays in to light pseudoscalar mesons in Section \ref{sec:PsiGamma} and the strong decays in to a vector and a pseudoscalar meson in Section \ref{sec:PsiVP}.  We discuss the strong decays of the $\chi_{c0}$ and $\chi_{c2}$ in to light pseudoscalars in Section \ref{sec:ChiC02PP}.  In Section \ref{sec:BDecays} we review what can be learnt from bottom and charmed meson semileptonic and hadronic decays.  We close with conclusions and general remarks in Section \ref{sec:Conclusions}.

\section{Radiative Decays of Light Mesons}
\label{sec:RadDecays}

One window on the composition of the $\eta$ and $\eta'$ are the radiative (magnetic dipole) decays of light vector and pseudoscalar mesons.  There are many studies of these decays in the literature (for example in Refs.\ \cite{O'Donnell:1981sj},\cite{rosner:1982ey},\cite{Gilman:1987ax},\cite{ball:1995zv},\cite{Feldmann:1998vh,Feldmann:1998sh},\cite{escribano:1999nh,escribano:2005qq,escribano:2005ci},\cite{bramon:2000fr},\cite{nasrallah:2004ms},\cite{kroll:2004rs,kroll:2005sd},\cite{Ambrosino:2006gk}).  The most recent study by Escribano and Nadal\cite{Escribano:2007cd} appeared while this work was under completion.

We study these decays using a quark model where the matrix elements of a transition $X \rightarrow Y \gamma$ are given by:
\begin{equation}
M(X \rightarrow Y \gamma) = <Y|\sum_{i=1}^{2}\mu_{qi} e_{qi} \sigma_i \cdot \epsilon|X>
\end{equation}
where $\mu_{qi}$ is the effective magnetic moment of the quark ($\mu_u = \mu_d$) \emph{after its relative charge has been removed}, $e_{qi}$ is its charge, $\sigma_i$ are the Pauli matrices and $\epsilon$ is the polarisation of the photon emitted.  Following Bramon et.\ al.\cite{bramon:2000fr} and Escribano and Nadal\cite{Escribano:2007cd} we introduce parameters $C_X$ to account for different wavefunction overlaps: $C_s$ for $\left<s\bar{s}|s\bar{s}\right>$, $C_q$ for $\left<n\bar{n}|u\bar{u}\right>=\left<n\bar{n}|d\bar{d}\right>$, $C_{\pi}$ for $\left<\pi|u\bar{u}\right>=-\left<\pi|d\bar{d}\right>$ and $C_K$ for $\left<K|K^*\right>$.  We only consider ratios of branching fractions and so need ratios of wavefunction overlap parameters.  We consider the kaon modes separately and so do not require $C_K$ explicitly.

The decay rate is then:
\begin{equation}
\Gamma \propto S |A|^2 \mvp^3
\end{equation}
where $\mvp$ is the recoil momentum of the final state meson in the rest frame of the initial meson, $S$ is a spin factor (1 for vector to pseudoscalar transitions and 3 for pseudoscalar to vector transitions), and $A$ is the matrix element with the momentum factors removed:  
\begin{eqnarray}
\label{equ:amplitudesstart}
A(\phi \rightarrow \pi^0 \gamma) &\propto& \frac{1}{2} C_{\pi} \mu_d \sin\theta_V  \nonumber \\
A(\phi \rightarrow \eta \gamma) &\propto& \frac{1}{3} C_s \mu_s Y_{\eta}\cos\theta_V - \frac{1}{6} C_q X_{\eta}\sin\theta_V \nonumber \\
A(\phi \rightarrow \eta' \gamma) &\propto& \frac{1}{3} C_s \mu_s Y_{\eta'}\cos\theta_V - \frac{1}{6} C_q X_{\eta'}\sin\theta_V \nonumber \\
A(\omega \rightarrow \pi^0 \gamma) &\propto& \frac{1}{2} C_{\pi} \mu_d \cos\theta_V \nonumber \\
A(\omega \rightarrow \eta \gamma) &\propto& - \frac{1}{6} C_q \mu_d X_{\eta}\cos\theta_V + \frac{1}{3} C_s \mu_s Y_{\eta}\sin\theta_V \nonumber \\
A(\rho^0 \rightarrow \pi^0 \gamma) &\propto& - \frac{1}{6} C_{\pi} \mu_d \nonumber \\
A(\rho^0 \rightarrow \eta \gamma) &\propto& \frac{1}{2} C_q \mu_d X_{\eta} \nonumber \\
A(\eta' \rightarrow \rho^0 \gamma) &\propto& \frac{1}{2} C_q \mu_d X_{\eta'} \nonumber \\
A(\eta' \rightarrow \omega \gamma) &\propto& -\frac{1}{6} C_q \mu_d X_{\eta'}\cos\theta_V + \frac{1}{3} C_s \mu_s Y_{\eta'}\sin\theta_V \nonumber \\
A(K^{0*} \rightarrow K^0 \gamma) &\propto& \frac{1}{6} C_K (\mu_s + \mu_d) \nonumber \\
A(K^{+*} \rightarrow K^+ \gamma) &\propto& \frac{1}{6} C_K (\mu_s - 2 \mu_d) 
\label{equ:amplitudesend}
\end{eqnarray}

Experimental data are given in Table \ref{table:RadDecayExperimentalData}.  KLOE\cite{Ambrosino:2006gk} have measured the ratio $\Gamma(\phi\rightarrow\eta'\gamma)/\Gamma(\phi\rightarrow\eta\gamma)$ with a smaller uncertainty than the current world average.  This can then be used with the PDG Review 2006 value for $\Gamma(\phi\rightarrow\eta\gamma)$ to obtain a more accurate $BR(\phi\rightarrow\eta'\gamma)$ consistent with the PDG value but with a smaller uncertainty.  They calculate  $BR(\phi\rightarrow\eta'\gamma) = (6.20\pm0.27)\times 10^{-5}$ (the uncertainty excludes that from $BR(\phi\rightarrow\eta\gamma)$) compared to the PDG value of $(6.2\pm0.7)\times 10^{-5}$.

There is also recent data from SND\cite{Achasov:2006dv} which are not included in the PDG Review 2006.  These SND results are not by themselves more constraining than the current world average.  We do not expect them to significantly improve the mixing parameter determinations, at least until they are included in the average.  Therefore we will not include these results but will comment on the effect of them later.

\begin{table}[tb]
\begin{center}
\begin{tabular}{|c|c|}
\hline
\textbf{Mode} & \textbf{$\rr \equiv \Gamma/\mvp^3$} \\
\hline
$\phi \rightarrow \pi^0 \gamma$ & $(4.24\pm0.24)\times 10^{-11}$ \\
$\phi \rightarrow \eta \gamma$ & $(1.159\pm0.025)\times 10^{-9}$ \\
$\phi \rightarrow \eta' \gamma$ & $(1.22\pm0.14)\times 10^{-9}$ \\
$\omega \rightarrow \pi^0 \gamma$ & $(1.377\pm0.041)\times 10^{-8}$ \\
$\omega \rightarrow \eta \gamma$ & $(5.20\pm0.53)\times 10^{-10}$ \\
$\rho^0 \rightarrow \pi^0 \gamma$ & $(1.69\pm0.23)\times 10^{-9}$ \\
$\rho^0 \rightarrow \eta \gamma$ & $(6.04\pm0.62)\times 10^{-9}$ \\
\hline
$\eta' \rightarrow \rho^0 \gamma$ & $(1.33\pm0.11)\times 10^{-8}$ \\
$\eta' \rightarrow \omega \gamma$ & $(1.53\pm0.20)\times 10^{-9}$ \\
\hline
$K^{*0} \rightarrow K^0 \gamma$ & $(4.06\pm0.36)\times 10^{-9}$ \\
$K^{*+} \rightarrow K^+ \gamma$ & $(1.69\pm0.15)\times 10^{-9}$ \\\hline
\hline
$\phi \rightarrow \eta' \gamma$ & $(1.22\pm0.06)\times 10^{-9}$\cite{Ambrosino:2006gk} \\
$\phi \rightarrow \eta \gamma$ & $(1.21\pm0.04)\times 10^{-9}$\cite{Achasov:2006dv} \\
$\omega \rightarrow \eta \gamma$ & $(4.91\pm0.51)\times 10^{-10}$\cite{Achasov:2006dv} \\
$\rho^0 \rightarrow \eta \gamma$ & $(4.91\pm0.53)\times 10^{-9}$\cite{Achasov:2006dv} \\
\hline
\end{tabular}
\caption{Experimental reduced partial widths ($\Gamma/\mvp^3$) for radiative transitions of light vector and pseudoscalar mesons.  (From the PDG Review 2006\cite{PDG06} unless otherwise stated.)}
\label{table:RadDecayExperimentalData}
\end{center}
\end{table}

\subsection{The Magnetic Moment Ratio $\mu_s/\mu_d$}
\label{sec:MagneticMomentRatio}

To calculate the partial widths we require the effective ratio of magnetic moments $\mu_s/\mu_d$ which we obtain from the radiative decays of the charged and neutral vector kaons.  This gives $\mu_s / \mu_d = 0.82 \pm 0.05$.  This value is consistent with the $0.805$ obtained by Gilman and Kauffman\cite{Gilman:1987ax}.  

The magnetic moments are inversely proportional to the (constituent) quark masses.  Our value is consistent with the effective mass ratio $\bar{m}/m_s = 0.81\pm0.05$ found in Ref.\ \cite{bramon:2000fr} and used by KLOE in Ref.\ \cite{Ambrosino:2006gk}.  It is interesting that Feldman et.\ al.\cite{Feldmann:1998vh} (Equ.\ 3.12) obtain an effective value of $\mu_s / \mu_d = 0.81$ using decay constants which is also consistent with our result.

We obtain a consistent result, $\approx 0.77$, if we use the $\phi$ and $\omega$ meson masses.  If we take $m_d = m_u = 350 \text{MeV}$ and $m_s  = 500 \text{MeV}$ we obtain $\mu_s / \mu_d \approx 0.7$.  Karliner and Lipkin \cite{Karliner:2006fr} consider baryon and meson masses and get $\mu_s / \mu_d \approx 0.62-0.65$ for baryons and $\mu_s / \mu_d \approx 0.56 - 0.63$ for mesons, significantly smaller than ours.  If we set $\mu_s / \mu_d = 0.63$ we obtain $\rr(K^{*0} \rightarrow K^0 \gamma) / \rr (K^{*+} \rightarrow K^+ \gamma) = 1.4$, significantly smaller than the experimental ratio of $2.4\pm0.3$\cite{PDG06}.  This might be expected: the strange-nonstrange effective mass ratio relevant for hadron masses is different from the effective magnetic moment ratio (effective mass ratio) which measures the relative couplings of strange/nonstrange quarks to a photon in a magnetic dipole transition.

We will take $\mu_s / \mu_d = 0.82 \pm 0.05$ from here onward.

\subsection{Results}
\label{sec:RadDecayResults}

Because of the lightness and ambiguous constitution of the pion, we choose first not to use pionic modes to determine the mixing parameters, but we shall return to this point later.

Before we fit the data, we investigate what we can learn from different ratios of branching ratios and give the results in Table \ref{table:MixingParamsRatios}.  We first choose ratios where the $C_X$ factors cancel.  The differences between (1) and (2) show that vector meson mixing produces a systematic change in the angle extracted but that this effect is not very significant.  The ratio  $\rho^0 \rightarrow \pi^0 \gamma / \omega \rightarrow \pi^0 \gamma$ is $0.122 \pm 0.017$ which is consistent with the predicted value of $1/9$ for ideal vector meson mixing.

\begin{table}[htb]
\begin{center}
\begin{tabular}{|c|c|c|c|}
\hline
 & \textbf{Assumptions} & \textbf{Modes} & \textbf{Mixing Parameters} \\
\hline
(1) & No gluonium & $\frac{\eta'\rightarrow\rho\gamma}{\rho\rightarrow\eta\gamma}$ & $\phi = (40.6 \pm 1.9)\degrees$ \\
\hline
(2) & No gluonium \& $\theta_V = 0$ & $\frac{\eta' \rightarrow \omega\gamma}{\omega\rightarrow\eta\gamma}$ & $\phi = (45 \pm 2)\degrees$ \\
\cline{3-4}
 & & $\frac{\phi\rightarrow\eta\gamma}{\phi\rightarrow\eta'\gamma}$ & $\phi = (44.2 \pm 1.7)\degrees$ \\
\cline{4-4}
 & & & (K) $\phi=44.2\degrees\pm0.7\degrees$ \\
\hline
(3) & Only glue in $\eta'$ \& $\theta_V = 0$ & $\frac{\eta' \rightarrow \omega\gamma}{\omega\rightarrow\eta\gamma}$ and $\frac{\phi\rightarrow\eta\gamma}{\phi\rightarrow\eta'\gamma}$ & $\phi=44.5\degrees$ \\
 & & & $\cos \phi_{G2} = 1.009\pm0.051$ \\
\cline{4-4}
 & & & (K) $\phi=44.5\degrees$ \\
 & & & $\cos\phi_{G2} = 1.009\pm0.043$ \\
\cline{3-4}
 & & $\frac{\eta' \rightarrow \rho\gamma}{\rho\rightarrow\eta\gamma}$ and $\frac{\phi\rightarrow\eta\gamma}{\phi\rightarrow\eta'\gamma}$ & $\phi=42.4\degrees$ \\
 & & & $\cos\phi_{G2} = 0.938\pm0.04$ \\
\cline{4-4}
 & & & (K) $\phi=42.4\degrees$ \\
 & & & $\cos\phi_{G2} = 0.938\pm0.03$ \\
\hline
(4) & $C_s = C_q$ \& $\theta_V = 0$ & $\frac{\eta' \rightarrow \rho^0\gamma}{\phi \rightarrow \eta'\gamma}$ & $\phi = (46.3 \pm 2.0)\degrees$ \\
\cline{4-4}
 & & & (K) $\phi = (46.3 \pm 1.4)\degrees$ \\
\cline{3-4}
 & & $\frac{\eta' \rightarrow \omega\gamma}{\phi \rightarrow \eta'\gamma}$ & $\phi = (46.8 \pm 2.5)\degrees$ \\
\cline{4-4}
 & & & (K) $\phi = (46.8 \pm 2.0)\degrees$ \\
\hline
(5) & $C_s = C_q$ \& Only glue in $\eta'$ \& $\theta_V = 0$ & $\frac{\phi \rightarrow \eta\gamma}{\rho^0 \rightarrow \eta\gamma}$ & $\phi = (38.6 \pm 1.5)\degrees$ \\
\cline{3-4}
 & & $\frac{\phi \rightarrow \eta\gamma}{\omega \rightarrow \eta\gamma}$ & $\phi = (42.2 \pm 1.5)\degrees$ \\
\hline
\end{tabular}
\caption{Extracting mixing parameters from ratios of branching ratios.  (K) indicates that the new KLOE result\cite{Ambrosino:2006gk} has been used.}
\label{table:MixingParamsRatios}
\end{center}
\end{table}

When we allow the $\eta'$ to have some gluonic component the data are consistent with no glue in the $\eta'$.  An estimate of the error due to ignoring vector meson mixing can be obtained by comparing the two sets of modes in (3).  These results are consistent and show that the mixing angle extracted is not very sensitive to any vector meson mixing.

If we allow gluonium mixing in both the $\eta$ and the $\eta'$, we can not extract the mixing angles without making an assumption about $C_s$ and $C_q$: there are at least four free parameters and only three independent ratios.

We now assume that $C_q = C_s$ which is consistent with the results of Bramon et.\ al.\cite{bramon:2000fr} where they find that $C_s = 0.89\pm0.07$ and $C_q=0.91\pm0.05$.  We also assume no vector meson mixing, consistent with the above results.  We can then extract the mixing angle where any possible gluonic mixing cancels as shown in Table \ref{table:MixingParamsRatios} (4) and from other modes if we only allow the $\eta'$ to have a gluonic component (5).  There is reasonable agreement between these four different determinations, but they do not all agree within the experimental uncertainties.  The new KLOE data decreases the experimental uncertainty and so increases the significance of this discrepancy.  The $\eta$ ratios produce a systematically lower angle than the $\eta'$ ratios.  This should not be due to gluonic mixing of the $\eta'$ because we have chosen ratios where this cancels but could be because of different wavefunction overlaps or due to form factors.

We now perform fits to the data and give the results in Table \ref{table:RadDecayResults}.  We first fit with no gluonic mixing (i.e. one mixing angle $\phi$ and one overall normalisation) and $C_q = C_s$.  The fit is not very good reflecting the discrepancy in the ratios seen above.  If we use the new KLOE result we get a significantly worse fit because of the smaller experimental uncertainty, and a large mixing angle.

We then repeat the fit keeping $C_q = C_s$ but allowing some glue in the $\eta'$.  With both sets of data the fit has improved slightly, but not significantly given that an additional parameter has been included.  The mixing angle has not changed significantly and $\cos{\phi_{G2}}$ is consistent with unity.  Then we do not allow any gluonic component but relax the constraint that $C_q = C_s$.  For the first time, we get a reasonable fit which shows the significance of the wavefunction overlaps.  The mixing angle has not been affected significantly by relaxing the constraint.  Finally, we relax both constraints and do not get a significantly better fit.  We again find that $\cos{\phi_{G2}}$ is consistent with unity.

\begin{table}
\begin{center}
\begin{tabular}{|c|c|c|c|c|}
\hline
\textbf{$\theta_V/\degrees$} & \textbf{$\phi/\degrees$} & \textbf{$\cos{\phi_{G2}}$} & \textbf{$C_s/C_q$} & $\chi^2/(\text{d.o.f.})$ \\
\hline
$0$ & $41.5 \pm 1.5$ & $(\equiv 1)$ & $(\equiv 1)$ & $11/4$ \\
$0$ & $42.7 \pm 1.8$ & $1.03 \pm 0.06$ & $(\equiv 1)$ & $10/3$ \\
$0$ & $43.5 \pm 1.1$ & $(\equiv 1)$ & $0.92 \pm 0.03$ & $3.4/3$ \\
$0$ & $43.6 \pm 1.4$ & $0.98 \pm 0.05$ & $0.91 \pm 0.04$ & $3.1/2$ \\
\hline
$3.4$ & $39.8 \pm 1.5$ & $(\equiv 1)$ & $(\equiv 1)$ & $11/4$ \\
$3.4$ & $40.0 \pm 1.9$ & $1.02 \pm 0.07$ & $(\equiv 1)$ & $11/3$ \\
$3.4$ & $41.3 \pm 0.8$ & $(\equiv )1$ & $0.90\pm 0.02$ & $1.9/3$ \\
$3.4$ & $41.3 \pm 0.9$ & $0.98 \pm 0.03$ & $0.89 \pm 0.03$ & $1.4/2$ \\
\hline
$2$ & $40.0 \pm 1.5$ & $(\equiv 1)$ & $(\equiv 1)$ & $11/4$ \\
$4$ & $39.4 \pm 1.6$ & $(\equiv 1)$ & $(\equiv 1)$ & $12/4$ \\
\hline
\hline
 & \multicolumn{3}{c}{Including new KLOE result:} & \\
\hline
$0$ & $45.2 \pm 1.5$ & $(\equiv 1)$ & $(\equiv 1)$ & $20/4$ \\
$0$ & $43.2 \pm 2.0$ & $0.99 \pm 0.07$ & $(\equiv 1)$ & $14/3$ \\
$0$ & $44.0 \pm 0.7$ & $(\equiv 1)$ & $0.91 \pm 0.03$ & $3.7/3$ \\
$0$ & $43.6 \pm 1.0$ & $ 0.98 \pm 0.03$ & $0.91\pm0.03$ & $3.1/2$ \\
\hline
$3.4$ & $41.1 \pm 1.2$ & $(\equiv 1)$ & $(\equiv 1)$ & $16/4$ \\
$3.4$ & $41.2 \pm 2.1$ & $0.99 \pm 0.07$ & $(\equiv 1)$ & $16/3$ \\
$3.4$ & $41.7 \pm 0.5$ & $(\equiv 1)$ & $0.90\pm 0.02$ & $2/3$ \\
$3.4$ & $41.3 \pm 0.7$ & $0.98 \pm 0.02$ & $0.89 \pm 0.02$ & $1.4/2$ \\
\hline
\end{tabular}
\caption{Fits to the data without a form factor}
\label{table:RadDecayResults}
\end{center}
\end{table}

We consider vector meson mixing and take $\theta_V = 3.4\degrees$ from Refs.\ \cite{bramon:2000fr,Ambrosino:2006gk} (compatible with a simple determination from the ratio $\phi \rightarrow \pi^0 \gamma$ to $\omega \rightarrow \pi^0 \gamma$ giving $\theta_V = 3.2 \pm 0.1\degrees$).  The results of the fits are shown in Table \ref{table:RadDecayResults}.  The same pattern as when we ignored vector meson mixing is observed.  The mixing angle $\phi$ is systematically a few degrees lower.  The fits for $C_s = C_q$ are poor and the fits for $C_s \neq C_q$ are good and slightly better with vector meson mixing included.  We again find results consistent with no glue in the $\eta'$. 

To further check the effect of vector meson mixing, we vary $\theta_V$ from $2\degrees$ to $4\degrees$ and require $\cos{\phi_{G1}} = 1$ and $C_s = C_q$.  The mixing angle and goodness of fit are not significantly affected by these changes, which show that our results are not sensitive to the exact vector meson mixing angle.

In all the cases considered the results are consistent with the $\eta'$ having no gluonic constituent and the wave function overlaps play a significant role.  The best fit is obtained with $\theta_V = 3.4\degrees$ and $\cos{\phi_{G2}} \equiv 1$: $\phi = (41.3 \pm 0.8)\degrees$ and $C_s/C_q = 0.90 \pm 0.02$ (or $\phi = (41.7 \pm 0.5)\degrees$ and $C_s/C_q = 0.90 \pm 0.02$ if the new KLOE result is included).  The effect of vector meson mixing does not alter our conclusions and systematically shifts the mixing angle slightly.  The new KLOE data decreases the experimental uncertainty and does not significantly change the results.

The recent SND results taken together with that from KLOE are also consistent with no glue in the $\eta'$ and there is no significant change in the results.  Performing a fit with $\theta_V = 3.4\degrees$ and $\cos{\phi_{G2}} \equiv 1$ gives $\phi = (42.6 \pm 0.5)\degrees$ and $C_s/C_q = 0.93 \pm 0.02 $ with $\chi^2/(\text{d.o.f.})=1.3/3$.

\subsection{Form Factors}

For exclusive processes there is a momentum dependent penalty.  We investigate the effect of this by using phenomenological Gaussians as from simple harmonic oscillator wave functions, for example Ref.\ \cite{Godfrey:1985xj}.  This adds an additional momentum dependent factor to the rate $\Gamma$ of:
\begin{equation}
\exp(-\mvp^2/(8\beta^2)) .
\end{equation}

Godfrey and Isgur\cite{Godfrey:1985xj} take $\beta=400\text{MeV}$ while Ref.\ \cite{JoThesis} find $\beta=344\text{MeV}$ for $n\bar{n}$ and $\beta=426\text{MeV}$ for $s\bar{s}$.  We use a few different values of $\beta$ to check the sensitivity to this parameter ($300\MeV$, $400\MeV$, $500\MeV$).

We fit the data as before and give the results in Table \ref{table:RadDecayResultsFormFactors}.  Again the fits are not very good if we require that $C_s = C_q$.  However, now the fits slightly favour $\cos{\phi_{G2}} \neq 1$.  The significance of this increases for smaller $\beta$ and with the new KLOE result.   It is difficult to tell whether this is some artifact of the form factor or the low number of degrees of freedom, particularly because it gets more significant for small $\beta$.  This gives an indication of the theoretical uncertainty due to the detailed dynamics.

For comparison, fitting the recent SND results taken together with that from KLOE ($\theta_V = 3.4\degrees$ and $\beta=400\MeV$) gives $\phi = (43.1 \pm 0.7)\degrees$, $\cos{\phi_{G2}} = 0.97 \pm 0.02$ and $C_s/C_q = 0.94 \pm 0.02$ with $\chi^2/(\text{d.o.f.})=1.3/2$.  There is no significant change in the mixing angle but $\phi_{G2}$ is sensitive to the experimental data as well as the form factor.

\begin{table}
\begin{center}
\begin{tabular}{|c|c|c|c|c|}
\hline
\textbf{$\beta/\text{MeV}$} & \textbf{$\phi/\degrees$} & \textbf{$\cos{\phi_{G2}}$} & $C_s/C_q$ & $\chi^2/(\text{d.o.f.})$ \\
\hline
No form factor & $39.8\pm1.5$ & $(\equiv 1)$ & $(\equiv 1)$ & $11/4$ \\
No form factor & $40.0\pm1.9$ & $1.02\pm0.07$ & $(\equiv 1)$ & $11/3$ \\
No form factor & $41.3 \pm 0.8$ & $(\equiv )1$ & $0.90\pm 0.02$ & $1.8/3$ \\
No form factor & $41.3 \pm 0.9$ & $0.98 \pm 0.03$ & $0.89 \pm 0.03$ & $1.4/2$ \\
\hline
$300$ & $40.9\pm1.4$ & $(\equiv 1)$ & $(\equiv 1)$ & $10/4$ \\
$300$ & $41.4\pm1.5$ & $0.96\pm0.05$ & $(\equiv 1)$ & $7/3$ \\
$300$ & $42.4\pm1.4$ & $(\equiv 1)$ & $0.95\pm0.04$ & $6/3$ \\
$300$ & $42.5\pm0.9$ & $0.93 \pm 0.03$ & $0.91 \pm 0.03$ & $1.3/2$ \\
\hline
$400$ & $40.0\pm1.4$ & $(\equiv 1)$ & $(\equiv 1)$ & $10/4$ \\
$400$ & $40.8\pm1.7$ & $0.99\pm0.06$ & $(\equiv 1)$ & $9/3$ \\
$400$ & $41.9\pm1.1$ & $(\equiv 1)$ & $0.93\pm0.03$ & $4/3$ \\
$400$ & $42.0\pm0.9$ & $0.95 \pm 0.03$ & $0.90 \pm 0.03$ & $1.3/2$ \\
\hline
$500$ & $40.5\pm1.4$ & $(\equiv 1)$ & $(\equiv 1)$ & $9/4$ \\
$500$ & $40.5\pm1.7$ & $1.00\pm0.06$ & $(\equiv 1)$ & $9/3$ \\
$500$ & $41.7\pm1.0$ & $(\equiv 1)$ & $0.92\pm0.03$ & $3/3$ \\
$500$ & $41.8\pm0.9$ & $0.96 \pm 0.03$ & $0.90 \pm 0.03$ & $1.4/2$ \\
\hline
\hline
 & \multicolumn{3}{c}{Including new KLOE result:} & \\
\hline
No form factor & $41.1\pm1.2$ & $(\equiv 1)$ & $(\equiv 1)$ & $16/4$ \\
No form factor & $41.1\pm2.1$ & $0.99\pm0.07$ & $(\equiv 1)$ & $16/3$ \\
No form factor & $41.7 \pm 0.5$ & $(\equiv 1)$ & $0.90\pm 0.02$ & $2/3$ \\
No form factor & $41.3 \pm 0.7$ & $0.98 \pm 0.02$ & $0.89 \pm 0.02$ & $1.4/2$ \\
\hline
$300$ & $43.7\pm1.2$ & $(\equiv 1)$ & $(\equiv 1)$ & $15/4$ \\
$300$ & $42.2\pm1.7$ & $0.93\pm0.05$ & $(\equiv 1)$ & $11/3$ \\
$300$ & $43.8\pm1.0$ & $(\equiv 1)$ & $0.93\pm0.04$ & $9/3$ \\
$300$ & $42.4\pm0.6$ & $0.93 \pm 0.02$ & $0.91 \pm 0.02$ & $1.3/2$ \\
\hline
$400$ & $42.7\pm1.2$ & $(\equiv 1)$ & $(\equiv 1)$ & $15/4$ \\
$400$ & $41.7\pm1.9$ & $0.95\pm0.06$ & $(\equiv 1)$ & $13/3$ \\
$400$ & $42.8\pm0.8$ & $(\equiv 1)$ & $0.91\pm0.03$ & $5/3$ \\
$400$ & $41.9\pm0.7$ & $0.95 \pm 0.02$ & $0.91 \pm 0.02$ & $1.3/2$ \\
\hline
$500$ & $41.9\pm1.2$ & $(\equiv 1)$ & $(\equiv 1)$ & $15/4$ \\
$500$ & $41.5\pm2.0$ & $0.96\pm0.07$ & $(\equiv 1)$ & $14/3$ \\
$500$ & $42.4\pm0.7$ & $(\equiv 1)$ & $0.91\pm0.03$ & $4/3$ \\
$500$ & $41.7\pm0.7$ & $0.96 \pm 0.02$ & $0.90 \pm 0.02$ & $1.4/2$ \\
\hline
\end{tabular}
\caption{Fits to the data when a form factor is included.  ($\theta_V=3.4\degrees$)}
\label{table:RadDecayResultsFormFactors}
\end{center}
\end{table}

\subsection{Pion Modes}
\label{sec:PionModes}

So far we have not discussed the pion modes and so have not been able to normalise both the $\eta$ and $\eta'$ mixing parameters.  Note that we can't simultaneously determine the glue content of the $\eta$ and the $\eta'$ without taking $C_{\pi}$ as an input.  With the assumption that $C_{\pi}=C_q$ we can normalise to the $\omega \rightarrow \pi^0 \gamma$ mode and hence extract the gluonic mixing.  Note that this mode has the largest recoil momentum of any of the modes we consider and so, for any form factor that decreases with increasing recoil momentum, this mode will be suppressed the most.  The apparent gluonic contribution will therefore be larger if we use such a form factor. 

We consider ratios of the $\eta$ and $\eta'$ decay modes with $\omega \rightarrow \pi^0 \gamma$ and then use $|Z_{\eta/\eta'}|^2 = 1 - |X_{\eta/\eta'}|^2 - |Y_{\eta/\eta'}|^2$ to extract the glue content of the mesons.  With ideal vector meson mixing, $C_s = C_q = C_{\pi}$ and no form factor we obtain: $Z_{\eta} = 0.586 \pm 0.016$ and $Z_{\eta'} = 0.62 \pm 0.02$ which translate to $\phi_{G1} = (48 \pm 2)\degrees$ and $\phi_{G2} = (38 \pm 2)\degrees$. 

These results appear to suggest that both the $\eta$ and the $\eta'$ have significant gluonic contributions.  However, we should not trust these determinations based on $C_{\pi} = C_q$ because of the pion's lightness and ambiguous constitution.  Bramon et.\ al.\cite{bramon:2000fr} and Escribano and Nadal\cite{Escribano:2007cd} both find that $C_s$ and $C_q$ are not consistent with $C_{\pi}$.

\subsection{Summary and Comparisons}

A summary of the results is shown in Table \ref{table:RadDecayResultsSummary}.

\begin{table}
\begin{center}
\begin{tabular}{|c|c|c|c|c|}
\hline
  & $\phi / \degrees$ & $\cos{\phi_{G2}}$ & $C_s/C_q$ & $\chi^2/(\text{d.o.f.})$ \\
\hline
No form factor & $41.3 \pm 0.8$ & $(\equiv )1$ & $0.90\pm 0.02$ & $1.8/3$ \\
 & $41.3 \pm 0.9$ & $0.98 \pm 0.03$ & $0.89 \pm 0.03$ & $1.4/2$ \\
\hline
$\beta = 400\MeV$ & $41.9\pm1.1$ & $(\equiv 1)$ & $0.93\pm0.03$ & $4/3$ \\
 & $42.0\pm0.9$ & $0.95 \pm 0.03$ & $0.90 \pm 0.03$ & $1.3/2$ \\
\hline
No form factor, new KLOE result & $41.7 \pm 0.5$ & $(\equiv 1)$ & $0.90\pm 0.02$ & $2/3$ \\ 
 & $41.3 \pm 0.7$ & $0.98 \pm 0.02$ & $0.89 \pm 0.02$ & $1.4/2$\\
\hline
$\beta = 400\MeV$, new KLOE result & $42.8\pm0.8$ & $(\equiv 1)$ & $0.91\pm0.03$ & $5/3$ \\
 & $41.9\pm0.7$ & $0.95 \pm 0.02$ & $0.91 \pm 0.02$ & $1.3/2$ \\
\hline
\end{tabular}
\caption{Summary of radiative decay results. ($\theta_V = 3.4\degrees$) }
\label{table:RadDecayResultsSummary}
\end{center}
\end{table}

From these decays we have found that the exact value of the $\eta$ and $\eta'$ mixing angle depends upon which approach is used, but that this variation is small.  We found that the vector meson mixing has a small effect on $\phi$ and increases the goodness of the fit but the results are not very sensitive to the exact value of $\phi_V$.  In general, the new KLOE and recent SND results give a mixing angle with smaller uncertainty but do not significantly change the results.

The mixing of gluonium in the $\eta$ or $\eta'$ is less clear.  If we assume that only the $\eta'$ has a gluonic component and do not include a form factor, our results are consistent with the gluonic component of the $\eta'$ being zero.  However, if we do include a gaussian form factor the fits slightly favour a small gluonic component in the $\eta'$.  The gluonic component grows when we decrease $\beta$ and with the new KLOE result.  If $\beta=400\MeV$ we get $\phi_{G2} = (18^{+5}_{-7})\degrees$, or $\phi_{G2} = (18\pm4)\degrees$ with the new KLOE result.  The conclusion as to whether there is any gluonic component in the $\eta'$ depends sensitively on the exclusive dynamics summarised by the form factor.

In their recent paper, KLOE\cite{Ambrosino:2006gk} obtained $\phi=(39.8\pm0.8)\degrees$ (no glueball mixing) and $\phi=(41.5^{+0.6}_{-0.7})\degrees$ (glueball mixing), which are consistent with our results.  They also estimated the gluonium fractional content of the $\eta'$ mesons to be $|\sin^2 \phi_{G2}|=0.14\pm0.04$.  This gives $\phi_{G2} = (22^{+3}_{-4})\degrees$, a value compatible with ours when a form factor is included ($\beta=400\MeV$).  However, they obtain this result without using a form factor.  We do not think that their method is consistent: they use wavefunction overlap parameters $C_s$ and $C_q$ from Bramon et.\ al.\cite{bramon:2000fr} where no glueball mixing was assumed and older data was used, and then use these to determine the gluonic component of the $\eta'$.

The recent paper by Escribano and Nadal\cite{Escribano:2007cd} also comments on the inconsistency in KLOE's approach.  They follow Bramon et.\ al.\cite{bramon:2000fr} in introducing the different parameters for the wavefunction overlaps.  They get a bad fit when all these parameters are equal, as we did, and a good fit when these parameters are allowed to vary.  If they don't allow for a gluonium component in the $\eta'$, they obtain $\phi= (41.5 \pm 1.2)\degrees$, $\theta_V = (3.2 \pm0.1)\degrees$ and $\mu_s / \mu_d = 0.81 \pm 0.05$ which are compatible with ours results.  They find that $C_q/C_{\pi}=0.86\pm0.03$ and $C_s/C_{\pi}=0.78\pm0.05$; these are not consistent with $C_s = C_q$ and give $C_s/C_q = 0.91$, again consistent with our results.  When they allow gluonic mixing they find that $\phi = (41.4\pm1.3)\degrees$ and $|\phi_{G2}| = (12\pm13)\degrees$ consistent with no gluonic mixing and our results.  

Although we have referred to the other constituent(s) of the $\eta'$ as being gluonic, they could equally well be anything invisible or suppressed in radiative decays such as a gluonic component, radially excited states or heavier quarks (for example, $c\bar{c}$).  For example, using harmonic oscillator wave functions, the $n \neq 1$ to $n=1$ transitions are suppressed by $>10^{-3}$ and hence the radially excited state contributions will be invisible to radiative decays.

In summary, from these radiative transitions there are tantilising hints of glue (or something else) in the $\eta'$.  However, there is a limit to how well the mixing parameters can be extracted from the data because of lack of knowledge of form factors and this is particularly significant for gluonic (or $c\bar{c}$, etc.) component.  This is exemplified by the different wavefunction overlaps:  previously Bramon et.\ al.\cite{bramon:2000fr} found that $C_q = 0.91 \pm 0.05$ and $C_s = 0.89 \pm 0.07$ which are consistent with being equal whereas with the current experimental data this is not the case.  Prediction of these form factors, for example by lattice QCD, coupled with more experimental data would more strongly constrain the constituents of the enigmatic light pseudoscalar mesons.

\section{Decays in to two photons: $\pi^0, \eta, \eta' \rightarrow \gamma \gamma$}
\label{sec:TwoPhoton}

Another set of modes that can be used to probe the structure of the $\eta$ and $\eta'$ are the decays of the light pseudoscalar mesons in to two photons.  There are many studies of these decays in the literature (for example in Refs.\ \cite{rosner:1982ey},\cite{Gilman:1987ax},\cite{Schechter:1992iz},\cite{ball:1995zv},\cite{Feldmann:1998vh,Feldmann:1998sh},\cite{Cao:1999fs},\cite{escribano:1999nh,escribano:2005qq,escribano:2005ci},\cite{Agaev:2003kb},\cite{kroll:2004rs,kroll:2005sd},\cite{Xiao:2005af},\cite{Kekez:2005ie},\cite{Huang:2006as}).

These decays can be parametrised in terms of dimensionless couplings $g_{P\gamma\gamma}$:
\begin{equation}
\Gamma(P\rightarrow\gamma\gamma) = \frac{g_{P\gamma\gamma}^2 M_P}{32\pi}
\end{equation}
where $M_P$ is the mass of the pseudoscalar meson $P$.

We use the PDG Review 2006\cite{PDG06} to calculate these couplings:
\begin{itemize}
\item $g_{\pi\gamma\gamma} = (2.402 \pm 0.086)\times10^{-3}$
\item $g_{\eta\gamma\gamma} = (9.70 \pm 0.26)\times10^{-3}$
\item $g_{\eta'\gamma\gamma} = (2.13 \pm 0.11)\times10^{-2}$
\end{itemize}

We can write these couplings in terms of decay constants:
\begin{eqnarray}
g_{\pi^0\gamma\gamma} &=& \frac{\alpha M_{\pi^0}}{\pi\sqrt{2}}\frac{1}{f_q} \\
g_{\eta\gamma\gamma} &=& \frac{\alpha M_{\eta}}{\pi\sqrt{2}}\frac{1}{3} \left(\frac{5}{f_q}\cos\phi - \frac{\sqrt{2}}{f_s}\sin\phi \right) \\
g_{\eta\gamma\gamma} &=& \frac{\alpha M_{\eta'}}{\pi\sqrt{2}}\frac{1}{3} \left(\frac{5}{f_q}\sin\phi + \frac{\sqrt{2}}{f_s}\cos\phi \right) \\
\end{eqnarray}
where $f_q$ is the non-strange decay constant, $f_s$ is the strange constant and $f_{\pi}=\sqrt{2} f_q$.

From the expression for $g_{\pi^0\gamma\gamma}$ we extract $f_{\pi}=131\pm5 \text{MeV}$.  This is consistent with $f_{\pi^0}=f_{\pi^+}=130.7\pm0.4 \text{MeV}$ given in the PDG Review 2006\cite{PDG06}.

In the quark model, $f_s/f_d = \mu_d/\mu_s$ and, using the results of Section \ref{sec:MagneticMomentRatio}, this is equal to $1.22 \pm 0.07$.  This value is consistent with the $1.23\pm0.05$ found by Feldman et.\ al.\cite{Feldmann:1998vh}, the $1.20\pm0.10$ found by Cao and Signal\cite{Cao:1999fs} and the $1.16\pm0.09$ found by Huang and Wu\cite{Huang:2006as} but is smaller than the $1.51\pm0.08$ found by Escribano and Frere\cite{escribano:2005qq}.

Assuming that there is no gluonic mixing, we extract the mixing angle $\phi$ from the $\eta'/\eta$, $\eta/\pi^0$ and $\eta'/\pi^0$ ratios.  To assess the robustness of our results, we also extract the mixing angle when $f_s/f_d=1$ and $f_s/f_d=1.5$ as shown in Table \ref{table:TwoPhotonMixingAngle}.  Following our comment above that we should not necessarily trust the pionic modes, we extract a mixing angle of $\phi = 38.3\pm1.8$ from the $\eta'/\eta$ ratio (the uncertainty includes that from $f_s/f_d$).  This is slightly smaller than the value extracted from radiative decays.

\begin{table}
\begin{center}
\begin{tabular}{|c|c|c|}
\hline
 & \textbf{$f_s/f_d$} & \textbf{$\phi/\degrees$} \\
\hline
$\eta'/\eta$ & $1.21\pm0.07$ & $38.3\pm1.8$ \\
 & $1$ & $35.7\pm1.6$ \\
 & $1.5$ & $40.8\pm1.7$ \\
\hline
$\eta/\pi^0$ & $1.21\pm0.07$ & $41.3\pm2.0$ \\
 & $1$ & $39.1\pm1.8$ \\
 & $1.5$ & $43.4\pm1.9$ \\
\hline
$\eta'/\pi^0$ & $1.21\pm0.07$ & $33.8\pm4.0$ \\
 & $1$ & $30.3\pm3.8$ \\
 & $1.5$ & $36.8\pm4.0$ \\
\hline
\end{tabular}
\caption{Mixing angle determinations}
\label{table:TwoPhotonMixingAngle}
\end{center}
\end{table}

If we include the pion mode, the mixing angles extracted from the $\eta/\pi^0$ and $\eta'/\pi^0$ show some tension: the angle obtained from the $\eta'/\pi^0$ ratio is lower than that obtained from $\eta/\pi^0$, although there is a large experimental uncertainty.  If we take this discrepancy seriously, assume that the $\eta$ has no gluonic component and take $\phi=(41.3\pm2.0)\degrees$ (from $\eta/\pi^0$ and compatible with that obtained from radiative decays) we obtain $\cos{\phi_{G2}} = 0.90 \pm 0.06$ ($\phi_{G2}=(26^{+7}_{-10})\degrees$).  This is consistent with the values obtained from radiative decays and hints at a small gluonic component.  However, the significance isn't very great and we can't exclude a form factor effect and/or effect related to the lightness of the pion.  Again, although we have referred to this as a gluonic component, it could actually be any component that has suppressed decays in to two photons.

\section{Radiative Decays of $J/\psi$ and $\psi'$  $\rightarrow \gamma \eta/\eta'/\pi^0$}
\label{sec:PsiGamma}

We now turn to the decays of charmonium and begin with the radiative decay of the vectors $J/\psi$ and $\psi'\equiv\psi(3686)$ to light pseudoscalar mesons.  These decays have been studied many times in the literature (for example in Refs.\ \cite{Kawarabayashi:1980dp,Kawarabayashi:1980uh}\cite{Gilman:1987ax},\cite{Seiden:1988rr},\cite{Kopke:1988cs},\cite{ball:1995zv},\cite{Feldmann:1998vh,Feldmann:1998sh},\cite{escribano:1999nh},\cite{Gerard:2004gx},\cite{kroll:2004rs,kroll:2005sd},\cite{Zhao:2006gw}).  The experimental data from the PDG Review 2006\cite{PDG06} are:
\begin{itemize}
\item $BR(J/\psi \rightarrow \pi^0 \gamma) = (3.3^{+0.6}_{-0.4})\times 10^{-5}$
\item $BR(J/\psi \rightarrow \eta \gamma) = (9.8\pm1.0)\times 10^{-4}$
\item $BR(J/\psi \rightarrow \eta' \gamma) = (4.71\pm0.27)\times 10^{-3}$
\item $BR(J/\psi \rightarrow \eta_c \gamma) = (1.3\pm0.4) \times 10^{-2}$
\item $BR(\psi' \rightarrow \eta \gamma) < 9 \times 10^{-5}$ (90\% CL)
\item $BR(\psi' \rightarrow \eta' \gamma) = (1.5\pm0.4)\times 10^{-4}$
\item $BR(\psi' \rightarrow \eta_c \gamma) = (2.6\pm0.4) \times 10^{-3}$
\end{itemize}

If there is any charmonium component in the $\eta$ and/or $\eta'$ (see for example Refs.\ \cite{Harari:1975ie,Feldmann:1998vh,Feldmann:1998sh}), we expect these processes to be dominated by the magnetic dipole transitions of charmonium (analogous to the processes considered in Section \ref{sec:RadDecays}).  In this case, we can extract the $c\bar{c}$ component by comparing with the $J/\psi \rightarrow \eta_c \gamma$ mode.  We obtain $|Z_{\eta}(c\bar{c})|^2 \approx (3 \pm 1) \times 10^{-5}$ and  $|Z_{\eta'}(c\bar{c})|^2 \approx (2.0 \pm 0.6) \times 10^{-4}$ where really these are the products of $|Z|^2$ with any form factor suppression.  If we use a form factor ($\beta=400\MeV$) we get $|Z_{\eta}(c\bar{c})|^2 \approx (1.9 \pm 0.6) \times 10^{-4}$ and  $|Z_{\eta'}(c\bar{c})|^2 \approx (9 \pm 3) \times 10^{-4}$.  Using the $\psi'$ results and no form factor gives $|Z_{\eta}(c\bar{c})|^2 \lesssim 1.5 \times 10^{-3}$ and $|Z_{\eta'}(c\bar{c})|^2 \approx (2.9 \pm 0.9) \times 10^{-3}$.  These results suggest that the amplitudes of the charmonium components of the $\eta$ and the $\eta'$ are both $\lesssim 5\%$.

\begin{figure}[htb]
\begin{center}
\includegraphics[width=7cm]{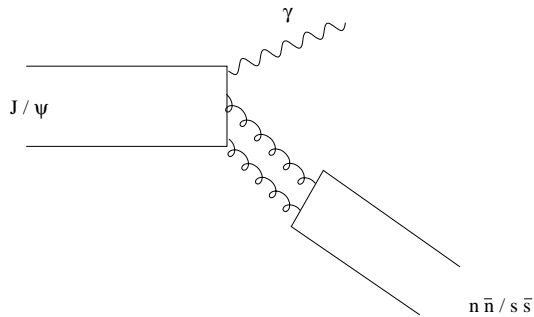}
\caption{Leading diagram for $J/\psi \rightarrow P \gamma$}
\label{fig:PsiPGammaDiagram}
\end{center}
\end{figure}

Justified by the above argument, we assume that there are only ground state light meson components and possibly glue in the $\eta$ and $\eta'$.  In this case, the diagram shown in Figure \ref{fig:PsiPGammaDiagram} dominates (see, for example, Ref. \cite{Zhao:2006gw}).  This diagram is related to the electromagnetic diagram which contributes in to the decays of $J/\psi$ in to pseudoscalar and vector mesons, see Section \ref{sec:PsiVP}.  The decay to $\pi^0$ is isospin violating and so the $\pi^0$ can not be produced via the same diagram.  This is consistent with the experimental data and provides justification for only considering one diagram.  We can't use the suppression of the $\pi^0$ mode to justify the charmonium or gluonic content of the $\eta$ or $\eta'$.

Again, we initially assume no gluonic component in either the $\eta$ or $\eta'$.  Following Seiden et.\ al.\cite{Seiden:1988rr}, we write the ratio in terms of a mixing angle and a strange/nonstrange factor $R$:
\begin{equation}
\frac{BR(J/\psi \rightarrow \eta\gamma)}{BR(J/\psi \rightarrow \eta'\gamma)} = 
\left(\frac{\mvp[\eta]}{\mvp[\eta']}\right)^3
\left|\frac{\sqrt{2}\cos\phi - R\sin\phi}{\sqrt{2}\sin\phi + R\cos\phi}\right|^2
\end{equation}
Note that, a priori, $R$ is \emph{not} the same as $\mu_s/\mu_d$ used in previous sections.  There we were considering the coupling of a photon to strange/nonstrange quarks whereas here we are considering the coupling of gluons to strange/nonstrange quarks.  However, assuming that the mass difference of the quarks is important, we expect the ratios to have similar magnitudes.

If we do assume that $R=\mu_s/\mu_d$ and take this from Section \ref{sec:MagneticMomentRatio} ($R=0.82\pm0.05$), we obtain $\phi=(37.4\pm1.9)\degrees$, a slightly lower value than our previous determinations.  The parameter $R$ should be the same as the equivalent parameter in hadronic $J/\psi$ decays (see Section \ref{sec:PsiVP} below).  If we use the results of that section ($R=(1-s)=0.708\pm0.024$) we obtain $\phi=(41.1\pm 1.4)\degrees$ (including the uncertainty from $R$), a value consistent with that extracted previously.  Alternatively, requiring $\phi=40.0\degrees$ we obtain $R=0.74\pm0.04$.  

Therefore, a reasonable value of $R$ gives a reasonable value of $\phi$ and vice-versa.  There is no discrepancy in the mixing angle obtained and so no evidence for gluonic mixing.  Because we don't have anything to normalise to, it is difficult to pin down the gluonic component here.  This is the first case we have considered where glue will not be invisible; indeed, it may even be enhanced relative to the $\state{q\bar{q}}$ components.

If we include some gluonic component, we can write:
\begin{equation}
\frac{BR(J/\psi \rightarrow \eta\gamma)}{BR(J/\psi \rightarrow \eta'\gamma)} = 
\left(\frac{\mvp[\eta]}{\mvp[\eta']}\right)^3
\left|\frac{\sqrt{2} X_{\eta} + R Y_{\eta} + G Z_{\eta} }{ \sqrt{2} X_{\eta'} + R Y_{\eta'} + G Z_{\eta'} } \right|^2
\end{equation}
where G is the relative amplitude for production of gluonium.  In the limit where production is only via the gluonium component, the ratio measures the relative glue content of the $\eta$ and $\eta'$ mesons.

We assume that there is no glue in the $\eta$ and consider a range of $G>1$.  For positive $\phi_{G2}$, the ratio is reduced and this reduction is greater for larger $G$ if $\phi_{G2}<30\degrees$.  For negative $\phi_{G2}$, the ratio is increased and the increase is greater for larger $G$.  The value of $\phi_{G2}$ favoured hence depends on $R$ and $G$ and so we can not extract the gluonic component from this ratio.

There are not enough experimental data to repeat the exercise for the $\psi'$ radiative decays.  We can however predict $BR(\psi' \rightarrow \eta \gamma)$.  Using $\phi=41\degrees$ and setting $R=1$ (see Section \ref{sec:PsiVP}) we predict $BR(\psi' \rightarrow \eta \gamma) = 1.0 \times 10^{-5}$.  Alternatively using $R = 0.708$ from $J/\psi$ we obtain $BR(\psi' \rightarrow \eta \gamma) = 2.9\times 10^{-5}$.  Both these are consistent with the experimental upper bound.

\section{Decays of $J/\psi$ and $\psi'$ in to Light Vector and Pseudoscalar Mesons}
\label{sec:PsiVP}

Having discussed the radiative decays of the $J/\psi$, we now turn to its strong decays in to light vector and pseudoscalar mesons.  The experimental situation for the $J/\psi$ and the radially excited $\psi'\equiv\psi(3686)$ is shown in Table \ref{table:JPsiPVExpData}.

\begin{table}
\begin{center}
\begin{tabular}{|c|c|c|}
\hline
\textbf{Mode} & $J/\psi \rightarrow$ & $\psi' \rightarrow$ \\
\hline
$K^{*+} K^-$ & $(2.5\pm0.2)\times10^{-3}$ & $(8.5\pm4.0)\times 10^{-5}$ \\
$K^{*0} \bar{K}^0$ & $(2.1\pm0.2)\times10^{-3}$ & $(5.5\pm1.0)\times 10^{-5}$ \\
$\rho^0 \pi^0$ & $(5.6\pm0.7)\times10^{-3}$ & - \\
$\rho^0 \eta$ & $(1.93\pm0.23)\times10^{-4}$ & $(2.2\pm0.6)\times 10^{-6}$ \\
$\rho^0 \eta'$ & $(1.05\pm0.18)\times10^{-4}$ & $(1.9^{+1.7}_{-1.2})\times 10^{-5}$ \\
$\omega \pi^0$ & $(4.5\pm0.5)\times10^{-4}$ & $(2.1\pm0.6)\times 10^{-5}$ \\
$\omega \eta$ & $(1.74\pm0.20)\times10^{-3}$ & $<1.1 \times 10^{-5}$ (90\% CL) \\
$\omega \eta'$ & $(1.82\pm0.21)\times10^{-4}$ & $(3.2^{+2.5}_{-2.1})\times 10^{-5}$ \\
$\phi \pi^0$ & $<6.4\times10^{-6}$ (90\% CL) & $<4\times 10^{-6}$ (90\% CL) \\
$\phi \eta$ & $(7.4\pm0.8)\times10^{-4}$ & $(2.8^{+1.0}_{-0.8})\times 10^{-5}$ \\
$\phi \eta'$ & $(4.0\pm0.7)\times10^{-4}$ & $(3.1\pm1.6)\times 10^{-5}$ \\
\hline
\end{tabular}
\caption{Experimental Data on $J/\psi$ and $\psi'$ $\rightarrow P V$\cite{PDG06}.  ($K^{*+} K^- = \frac{1}{2} (K^{*+} K^- + \text{c.c.})$ and $K^{*0} \bar{K}^0 = \frac{1}{2} (K^{*0} \bar{K}^0 + \text{c.c.})$) }
\label{table:JPsiPVExpData}
\end{center}
\end{table}

\begin{figure}[htb]
\begin{center}
\includegraphics[width=15cm]{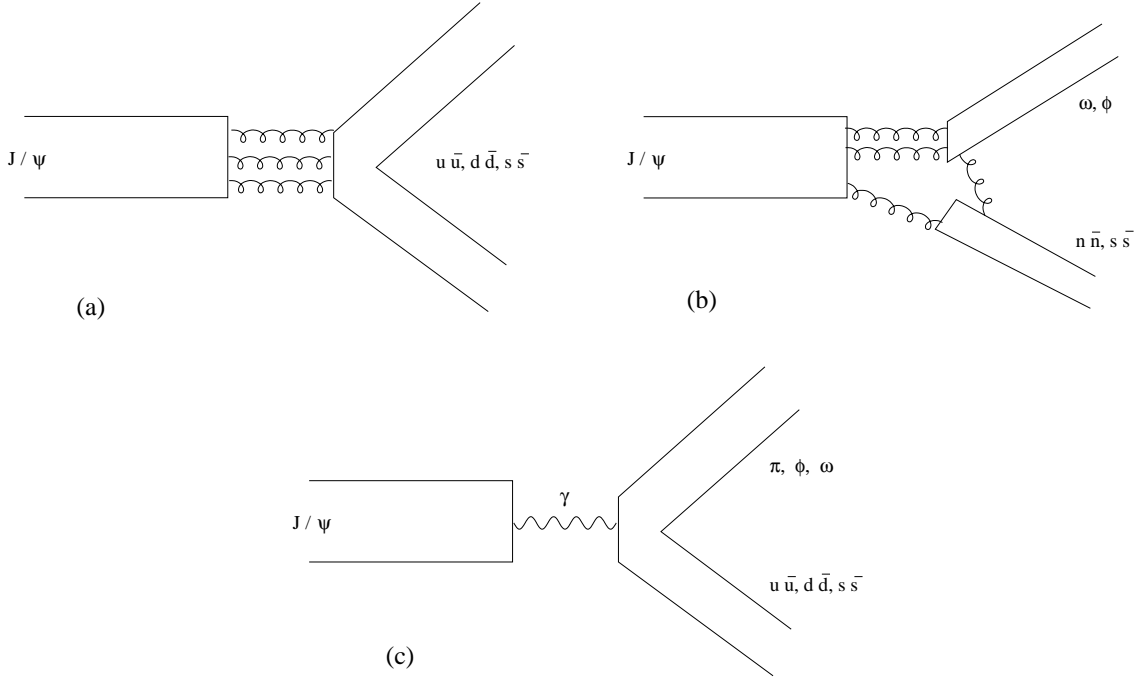}
\caption{Diagrams contributing to $J/\psi \rightarrow P V$.  (a) SOZI, (b) DOZI, (c) SOZI EM}
\label{fig:PsiPVDiagrams}
\end{center}
\end{figure}

These decays have been studied in the literature (for example in Refs.\ \cite{Kawarabayashi:1980uh}\cite{Baltrusaitis:1984rz}, \cite{Haber:1985cv}, \cite{Gilman:1987ax}, \cite{Seiden:1988rr}, \cite{Kopke:1988cs}, \cite{Jousset:1988ni}, \cite{Bramon:1997mf}, \cite{Feldmann:1998vh}, \cite{escribano:1999nh}, \cite{Zhao:2006gw,Li:2007ky}).  We follow the approach of Seiden et.\ al.\cite{Seiden:1988rr} by parameterising the amplitudes.  The relevant diagrams are shown in Figure \ref{fig:PsiPVDiagrams}: all the diagrams are disconnected (OZI rule violating), some are singly disconnected (SOZI) and some are double disconnected (DOZI).  Following that reference, we ignore DOZI EM diagrams and ignore terms second order in $r$, $s$ or $e$.  Ignoring the DOZI EM diagrams means that we predict no decay to $\phi \pi^0$ which is consistent with the current experimental data.  For simplicity we ignore vector meson mixing which is justified by our results above and will be commented on below.  Our approach is similar to that used by Li et.\ al.\cite{Li:2007ky}, the main difference is in the way the electromagnetic diagrams are calculated.  

As well as the $\eta-\eta'$ mixing parameters described above, we use the following decay parameters:
\begin{itemize}
\item $g$ - Overall normalisation.
\item $r$ - Relative amplitude for a DOZI diagram compared with a SOZI diagram.
\item $|e|$ - Magnitude of the electromagnetic (EM) diagrams.
\item $\arg{e}$ - Relative phase of the EM diagrams.
\item $R\equiv(1-s)$ - Amplitude for production of strange quarks compared with non-strange quarks.
\item $r_g$ - Relative gluonic production amplitude.
\end{itemize}

The amplitudes with momentum factors removed, $A$, are presented in Table \ref{table:JPsiPVAmps}. The branching ratio is given by 
\begin{equation}
\label{equ:JPsiPVRate}
BR \propto C |\text{A}|^2 \mvp^{2L+1} F(\mvp)
\end{equation}
where $C$ is a factor which takes in to account the combinations of particles in the final state, $L$ is the relative orbital angular momentum of the two mesons in the final state, $\mvp$ is the final state recoil momentum in the rest frame of the $J/\psi$ and $F(\mvp)$ is a form factor.  In these decays $L=1$ and $C=1$.

\begin{table}
\begin{center}
\begin{tabular}{|c|c|c|}
\hline
\textbf{Mode} & \textbf{Amplitude A} \\
\hline
$K^{*+} K^-$ & $g(1-s)+e$ \\
$K^{*0} \bar{K}^0$ & $g(1-s)-2e$ \\
$\rho^0 \pi^0$ & $g+e$ \\
$\rho^0 \eta$ & $-3e X_{\eta}$ \\
$\rho^0 \eta'$ & $-3e X_{\eta'}$ \\
$\omega \pi^0$ & $-3e$ \\
$\omega \eta$ & $\left((g+e)X_{\eta} + \sqrt{2}rg(\sqrt{2}X_{\eta} + Y_{\eta}\right) + \sqrt{2} g r_g Z_{\eta}$ \\
$\omega \eta'$ & $\left((g+e)X_{\eta'} + \sqrt{2}rg(\sqrt{2}X_{\eta'} + Y_{\eta'}\right) + \sqrt{2} g r_g Z_{\eta'}$ \\
$\phi \pi^0$ & $0$ \\
$\phi \eta$ & $\left(\left[g(1-2s)-2e\right] Y_{\eta} + rg\left[\sqrt{2}X_{\eta} + Y_{\eta}\right]\right) + g r_g Z_{\eta}$ \\
$\phi \eta'$ & $\left(\left[g(1-2s)-2e\right] Y_{\eta'} + rg\left[\sqrt{2}X_{\eta'} + Y_{\eta'}\right]\right) + g r_g Z_{\eta'}$ \\
\hline
\end{tabular}
\caption{Amplitudes for $J/\psi \rightarrow P V$ (no vector meson mixing)}
\label{table:JPsiPVAmps}
\end{center}
\end{table}

We initially assume no gluonic mixing which leaves us with 10 data points and 6 parameters and hence 4 degrees of freedom.  We fit the data without a form factor ($F(\mvp)=1$) and using a gaussian ($F(\mvp)=\exp(-\mvp^2/8\beta^2)$) with $\beta=500\text{MeV}$ (the value used in Ref.\ \cite{Li:2007ky}) and $\beta=400\text{MeV}$.  The results are shown in Table \ref{table:JPsiPVResults}.

\begin{table}
\begin{center}
\begin{tabular}{|c|c|c|c|}
\hline
\textbf{Parameter} & \multicolumn{3}{c}{\textbf{Fitted Value}} \vline \\
 \cline{2-4}
 & \textbf{No Form Factor} & \textbf{$\beta=500\MeV$}  & \textbf{$\beta=400\MeV$}\\
\hline
$g$ & $(1.35\pm0.04)\times 10^{-6}$ & $(2.20\pm0.10)\times 10^{-6}$ & $(2.88\pm0.15)\times 10^{-6}$\\
$|e|$ & $(1.20\pm0.04)\times 10^{-7}$ & $(1.93\pm0.09)\times 10^{-7}$ & $(2.51\pm0.14)\times 10^{-7}$ \\
$\arg{e}$ & $1.28\pm0.15$ & $1.27\pm0.20$ & $1.27\pm0.25$ \\
$s$ & $0.292\pm0.024$ & $0.304\pm0.031$ & $0.31\pm0.04$ \\
$r$ & $-0.368\pm0.012$ & $-0.347\pm0.016$ & $-0.335 \pm 0.019$ \\
$\phi$ & $(40 \pm 2)\degrees$ & $(37 \pm 3)\degrees$ & $(35 \pm 4)\degrees$ \\
\hline
$\chi^2/(\text{d.o.f.})$ & $3.3/4$ & $6.3/4$ & $9.1/4$ \\
\hline
\end{tabular}
\caption{Results of fit to $J/\psi \rightarrow PV$ data}
\label{table:JPsiPVResults}
\end{center}
\end{table}

The results show that there is a good fit to the data when no form factor is used; using a form factor decreases the goodness of fit slightly.  The EM diagrams are suppressed by $\sim \alpha^2$ as expected and this is true whether or not a form factor is used.  The DOZI diagrams are significant and the form factor does not significantly change the mixing angle or parameters.  The three values of $\phi$ are all consistent and generally consistent with those obtained above.  The values of $R$ obtained are consistent with the ratio of constituent quark masses (see Section \ref{sec:MagneticMomentRatio}) and slightly lower than the value from the ratio $\mu_s/\mu_d$ from Kaon decays (as noted above, we do not expect these ratios to be equal).

To check the effect of vector meson mixing we set $\theta_V = 3.4\degrees$ and repeat the fits.  The amplitude parameters are consistent with those already obtained, the goodness of fit is approximately the same and the mixing angle is unchanged.  This further justifies ignoring vector meson mixing.

If we use the same approach to study the decays of the $\psi'$ we should get the same mixing angle but allow for different decay parameters $g$, $r$, $|e|$, $\arg{e}$ and $s$.  The experimental data have larger uncertainties for these decays compared to the $J/\psi$ and for this reason we chose to fit the decays of these two mesons separately.  Here we have 8 data points and 6 parameters which gives 2 degrees of freedom.  The results in Table \ref{table:PsipPVResults} show good fits but that there are large uncertainties on the parameters.

\begin{table}
\begin{center}
\begin{tabular}{|c|c|c|c|}
\hline
\textbf{Parameter} & \multicolumn{3}{c}{\textbf{Fitted Value}} \vline \\
 \cline{2-4}
 & \textbf{No Form Factor} & \textbf{$\beta=500\text{MeV}$}  & \textbf{$\beta=400\text{MeV}$}\\
\hline
$g$ & $(-1.16\pm0.25)\times 10^{-7}$ & $(2.41\pm0.38)\times 10^{-7}$ & $(-3.65\pm0.49)\times 10^{-7}$\\
$|e|$ & $(2.14\pm0.17)\times 10^{-8}$ & $(4.64\pm0.35)\times 10^{-8}$ & $(7.15\pm0.49)\times 10^{-8}$ \\
$\arg{e}$ & $1.92\pm0.44$ & $1.98\pm0.33$ & $-2.02\pm0.28$ \\
$s$ & $4\times10^{-13}\pm0.15$ & $4\times10^{-14}\pm0.11$ & $4\times10^{-14}\pm0.09$ \\
$r$ & $0.02\pm0.09$ & $-0.01\pm0.06$ & $-0.03 \pm 0.05$ \\
$\phi$ & $(44^{+7}_{-8})\degrees$ & $(43^{+5}_{-6})\degrees$ & $(42 \pm 5)\degrees$ \\
\hline
$\chi^2/(\text{d.o.f.})$ & $1.4/2$ & $0.9/2$ & $0.7/2$ \\
\hline
\end{tabular}
\caption{Results of fit to $\psi' \rightarrow PV$ data}
\label{table:PsipPVResults}
\end{center}
\end{table}

The overall scale in $\psi'$ decays is about one order of magnitude smaller that in $J/\psi$ decays.  EM diagrams are slightly more important and the phase is different, but they are still suppressed by approximately an order of magnitude.  DOZI diagrams are much less important here and consistent with zero.  In addition, the strange - nonstrange difference is insignificant.  The mixing angle $\phi$ is consistent with the $J/\psi$ results but the values obtained here have large uncertainties and these decays do not strongly constrain it.  Li et.\ al.\cite{Li:2007ky} have fit both the $J/\psi$ and the $\psi'$ together - we think that this obscures that the $J/\psi$ decays are the source of the strong constraint on the mixing angle.

In the $J/\psi$ and $\psi'$ decays any gluonium component of the $\eta$ or $\eta'$ mesons will be `active' with a relative production amplitude expected to be of the same order as the DOZI diagrams.  We assume that only the $\eta'$ has any gluonic component and include a parameter $r_g$ which is the relative amplitude for gluonium production.  The results are given in Table \ref{table:JPsiPVGluonicResults}.  (We do not use the $\psi'$ data for this determination due to their limited accuracy and the fewer decay modes measured.)

\begin{table}
\begin{center}
\begin{tabular}{|c|c|c|c|}
\hline
\textbf{Parameter} & \multicolumn{3}{c}{\textbf{Fitted Value}} \vline \\
 \cline{2-4}
 & \textbf{No Form Factor} & \textbf{$\beta=500\text{MeV}$}  & \textbf{$\beta=400\text{MeV}$}\\
\hline
$g$ & $(1.32\pm0.06)\times 10^{-6}$ & $(2.11\pm0.10)\times 10^{-6}$ & $(2.74\pm0.16)\times 10^{-6}$\\
$|e|$ & $(1.27\pm0.06)\times 10^{-7}$ & $(2.13\pm0.12)\times 10^{-7}$ & $(2.86\pm0.20)\times 10^{-7}$ \\
$\arg{e}$ & $1.30\pm0.15$ & $1.30\pm0.16$ & $1.31\pm0.18$ \\
$s$ & $0.27\pm0.03$ & $0.27\pm0.04$ & $0.27\pm0.04$ \\
$r$ & $-0.36\pm0.08$ & $-0.33\pm0.11$ & $-0.30 \pm 0.15$ \\
$\phi$ & $(45 \pm 4)\degrees$ & $(46^{+4}_{-5})\degrees$ & $(47^{+5}_{-6})\degrees$ \\
$r_g$ & $\mp(0.13\pm0.23)$ & $\mp(0.17\pm0.21)$ & $\mp(0.20\pm0.25)$ \\
$\phi_{G2}$ & $\pm(33 \pm 13)\degrees$ & $\pm(44 \pm 9)\degrees$ & $\pm(48 \pm 10)\degrees$ \\
\hline
$\chi^2/(\text{d.o.f.})$ & $1.9/2$ & $2.3/2$ & $3.3/2$ \\
\hline
\end{tabular}
\caption{Results of fit to $J/\psi \rightarrow PV$ data, including a gluonic component}
\label{table:JPsiPVGluonicResults}
\end{center}
\end{table}

The mixing angle and most of the other parameters are consistent with those when no gluonic component was allowed but the uncertainties are larger here because there are fewer constraints.  The $\chi^2$ has decreased slightly but the fit is not significantly better taking in to account the smaller number of degrees of freedom.  The fit favours a small gluonic component in the $\eta'$, compatible with the previous sections.  Again the significance of this is not great.  The parameter $r_g$ is consistent with both $r$ and zero and has a large uncertainty.

There is an interesting question as to whether a large DOZI amplitude ($r$) is phenomenologically equivalent to a gluonic component of the $\eta'$.  To test this we set $r=0$ and fit again.  Regardless of whether we include a form factor, we get a poorer fit with $\chi^2/(\text{d.o.f.}) = (5-6)/3$ with a significantly larger mixing angle $\phi \approx (52\pm3)\degrees$.  The factor $r_g$ increases in magnitude significantly to $(-0.6) - (-0.9)$ and $\phi_{G2}$ increases slightly to $43\degrees-52\degrees$.  Therefore the data appear to favour a non-zero DOZI amplitude with zero (or at most, small) gluonic component of the $\eta'$.

As a check of our method, we fit the $J/\psi$ data again (both with and without glue) and include higher order corrections in combinations of the parameters $r$, $s$ and $e$ (for example, $s^2$) but no additional diagrams.  When these higher order corrections are included the fit is poor ($\chi^2/\text{d.o.f.} \approx 12/4$) and we obtain $\phi = (36 \pm 3)\degrees$.  We find that the higher order corrections including $e$ are insignificant.  Including a form factor does not improve the fit.   If we remove the pion modes from the fit the goodness of fit is increased ($\chi^2/\text{d.o.f.} = 0.34/2$), the normalisation is very weekly constrained and we obtain $\phi = (39.0 \pm 0.8)\degrees$.  When glue is included the fit becomes more reasonable $\chi^2/(\text{d.o.f.}) = 3.5/2$ (no form factor) with mixing angles of $\phi = (52 \pm 3)\degrees$ and $\phi_{G2} = \pm(43 \pm 10)\degrees$ and $r$ is consistent with zero.  With a form factor ($\beta=400\MeV$) we get $\chi^2/\text{d.o.f.} = 0.49/2$ with $\phi = (50.6 \pm 1.4)\degrees$ and $\phi_{G2} = \pm(51 \pm 3)\degrees$ and $r$ not consistent with zero.  These are inconsistent with the other determinations and in particular give a significantly larger mixing angle similar to when we set $r=0$ above.  From these results we can either conclude that we are being consistent if we only stay at first order in these corrections: there are other effects that we are missing at higher orders, that we should exclude the pion modes, or that this is a sign of the approach breaking down.

As commented above, our approach is similar to that used by Li et.\ al.\cite{Li:2007ky}, the main difference is in the way the electromagnetic diagrams are calculated.  They use a number of different mixing schemes and comparison is complicated by the different mixing parameters used.  In their `CKM' approach (I) (which in the limit $\theta_{13}=0$ translates to our approach with $\phi_{G1}=0$, $\phi \rightarrow - \theta_{21}$ and $\phi_{G2} \rightarrow \theta_{23}$) they find that $\phi_{G1} = 0\degrees - 0.9\degrees$, $\phi_{G2} \approx -10\degrees$ and $\phi = 24\degrees - 26\degrees \pm (2\degrees-3\degrees)$ (the range we give is that from their three approaches to calculating or ignoring the EM diagrams).  In their `mixing due to higher Fock state' approach (II) they find that $\phi = 30\degrees - 31\degrees$.  In both these approaches they find $\phi_{G1}$ close to $0$, consistent with our assumption, and a small $\phi_{G2}$ consistent with ours.  However, they find a significantly smaller $\phi$.  In their `old perturbation theory' approach (III) they find $\phi_{G1} \approx \phi_{G2}$, inconsistent with ours.

In summary, from the hadronic $J/\psi$ and $\psi'$ decays we have again found a consistent description in terms of one mixing angle with a suggestion that there is some small gluonic component of the $\eta'$.  There is some theoretical uncertainty if we include higher order effects.

\section{Decays of $\chi_{c0,2}$ in to Light Pseudoscalar Mesons}
\label{sec:ChiC02PP}

We analyse the decays of the $\chi_{c0}$ and $\chi_{c2}$ in a similar way to those of the $J/\psi$.  The experimental situation is shown in Table \ref{table:ChiC02PPExpData}.  These decays have been studied by Zhao\cite{Zhao:2005im}.

\begin{table}
\begin{center}
\begin{tabular}{|c|c|c|}
\hline
\textbf{Mode} & $\chi_{C0} \rightarrow$ & $\chi_{C2} \rightarrow$ \\
\hline
$\pi \pi$ & $(7.2\pm0.6)\times 10^{-3}$ & $(2.14\pm0.25)\times 10^{-3}$ \\
$\eta \eta$ & $(1.9\pm0.5)\times 10^{-3}$ & $<1.2\times 10^{-3}$ (90\% CL)  \\
 & $(3.1\pm0.7)\times 10^{-3}$\cite{Adams:2006na} & $<4.7\times 10^{-4}$ (90\% CL)\cite{Adams:2006na} \\
$\eta' \eta'$ & $(1.7\pm0.5)\times 10^{-3}$\cite{Adams:2006na} & $<3.1\times 10^{-4}$ (90\% CL)\cite{Adams:2006na} \\
$\eta \eta'$ & $<5\times 10^{-4}$ (90\% CL)\cite{Adams:2006na} & $<2.3\times 10^{-4}$ (90\% CL)\cite{Adams:2006na} \\
$K^+ K^-$ & $(5.4\pm0.6)\times 10^{-3}$ & $(7.7\pm1.4)\times 10^{-4}$ \\
$K_S^0 K_S^0$ & $(2.8\pm0.7)\times 10^{-3}$ & $(6.7\pm1.1)\times 10^{-4}$ \\
\hline
\end{tabular}
\caption{Experimental Data on $\chi_{c0}$ and $\chi_{c2}$ $\rightarrow P P$, from the PDG Review 2006\cite{PDG06} unless otherwise stated.}
\label{table:ChiC02PPExpData}
\end{center}
\end{table}

The amplitudes with momentum factor removed, $A$, are given in Table \ref{table:ChiC02Amps} where the notation is the same as in Section \ref{sec:PsiVP} although the parameters will in general have different values.  The electromagnetic (EM) diagrams (parametrised by $e$ above) are expected to be even more suppressed in $\chi_{c0,2}$ decays because two photons or a photon and gluons would be required in such a transition.  In $J/\psi$ decays we found that the electromagnetic diagram was strongly suppressed ($\sim \alpha$ in rate) and so, given that it should be more strongly suppressed here ($\sim \alpha^2$ in rate) and the current experimental uncertainties, we ignore any EM contribution in this analysis.  We assume no gluonic mixing because there are not enough constraints to pin down any gluonic component with these modes.  A consistent mixing angle would support this assumption.

\begin{table}
\begin{center}
\begin{tabular}{|c|c|c|}
\hline
\textbf{Mode} & \textbf{Amplitude $A$} \\
\hline
$K^{+} K^-$ & $g(1-s)$ \\
$K^{0} \bar{K}^0$ & $g(1-s)$ \\
$\pi^0 \pi^0$ & $g$ \\
$(s\bar{s}) (s\bar{s})$ & $g(1+r)(1-s)^2$ \\
$(s\bar{s}) (n\bar{n})$ & $rg\sqrt{2}(1-s)$ \\
$(n\bar{n}) (n\bar{n})$ & $g+2rg$ \\
$\pi^0 (n\bar{n})$ & $0$ \\
$\pi^0 (s\bar{s})$ & $0$ \\
\hline
\end{tabular}
\caption{Amplitudes for $\chi_{c0,2} \rightarrow P P$ }
\label{table:ChiC02Amps}
\end{center}
\end{table}

Equ.\ \ref{equ:JPsiPVRate} above gives the rate in terms of the amplitude $A$.  The $\chi_{c0}$ decay has $L=0$ and the $\chi_{c2}$ decay $L=2$.  The decay is to two light pseudoscalars and we have to account for the combinations of mesons in the final state: $C$ is $3$ for $\pi\pi$ ($\pi^0 \pi^0 + \pi^+ \pi^- + \pi^- \pi^+$), $2$ for $K^+ K^-$ and $\eta \eta'$, and $1$ for $\eta \eta$, $\eta' \eta'$ and $K_S^0 K_S^0$.

\begin{figure}[tb]
\begin{center}
\includegraphics[width=7cm,angle=270]{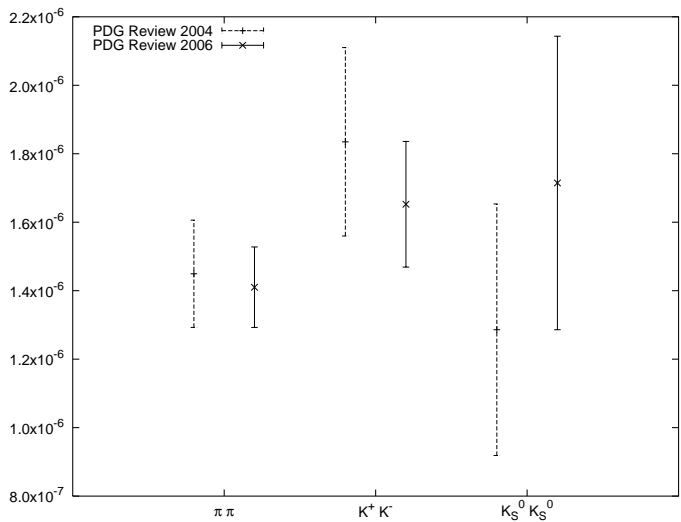}
\caption{Branching ratios of $\chi_{c0}$ with the $C$ factor and momentum dependence removed (no form factor).}
\label{fig:Chic0_naive}
\end{center}
\end{figure}

In Figure \ref{fig:Chic0_naive} we plot the reduced branching ratios of $\chi_{c0}$ (after the $C$ factor and momentum dependence have been removed, no form factor has been used).  If $r=s=0$ then all these should be equal.  For comparison, we plot both data from the PDG Review 2006\cite{PDG06} and from the previous PDG Review in 2004.  The experimental data has moved around a bit between the two reviews, most of the error bars have decreased and both sets of data are compatible with $r=s=0$.

\begin{figure}[tb]
\begin{center}
\includegraphics[width=7cm,angle=270]{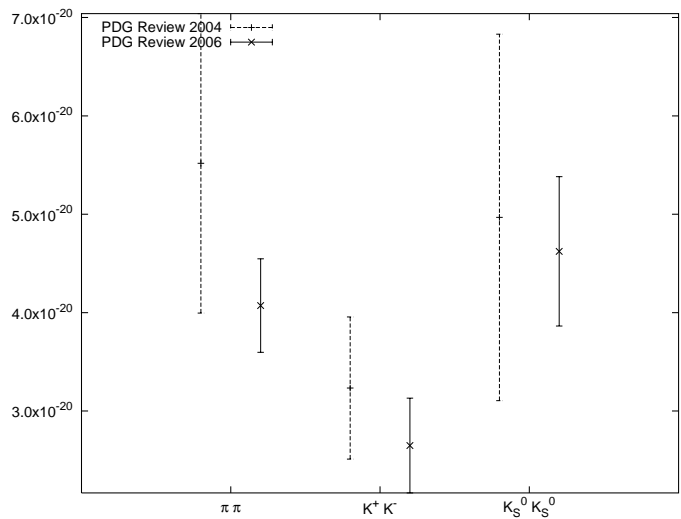}
\caption{Branching ratios of $\chi_{c2}$ with the $C$ factor and momentum dependence removed (no form factor).}
\label{fig:Chic2_naive}
\end{center}
\end{figure}

We repeat the exercise for the $\chi_{c2}$ decays and plot the results in Figure \ref{fig:Chic2_naive}.  This shows that the experimental uncertainties have decreased significantly between the 2004 and 2006 PDG Reviews.  The 2004 data are compatible with $r=s=0$.  However, the 2006 data shows some discrepancy: whereas the $\pi \pi$ and $K^0_S K^0_S$ modes are consistent with each other, the $K^+ K^-$ mode is significantly smaller.  This is particularly interesting because the difference between $K^0_S K^0_S$ and $K^+ K^-$ can only be due to an electromagnetic correction and not a non-zero $r$ and/or $s$.

We can't perform a fit because there are only 5 data points but we have 6 parameters (including $\phi$) for $\chi_{c0} \rightarrow PP$ and only 3 data points with 6 parameters (including $\phi$) for $\chi_{c2} \rightarrow PP$.  However, we can extract some of the parameters in a consistent way and these are shown in Table \ref{table:ChiC02Params}.  From the $\pi \pi$ mode we extract $g$ and we then extract $s$ from $K^+ K^-$.  Note that experimentally the ratio of branching ratios $(K^+ K^-)/(K_S^0 K_S^0)$ is consistent with $2$ for the $\chi_{c0}$ decays which adds additional justification for ignoring the EM diagrams - the amplitudes for these two decays would have different corrections if these diagrams were included.  However, the same ratio for $\chi_{c2}$ decays is not consistent with two.  The decay parameters (or ratios) are not consistent between the $\chi_{c0}$ and $\chi_{c2}$ decays. 

\begin{table}
\begin{center}
\begin{tabular}{|c|c|c|c|}
\hline
\textbf{Parameter} & \textbf{No Form Factor} & $\beta=500\text{MeV}$ & $\beta=400\text{MeV}$ \\
\hline
 \multicolumn{3}{c}{$\chi_{c0} \rightarrow P P$}   \\
\hline
$g$ & $(1.19\pm0.05)\times 10^{-3}$ & $(2.45\pm0.10)\times 10^{-3}$ & $(3.68\pm0.15)\times 10^{-3}$ \\
$s$ & $-0.08\pm0.08$ & $-0.02\pm0.07$ & $-0.009\pm0.069$ \\
\hline
\hline
 \multicolumn{3}{c}{$\chi_{c2} \rightarrow P P$}   \\
\hline
$g$ & $(2.02\pm0.12)\times 10^{-10}$ & $(4.4\pm0.3)\times 10^{-10}$ & $(6.89\pm0.04)\times 10^{-10}$ \\
$s$ & $0.19\pm0.09$ & $0.24\pm0.08$ & $0.26\pm0.08$ \\
\hline
\end{tabular}
\caption{Extracted parameters, see the text for details}
\label{table:ChiC02Params}
\end{center}
\end{table}

Turning to the $\eta$ and $\eta'$ modes, we begin with the $\chi_{c0}$ decays.  $s$ is consistent with zero for these modes and if we set $s=0$ and $r=0$ (no DOZI diagrams) the $\eta \eta$ and $\eta' \eta'$ amplitudes would both be equal to $g$ and \emph{not depend on the mixing angle}.  The $\eta \eta'$ amplitude would be zero.  Although we do not expect $r$ to be zero, this suggests that a small $r$ would lead to difficulty in extracting the mixing angle from these decays.

\begin{figure}[tb]
\begin{center}
\includegraphics[width=15cm]{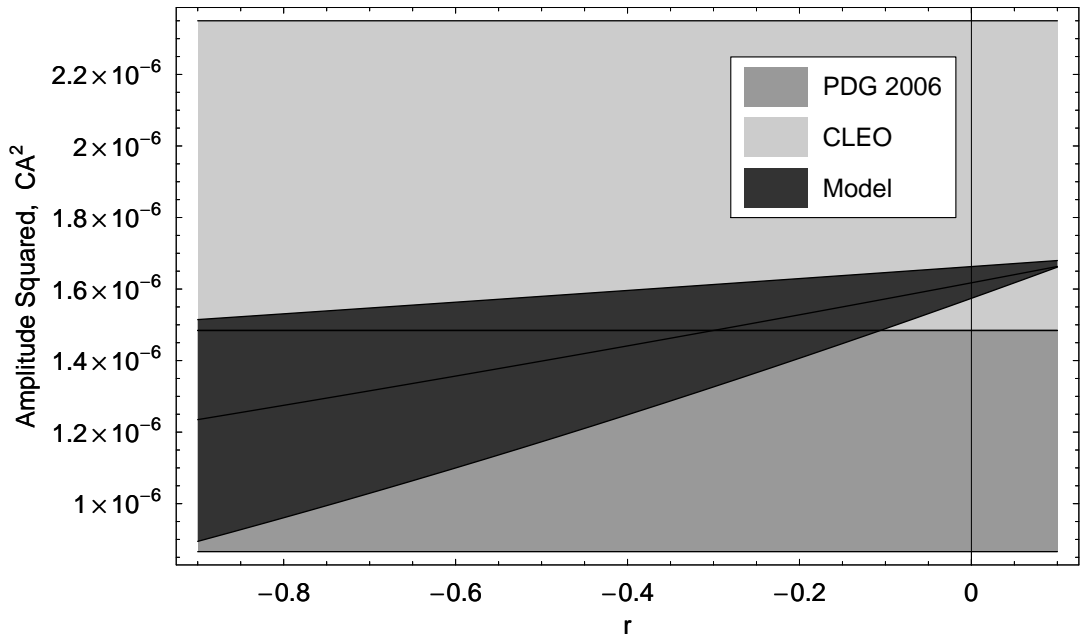}
\caption{$\chi_{c0} \rightarrow \eta\eta$ with no form factor.  Horizontal bands show the experimental data ($1\sigma$ range) from the PDG\cite{PDG06} and CLEO\cite{Adams:2006na}.  Curves show predictions for $\phi=40\degrees$ and the range $35\degrees$ to $45\degrees$.}
\label{fig:ChiC0plot_etaeta_r_noff}
\end{center}
\end{figure}

\begin{figure}[tb]
\begin{center}
\includegraphics[width=15cm]{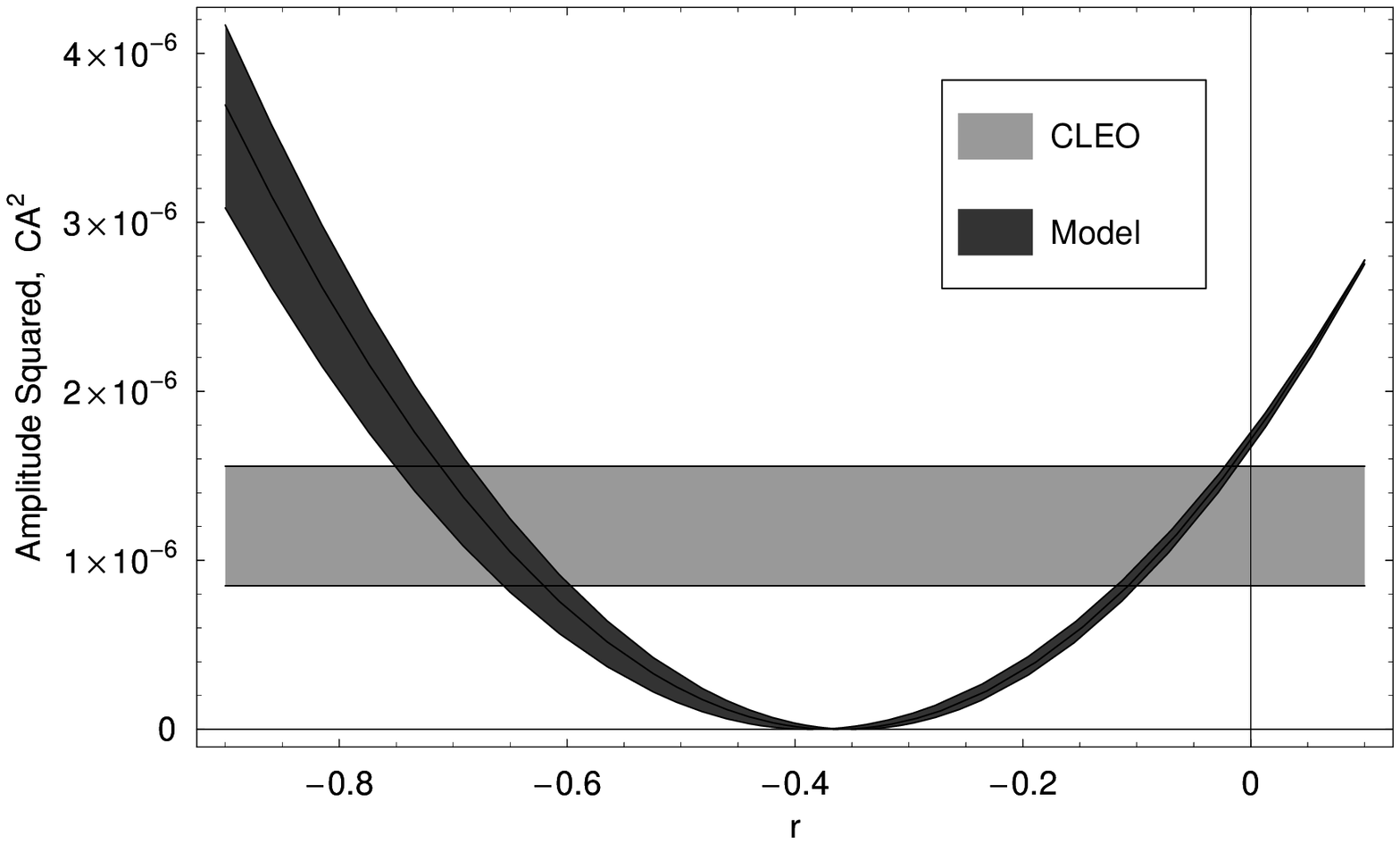}
\caption{$\chi_{c0} \rightarrow \eta'\eta'$ with no form factor.  Horizontal band shows the experimental data ($1\sigma$ range) from CLEO\cite{Adams:2006na}.  Curves show predictions for $\phi=40\degrees$ and the range $35\degrees$ to $45\degrees$.}
\label{fig:ChiC0plot_etapetap_r_noff}
\end{center}
\end{figure}

\begin{figure}[tb]
\begin{center}
\includegraphics[width=15cm]{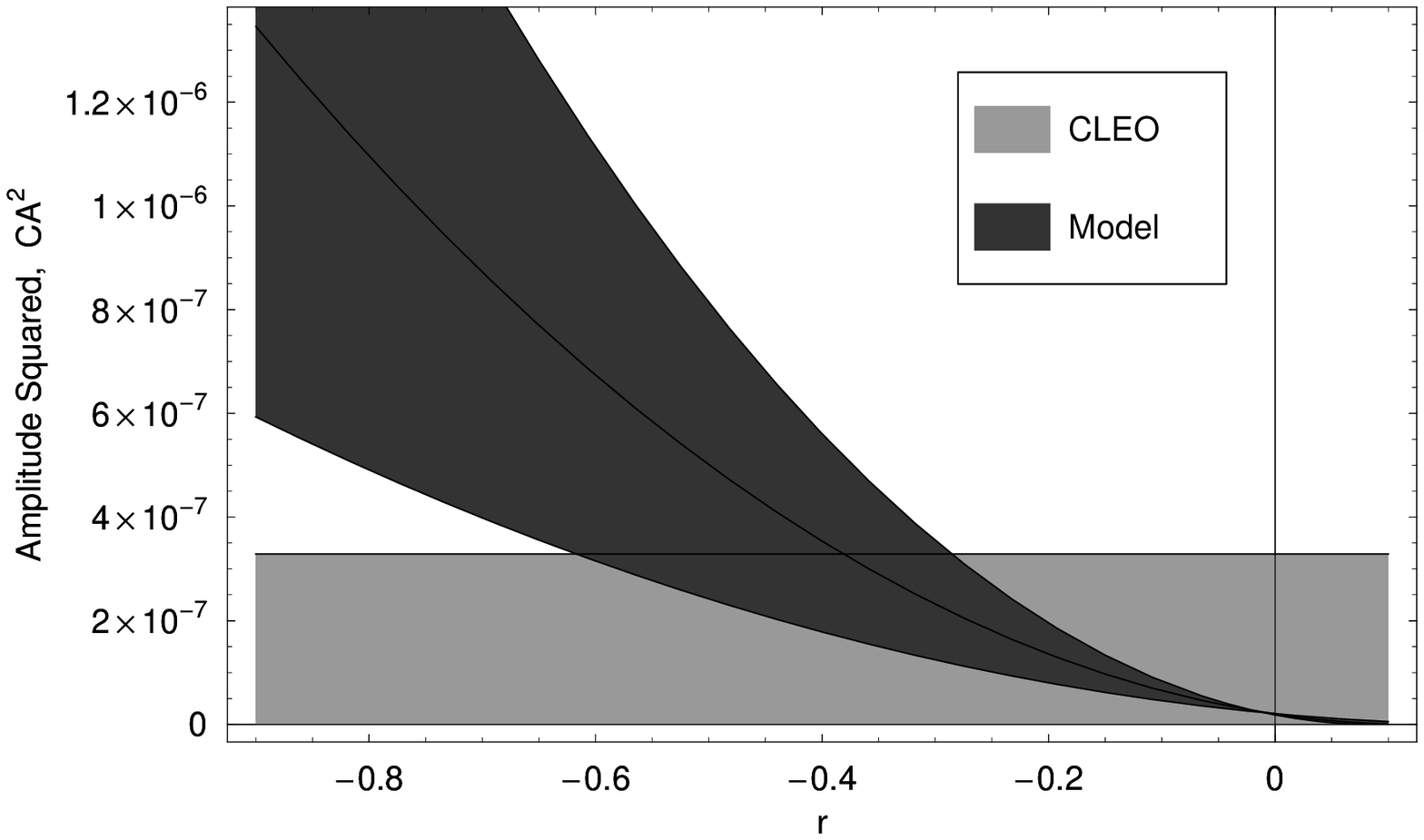}
\caption{$\chi_{c0} \rightarrow \eta\eta'$ with no form factor.  Horizontal band shows the experimentally allowed region (upper limit at $90\%$ C.L.) from CLEO\cite{Adams:2006na}.  Curves show predictions for $\phi=40\degrees$ and the range $35\degrees$ to $45\degrees$.}
\label{fig:ChiC0plot_etaetap_r_noff}
\end{center}
\end{figure}

We plot the reduced branching ratios (after the momentum dependence has been removed but including the $C$ factor) as a function of $r$ for $\phi=40\degrees$ and the range $35\degrees$ to $45\degrees$ with no form factor for the $\eta \eta$, $\eta' \eta'$, and $\eta \eta'$ modes in Figures \ref{fig:ChiC0plot_etaeta_r_noff}, \ref{fig:ChiC0plot_etapetap_r_noff}, and \ref{fig:ChiC0plot_etaetap_r_noff} respectively.  The horizontal bands show the experimental data from the PDG Review 2006\cite{PDG06} and CLEO\cite{Adams:2006na}.  If a form factor is included ($\beta=400\MeV$ and $\beta=500\MeV$) the general features are the same although the quantitative details change.  As expected, in all cases the branching ratios vary slowly with $\phi$ for small $r$.  

The $\eta' \eta'$ modes provide the most constraint.  With no form factor we can constrain $-0.12<r<-0.01$.  However, there is also a region only just excluded by the $\eta \eta'$ limit close to $r=-0.65$.  Taking these ranges of $\phi$ and $r$ ($-0.12<r<-0.01$) leads to a predicted $\eta \eta'$ branching ratio of $3.5\times 10^{-5} - 1.6\times 10^{-4}$.  Using a form factor with $\beta=500\text{MeV}$ we can constrain $-0.15<r<-0.07$.  (There is now also an allowed region close to $r=-0.55$)  Taking these ranges of $r$ and $\phi$ gives predicted $\eta \eta'$ branching ratio of $1.8\times 10^{-5} - 1.5\times 10^{-4}$.  Finally, with $\beta=400\text{MeV}$ we can constrain  $-0.17<r<-0.08$.  (Again there is also an allowed region closer to $r=-0.55$.)  Taking these ranges of $r$ and $\phi$ leads to a predicted $\eta \eta'$ branching ratio of $6.7\times 10^{-6} - 1.6\times 10^{-4}$.

\begin{figure}[tb]
\begin{center}
\includegraphics[width=15cm]{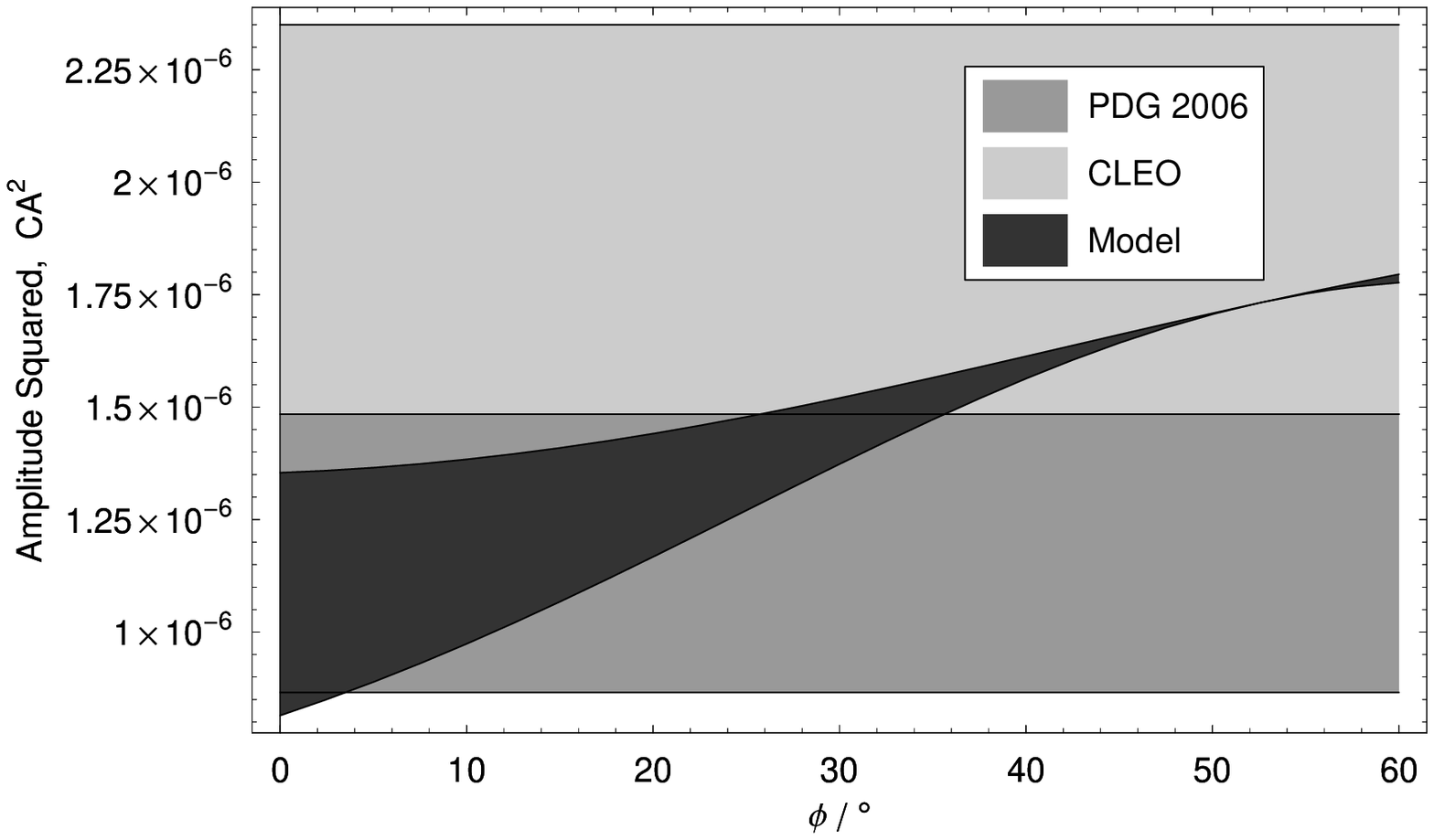}
\caption{$\chi_{c0} \rightarrow \eta\eta$ with no form factor.  Horizontal bands show the experimental data ($1\sigma$ range) from the PDG\cite{PDG06} and CLEO\cite{Adams:2006na}.  Curves are predictions for the range $r=-0.12$ to $-0.01$.}
\label{fig:ChiC0plot_etaeta_phi_noff}
\end{center}
\end{figure}

\begin{figure}[tb]
\begin{center}
\includegraphics[width=15cm]{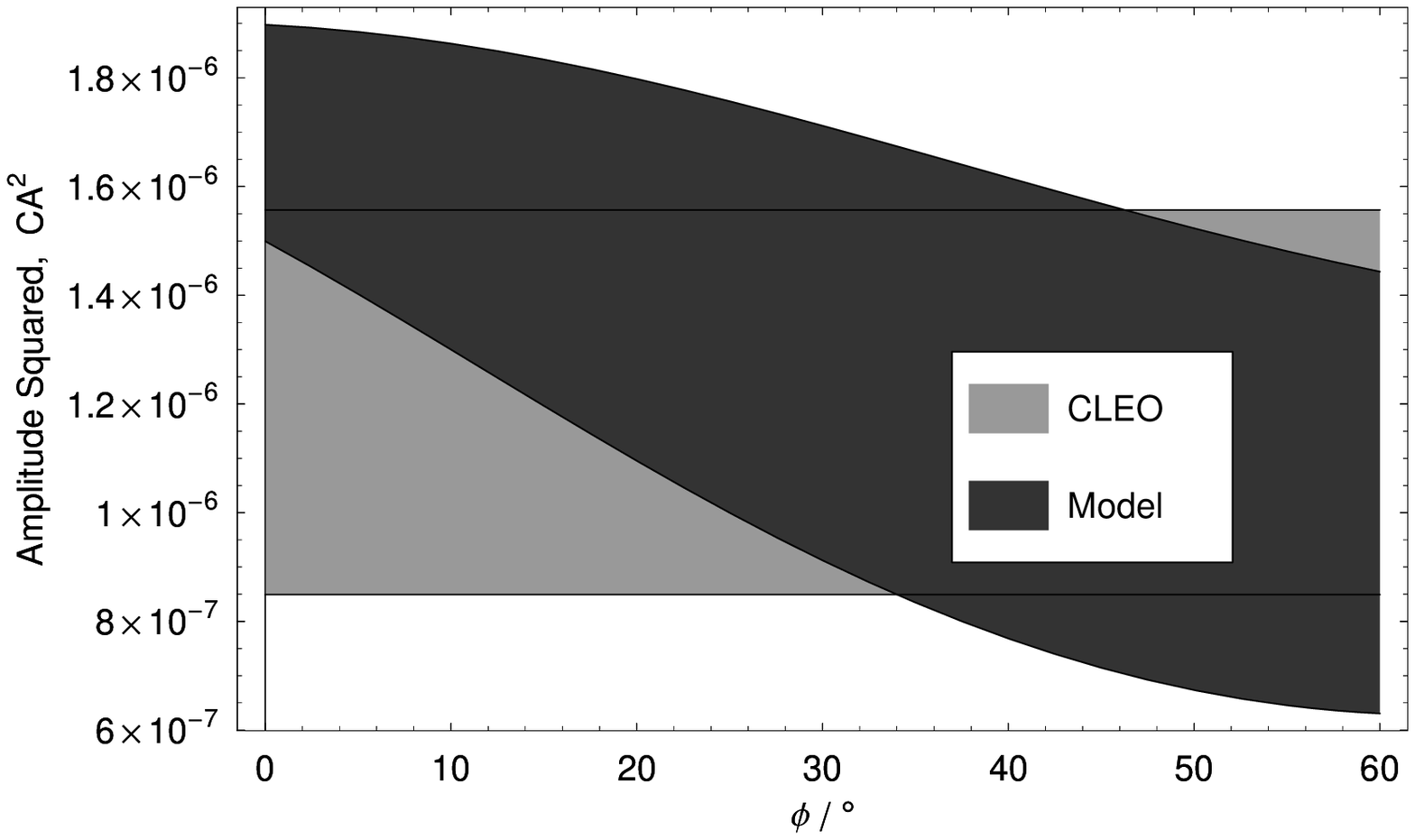}
\caption{$\chi_{c0} \rightarrow \eta'\eta'$ with no form factor.  Horizontal band shows the experimental data ($1\sigma$ range) from CLEO\cite{Adams:2006na}.  Curves are predictions for the range $r=-0.12$ to $-0.01$.}
\label{fig:ChiC0plot_etapetap_phi_noff}
\end{center}
\end{figure}

\begin{figure}[tb]
\begin{center}
\includegraphics[width=15cm]{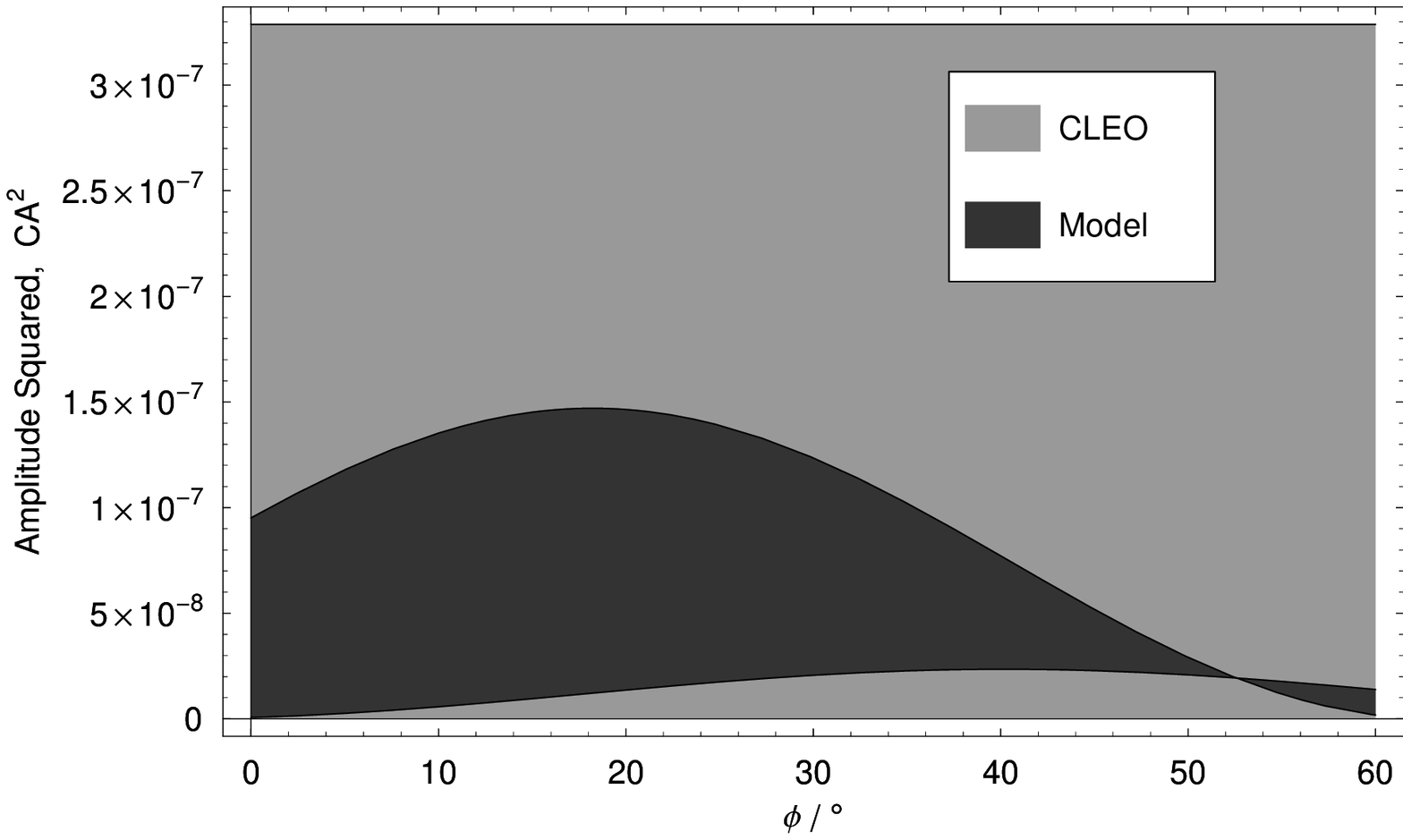}
\caption{$\chi_{c0} \rightarrow \eta\eta'$ with no form factor.  Horizontal band shows the experimentally allowed region (upper limit at $90\%$ C.L.) from CLEO\cite{Adams:2006na}.  Curves are predictions for the range $r=-0.12$ to $-0.01$.}
\label{fig:ChiC0plot_etaetap_phi_noff}
\end{center}
\end{figure}

We now turn to the dependence on the mixing angle $\phi$ for the allowed values of $r$. We plot reduced branching ratios (after the momentum dependence has been removed but including the $C$ factor) for $r=-0.12$ and $r=-0.01$ with no form factor for the $\eta \eta$, $\eta' \eta'$, and $\eta \eta'$ modes in Figures \ref{fig:ChiC0plot_etaeta_phi_noff}, \ref{fig:ChiC0plot_etapetap_phi_noff}, and \ref{fig:ChiC0plot_etaetap_phi_noff} respectively.  As expected with small $r$, $\phi$ is not very well constrained at all.  This situation does not change significantly when we include a form factor.

We can not extract or predict anything from the $\chi_{c2}$ decays because we have no constraint on $r$.  However, once experimental data exists for one of the $\eta$ and $\eta'$ modes, the others can be predicted using a value of $\phi$ and the method used for $\chi_{c0}$ decays.

In summary, we get consistent results for $\phi \approx 35\degrees - 45\degrees$ but can not strongly constrain $\phi$ from these modes.  However, we can predict $BR(\chi_{c0} \rightarrow \eta \eta') = 7\times 10^{-6} - 1.6\times 10^{-4}$ for this range of $\phi$, or $2\times 10^{-5} - 1.2\times 10^{-4}$ for $\phi=40\degrees$.

\section{Semileptonic and Hadronic B and D Decays}
\label{sec:BDecays}

In this section we investigate what can be learnt on the composition of the $\eta$ and $\eta'$ mesons from their production in electroweak decays of bottom and charmed mesons.  Here we do not attempt a detailed analysis of all the decays.  Detailed discussion of form factors and interfering amplitudes is deferred to future work.  We summarise what can be learnt in a theoretically clean way from these decays with few assumptions and will also review work in the literature.

\subsection{Semileptonic Decays}
\label{sec:SemileptonicDecays}

Semileptonic decays are simpler that hadronic decays because they contain only one strongly interacting meson in the final state.  However, mass differences between the $\eta$ and $\eta'$ require a knowledge of the form factors (including momentum dependence) which adds complication and model dependence.  

Bottom meson semileptonic decays (including the weak neutral current decays) are discussed in the literature.  See, for example, Refs.\ \cite{Kim:2003ka,Skands:2000ru,Bashiry:2005nq,Chen:2006qr,Akeroyd:2007fy}.  Form factors have been calculated in light cone QCD sum rules (LCSRs)\cite{Aliev:2002tra,Aliev:2003ge,Ball:2004ye,Ball:2007hb} and perturbative QCD (pQCD)\cite{Charng:2006zj} and these can also be used as inputs to hadronic decays.

The only relevant measured semileptonic branching ratio of the $B$ meson is $B^+ \rightarrow \eta l^+ \nu$ and this has large uncertainties\cite{PDG07,Aubert:2006gba}.

Charmed meson semileptonic decays have been studied, for example, in Refs.\ \cite{Lipkin:1992fd,Anisovich:1997dz,Datta:2001ir}.  Datta et. al.\cite{Datta:2001ir} observe that in the ratios $r_{\eta} \equiv \Gamma(D \rightarrow \eta l \nu)/\Gamma(D_s \rightarrow \eta l \nu)$ and $r_{\eta'} \equiv \Gamma(D \rightarrow \eta' l \nu)/\Gamma(D_s \rightarrow \eta' l \nu)$, proportional to $\cot^2 \phi$ and $\tan^2 \phi$ respectively, the form factors should cancel in the U-spin limit (after allowing for different CKM factors).  These Cabibbo favoured semileptonic branching ratios of $D_s^+$ have been measured but the corresponding Cabibbo suppressed branching ratios of $D^+$ only have limits on branching ratios in the PDG 2007 Update\cite{PDG07}.  Datta et. al. also suggest how to extract the mixing angle from differential decay rates.

In principle, the weak neutral current decays ($B^0_{(s)} / D^0 \rightarrow \eta/\eta' l^+ l^-$ and $\eta/\eta' \nu \bar{\nu}$) can be analysed in a similar way to charged current semileptonic decays.

In summary, semileptonic decays of bottom and charmed mesons are in many ways simpler than hadronic decays.  However, theoretical uncertainties in the form factors and the lack of precise experimental measurements mean that we can not currently cleanly extract the $\eta$ - $\eta'$ mixing from these processes.

\subsection{Hadronic Bottom Meson Decays}
\label{sec:HadronicBDecays}

We now examine the hadronic decays of bottom mesons using a topological/diagrammatic analysis\cite{Gronau:1994rj,Lipkin:2005sn}.  We ignore any momentum dependence in the form factors except for simple phase space ($\propto \mvp^{2L+1}$ where $L$ is the relative orbital angular momentum of the two mesons in the final state and $\mvp$ is the final state recoil momentum in the rest frame of the $B$ meson) and defer any detailed discussion of form factors to future work.  To be concise, we will refer to decays to two pseudoscalars but implicitly include any allowed combination of excited mesons where appropriate.

We consider the following types of diagram (figures are given in Appendix \ref{sec:EWDecayAmps}): colour favoured tree ($T$), colour suppressed tree ($C$), penguin ($P2_{u,d,s}$), `gluon hairpin' penguin ($P1_{u,d,s}$), weak-exchange ($E2_{u,d,s}$), `gluon hairpin' weak-exchange ($E1_{u,d,s}$), weak-annihilation ($A2_{u,d,s}$), and `gluon hairpin' weak-annihilation ($A1_{u,d,s}$).  The subscript on the penguin, weak-exchange and weak-annihilation amplitudes refers to the quark-antiquark pair produced at the gluon vertex.   By default the second meson (i.e. $Y$ in $B \rightarrow X + Y$) contains the spectator quark and we denote with a $'$ those amplitudes where the other meson contains the spectator quark.  We refer to the lowest order diagrams but these are just an example of the topologies: any topologically equivalent correction can be incorporated in to these amplitudes.  We give details of the diagrams contributing to each mode along with the CKM factors in Appendix \ref{sec:EWDecayAmps}; here we will summarise the results.

Where the transitions only produce either the non-strange or the strange component of the $\eta$ and $\eta'$, it is relatively straightforward to extract the mixing angle.  In these cases the branching ratios will be proportional to $|X_{\eta/\eta'}|^2$ or $|Y_{\eta/\eta'}|^2$ and the $\eta/\eta'$ ratio of branching ratios will be proportional to $\cot^2 \phi$ or $\tan^2 \phi$ respectively.  The isovector $\pi(u\bar{u} - d\bar{d})$ prevents simple comparison with the isoscalar $\eta/\eta'(u\bar{u} + d\bar{d})$ except where the mesons are produced from only one of either the $u\bar{u}$ or the $d\bar{d}$ components.

Following, for example, Lipkin\cite{Lipkin:2005sn}, we assume that `gluon hairpin' diagrams ($P1$, $E1$ and $A1$) can be ignored.  We assume no gluonic component in the $\eta$ and $\eta'$ mesons and restrict ourselves to extracting the $\eta-\eta'$ mixing angle and checking for consistency.  Any gluonic component could enhance these `gluon hairpin' diagrams and change the pattern of the branching ratios.  Any intrinsic charm component would also distort the pattern.

\begin{table}
\begin{center}
\begin{tabular}{|c|c|c|}
\hline
\textbf{Mode} & \textbf{Branching Ratio} & $\mvp / \MeV$ \\
\hline
$B^+ \rightarrow \pi^+ \eta$ & $(4.9 \pm 0.5)\times10^{-6}$ & $2609$ \\
$B^+ \rightarrow \pi^+ \eta'$ & $(2.6 \pm 1.1)\times10^{-6}$ & $2551$ \\
$B^+ \rightarrow \pi^+ \eta$ & $(5.0 \pm 0.6)\times10^{-6}$\cite{Aubert:2007si} & $2609$ \\
$B^+ \rightarrow \pi^+ \eta'$ & $(3.9 \pm 0.8)\times10^{-6}$\cite{Aubert:2007si} & $2551$ \\
$B^+ \rightarrow \rho^+ \eta$ & $(8.4 \pm 2.2)\times10^{-6}$ & $2553$ \\
$B^+ \rightarrow \rho^+ \eta'$ & $(8.7^{+3.9}_{-3.1})\times10^{-6}$ & $2492$ \\
$B^+ \rightarrow \rho^+ \eta$ & $(4.1 \pm 1.4)\times10^{-6}$\cite{Wang:2007rzb} & $2553$ \\
$B^+ \rightarrow \bar{K}^0 K^+$ & $(1.28 \pm 0.30)\times10^{-6}$ & $2593$ \\
\hline
$B^0 \rightarrow J/\psi \pi^0$ & $(2.05\pm0.24)\times10^{-5}$ & $1728$ \\
$B^0 \rightarrow J/\psi \eta$ & $<2.7\times10^{-5}$ (90\% CL) & $1672$ \\
$B^0 \rightarrow J/\psi \eta'$ & $<6.3 \times10^{-5}$ (90\% CL) & $1546$ \\
$B^0 \rightarrow J/\psi \eta$ & $(9.5\pm1.9)\times10^{-6}$\cite{Chang:2006sd} & $1672$ \\
\hline
$B^0 \rightarrow \bar{D}^0 \pi^0$ & $(2.61\pm0.24)\times10^{-4}$ & $2308$ \\
$B^0 \rightarrow \bar{D}^0 \eta$ & $(2.02\pm0.35)\times10^{-4}$ & $2274$ \\
$B^0 \rightarrow \bar{D}^0 \eta'$ & $(1.25\pm0.23)\times10^{-4}$ &  $2198$ \\
$B^0 \rightarrow \bar{D}^{*0} \pi^0$ & $(1.7\pm0.4)\times10^{-4}$ & $2256$ \\
$B^0 \rightarrow \bar{D}^{*0} \eta$ & $(1.8\pm0.6)\times10^{-4}$ & $2220$ \\
$B^0 \rightarrow \bar{D}^{*0} \eta'$ & $(1.23\pm0.35)\times10^{-4}$ & $2141$ \\
$B^0 \rightarrow D_s^- K^+$ & $(2.8\pm0.5)\times10^{-5}$ & $2242$ \\
\hline
$B^0 \rightarrow D_s^+ \pi^-$ & $(1.50\pm0.35)\times10^{-5}$ & $2270$ \\
$B^+ \rightarrow D^+ K^0$ & $<5.0\times10^{-6}$ (90\% CL) & $2278$ \\
\hline
\end{tabular}
\caption{Experimental Data on $B$, $B_s$ and $B_c$ decays used in this section.  From the PDG 2007 Update\cite{PDG07} unless otherwise stated.}
\label{table:BDecayExpData}
\end{center}
\end{table}

\begin{table}[htb]
\begin{center}
\begin{tabular}{|c|c|c|}
\hline
\textbf{Mode} & \textbf{$\phi/\degrees$} & \textbf{Notes} \\
\hline
$B^+ \rightarrow \pi^+ + \eta'/\eta$ & $36\pm6$ &  \\
$B^+ \rightarrow \pi^+ + \eta'/\eta$ & $42^{+3}_{-4}$ &  Using new BABAR results\cite{Aubert:2007si} \\
$B^+ \rightarrow \rho^+ + \eta'/\eta$ & $47^{+6}_{-8}$ &   \\
$B^+ \rightarrow \rho^+ + \eta'/\eta$ & $57^{+6}_{-9}$ &  Using new Belle result\cite{Wang:2007rzb} \\
\hline
$B^0 \rightarrow J/\psi + \eta/\pi$ & $44^{+6}_{-7}$ &  \\
\hline
$B^0 \rightarrow \bar{D}^0 + \eta'/\eta$ & $39^{+3}_{-4}$ &  \\
$B^0 \rightarrow \bar{D}^{*0} + \eta'/\eta$ & $41^{+6}_{-7}$ &  \\
\hline
\end{tabular}
\caption{Mixing angle determinations from $B$ meson decays}
\label{table:BDecayMixingAngles}
\end{center}
\end{table}

The following modes only couple to the non-strange component when we ignore `gluon hairpins'.  Hence they should all allow clean determinations, although in some cases they are highly suppressed by CKM factors.  There have not been experimental measurements of many of the branching ratios but some have been measured and some should be accessible.  Relevant experimental data are given in Table \ref{table:BDecayExpData} and the mixing angles extracted are given in Table \ref{table:BDecayMixingAngles}.

\begin{itemize}

\item $B^+ \rightarrow \pi^+ + \eta/\eta'/\pi^0$:  There has been much discussion in the literature on these modes, to which we give references below, and we refer to this for a detailed discussion of the contributing amplitudes.  The tree and annihilation diagrams have CKM factor $\approx 4\times10^{-3}$ but the penguin has leading order CKM factor $\approx 1\times10^{-2}$ and so the penguin diagrams are expected to be significant.  This can be seen empirically from the observation that the branching ratio to $\bar{K}^0 K^+$ is of same order as the branching ratios to $\pi^+ + \eta/\eta'/\pi^0$.  The mixing angle extracted from simple ratios (Table \ref{table:BDecayMixingAngles}) is consistent with other determinations, except when we use the new Belle measurement\cite{Wang:2007rzb} of the branching ratio to $\rho^+ \eta$; this will lead to a puzzle if confirmed.

\item $B^0 \rightarrow (\eta/\eta'/\pi^0)+(\eta/\eta'/\pi^0)$:  Again, there has been a lot of discussion in the literature and we refer to references below.  As for $B^+ \rightarrow \pi^+ + \eta/\eta'/\pi^0$, more than one type of diagram is expected to be important.  Empirically, the penguin diagrams are important (from the $\bar{K}^0 K^0$ branching ratio measurement) but the weak-exchange diagrams are less so (from the $K^- K^+$ branching ratio limit).  There are not enough data to be able to extract the mixing angle cleanly, but the fact that data for $\pi^0 \eta'$ exist suggests that measurement of the $\pi^+ \eta$ branching ratio should be forthcoming.  Extracting anything from only the $\pi^0 \eta'$ and $\pi^0 \pi^0$ modes would require a knowledge of the different diagrams' amplitudes and phases.

\item $B^0 \rightarrow (c\bar{c}) + \eta/\eta'/\pi^0$ (leading CKM factor $\approx 1\times10^{-2}$).  There is only coupling to $d\bar{d}$ and so the $\pi^0$ mode can be used as normalisation.  There are new data from Belle\cite{Chang:2006sd} on the $J/\psi \eta$ mode which, along with the $J/\psi \pi^0$ measurement in the PDG 2007 Update (see Table \ref{table:BDecayExpData}), allows us to extract the mixing angle (Table \ref{table:BDecayMixingAngles}).  In addition we predict the $J/\psi \eta'$ branching ratio to be $\approx (4 - 7) \times10^{-6}$ depending on whether we take the $\phi$ extracted from this mode, or we use $\phi \approx 40\degrees$ with either the $J/\psi \pi^0$ mode or the $J/\psi \eta$ mode.  This prediction satisfies the current experimental upper limit. 

Datta et. al.\cite{Datta:2001ir} have suggested a way to use the $J/\psi K^0$ mode (with a different CKM factor) along with the $B_s \rightarrow (c\bar{c}) + \eta/\eta'/\pi^0/\bar{K}^0$ (see below) decays to overcome kinematic factors and test the $\eta-\eta'$ mixing.  (This CKM factor should be the as that in $B^+ \rightarrow (c\bar{c}) + K^+$ compared with $B^+ \rightarrow (c\bar{c}) + \pi^+$ which proceed through a similar colour suppressed tree diagram.)  These decays have also been studied in the literature\cite{Skands:2000ru}, using the pQCD approach\cite{Liu:2007yg} and using the QCD improved factorisation approach (QCDF) with the non-factorisable hard spectator correction from pQCD\cite{Li:2007xf}.  The branching ratios are generally found to be consistent with a mixing angle $\approx 40\degrees$, although some of these predicted branching ratios have large uncertainties.

\item $B^0 \rightarrow \bar{D}^0 + \eta/\eta'/\pi^0$ (CKM factor $\approx 4\times10^{-2}$):  There are data in the PDG 2007 Update\cite{PDG07} (see Table \ref{table:BDecayExpData}).  The mixing angle extracted (Table \ref{table:BDecayMixingAngles}) is consistent with other determinations.  These modes have been studied in the pQCD approach\cite{Lu:2003xc}.  The measurement of the branching ratio to $D_s^-K^+$ (approximately an order of magnitude smaller than these $\bar{D}^0 + \eta/\eta'/\pi^0$ branching ratios) gives a handle on the weak-exchange diagram $E2$.  This can enable the different contributions to be disentangled and hence allow the $\pi^0$ mode to be used in the analysis.  

\item $B^0 \rightarrow D^0 + \eta/\eta'/\pi^0$:  These modes have the same structure as $B^0 \rightarrow \bar{D}^0 + \eta/\eta'/\pi^0$ modes but with a smaller CKM factor ($\approx 9\times10^{-4}$).  There are no data in the PDG 2007 Update.

\item $B^+ \rightarrow D^+ + \eta/\eta'/\pi^0$:  These modes have a very small CKM factor ($\approx 9\times10^{-4}$) and there are no data in the PDG 2007 Update.  The $D_s^+ \bar{K}^0$ mode gives a handle on the weak-annihilation diagram ($A2$) and this enables the different contributions to be disentangled.  If $A2$ is negligible compared to the colour favoured tree ($T$), the transition is only to $u\bar{u}$ and so the $\pi^0$ mode could easily be used when extracting the mixing parameters.

\item $B_c^+ \rightarrow D^+ + \eta/\eta'/\pi^0$:  There are no data in the PDG 2007 Update.  Comparison with the $\pi^0$ mode is complicated by interference between $d\bar{d}$ and $u\bar{u}$ components from respectively the colour suppressed tree (CKM factor $\approx 4\times10^{-3}$) and penguins and weak-annihilation diagrams (leading CKM factor $\approx 1\times 10^{-2}$).  The $D_s^+ \bar{K}^0$ mode can be used to get a handle on the penguin and weak-annihilation contributions and so to disentangle the different contributions.
  
\item $B_c^+ \rightarrow \pi^+ + \eta/\eta'/\pi^0$ (CKM factor $\approx 4\times10^{-2}$):  There are no data in the PDG 2007 Update.  These decays only occur by weak-annihilation and so could be suppressed; they have a relatively large CKM factor.

\end{itemize}

The following mode only couples to the strange component when we ignore `gluon hairpin' diagrams:

\begin{itemize}

\item $B_s^0 \rightarrow (c\bar{c}) + \eta/\eta'/\pi^0$ (leading CKM factor $\approx 4\times10^{-2}$):  There are no data on these processes in the PDG 2007 Update, but the measurement of the $J/\psi\ \phi$ branching ratio suggests that these modes should be accessible.  These modes have been studied in the literature by Datta et. al.\cite{Datta:2001ir} who suggest a method of combining with the related $B^0$ decays (see above) to test whether the $\eta$ and $\eta'$ have any additional constituents beyond $n\bar{n}$ and $s\bar{s}$.

\end{itemize}

The following modes couple to both the non-strange and the strange components and hence do not allow a clean extraction of the mixing angles.  They require further approximations and/or a more detailed analysis.

\begin{itemize}

\item $B \rightarrow K + \eta/\eta'/\pi$:  These are expected to be penguin dominated (leading CKM factor $\approx 4\times10^{-2}$) because of the relative CKM factors.  Many of these branching ratios have been measured.  We do not attempt to extract the mixing angle here because of the complicated pattern of interfering amplitudes and because there are open theoretical questions regarding these modes.  There has been extensive discussion in the literature on these modes and we review this below.

\item $B^+ \rightarrow D_s^+ + \eta/\eta'/\pi^0$:  These modes have a small CKM factor ($\approx 4\times10^{-3}$) and there are currently only limits on branching ratios in the PDG 2007 Update.  If the weak-annihilation diagram ($A2$) is negligible compared to the colour favoured tree diagram ($T$), the transition is only to $u\bar{u}$ and hence the mixing parameters can be extracted cleanly.  The measurement of the $B^0 \rightarrow D_s^+ \pi^-$ branching ratio and the limit on the $B^+ \rightarrow D^+ K^0$ branching ratio (see Table \ref{table:BDecayExpData}) from the PDG 2007 Update decays can be used to disentangle the $T$ and the $A2$ contributions.  These results suggest that ignoring $A2$ may be reasonable.

\item $B_s^0 \rightarrow \bar{K}^0 + \eta/\eta'/\pi^0$:  The tree diagram has CKM factor $\approx 4\times10^{-3}$ but the penguin has leading order CKM factor $\approx 1\times10^{-2}$ and so more than one type of diagram is expected to be significant.  It may be possible to disentangle these modes with a full set of data; there is no data in the PDG 2007 Update.  These modes have been studied in the literature, for example, Ref.\ \cite{Sun:2002rn}.

\item $B_s^0 \rightarrow (\eta/\eta'/\pi) + (\eta/\eta'/\pi)$:  Here the tree and weak-exchange diagrams have CKM factor $\approx 9\times10^{-4}$ but the penguins have leading order CKM factor $\approx 4\times10^{-2}$.  The penguin diagrams may therefore dominate or more than one type of diagram could be significant (c.f. $B^0 \rightarrow K^+ + \eta/\eta/\pi$ above).  The different components could be disentangled in principle with sufficient data; there are only limits on branching ratios in the PDG 2007 Update.  These modes have been studied in the literature, for example, Refs.\ \cite{Sun:2002rn,Xiao:2006gf,Chen:2007qm,Liu:2007sh}.    

\item $B_s^0 \rightarrow \bar{D}^0 + \eta/\eta'/\pi^0$ (CKM factor $\approx 1\times10^{-2}$):  There are no data on these modes in the PDG 2007 Update.  The $D^- \pi^+$ and $\bar{D}^0 \pi^0$ modes can be used to get a handle on the weak-exchange diagram ($E2$).  If $E2$ can be ignored compared to the colour suppressed tree ($C$), the transition is only to the strange components and so the mixing parameters can be extracted straightforwardly.  

\item $B_s^0 \rightarrow D^0 + \eta/\eta'/\pi^0$ (CKM factor $\approx 4\times10^{-3}$): these processes have the same structure as $B_s^0 \rightarrow \bar{D}^0 + \eta/\eta'/\pi^0$ and $D^- \pi^+$ decays but with a smaller CKM factor.  There are no data on these modes in the PDG 2007 Update.

\item $B_c^+ \rightarrow D_s^+ + \eta/\eta'/\pi^0$:  There are no data in the PDG 2007 Update.  The colour suppressed tree (CKM factor $\approx 9\times10^{-4}$) couples to the $u\bar{u}$ component and the penguins and weak-annihilation diagrams (leading CKM factor $\approx 4\times 10^{-2}$) couple to the $s\bar{s}$ component.  If only the leading CKM part is significant, then only the strange components contribute and the mixing angle can be extracted simply.  This approximation can be tested using the $D_s^+ \pi^0$ and $D^+ K^0$ modes.

\item $B_c^+ \rightarrow K^+ + \eta/\eta'/\pi^0$ (CKM factor $\approx 1\times10^{-2}$):  There are no data in the PDG 2007 Update.  These decays only occur by weak-annihilation ($A2$) and so could be small.  The transition is to both the $u\bar{u}$ and $s\bar{s}$ components.  The $K^0 \pi^+$ mode can be used to help extract the mixing angle.

\end{itemize}

Now that we have surveyed the relevant processes, we review in more detail the decays of $B$ to light mesons ($B \rightarrow K + \eta/\eta'/\pi$ and $B \rightarrow \eta/\eta'/\pi + \eta/\eta'/\pi$) and the related semi-inclusive processes $B \rightarrow \eta/\eta' + X_{s}$.  There has been extensive discussion in the literature on these processes: a range of approaches have been used to calculate the relevant amplitudes, such as QCDF\cite{Du:2001hr,Du:2002up,Beneke:2002jn,Beneke:2003zv,Dutta:2003ha,Yang:2000ce}, pQCD\cite{Liu:2005mm,Wang:2005bk,Xiao:2006mg,Guo:2007vw}, and soft collinear effective theory (SCET) \cite{Williamson:2006hb}.  The $B \rightarrow \eta/\eta'$ form factors have been calculated, for example, using LCSRs and pQCD (see Section \ref{sec:SemileptonicDecays} above).  In addition, there have been many topological, diagrammatic and symmetry analyses (for example, Refs.\ \cite{Lipkin:1990us,Gronau:1994rj,Dighe:1995gq,Dighe:1997hm,Gronau:1999hq,Lipkin:2000sf,Datta:2002nm,Lipkin:2005sn,Gronau:2006eb,Lipkin:2007yi}), some of which also consider CP asymmetries (for example, Refs.\ \cite{Chiang:2001ir,Chiang:2003rb,Chiang:2003rb,Gronau:2006qh,Escribano:2007mq}).

Various mechanisms have been proposed to explain the relatively large $B\rightarrow K \eta'$ and $B \rightarrow \eta' X_s$ branching ratios such as anomaly contributions, enhanced singlet penguin amplitudes, spectator scattering and weak-annihilation contributions (see for example Refs. \cite{Atwood:1997bn,Hou:1997wy,Kagan:1997qn,Ahmady:1997fa,Fritzsch:1997ps,Du:1997hs,He:1997xk,Ali:1997ex,Du:1998rp,Cheng:1997if,Kou:2001pm,Gerard:2006ch,Charng:2006zj}), intrinsic charm in the $\eta'$ (for example Refs. \cite{Halperin:1997as,Yuan:1997ts}), radial mixing in the $\eta$-$\eta'$ system\cite{Datta:2002pk} (this reference also includes comments on the related $\Lambda_b \rightarrow \Lambda \eta/\eta'$ decays), or anomalous tensor operators\cite{Chang:2006dh}.  

The pattern of exclusive modes $K \eta$, $K \eta'$, $K^* \eta$ and $K^* \eta'$ can be explained qualitatively by the interference of strange and non-strange contributions\cite{Lipkin:1990us,Lipkin:2000sf}.  The relative strange penguin ($P2'_s$) to non-strange penguin ($P2_{u,d}$) phase is reversed in the vector $K^*$ modes compared with the pseudoscalar $K$ modes leading to a `parity selection rule'.  (Note that the spectator quark is in a different meson in the strange penguin compared to the non-strange penguin).  This leads to constructive (destructive) interference for the $\eta$ ($\eta'$) in the $K^*$ mode and vice-versa in the $K$ mode.  The `parity selection rule' can be used to probe the nature of the $\eta'$ enhancement\cite{Lipkin:2000sf,Lipkin:2005sn}: any additional diagram such as the `gluon hairpin' diagrams should appear in all similar final states and be independent of the parity of the final state and so violate this rule.

This explanation is supported by calculations in QCDF by Beneke and Neubert\cite{Beneke:2002jn} where they find that the observed branching ratios are due to constructive or destructive interference of non-singlet penguin amplitudes rather than enhanced singlet (i.e. `gluonic penguin' type) amplitudes.  However, the calculated branching ratios have large theoretical uncertainties.  Charng et. al.\cite{Charng:2006zj} use the pQCD approach and find the singlet gluonic contribution to be negligible in $B \rightarrow \eta$, and giving at most a few percent contribution to $B \rightarrow \eta'$.

Lipkin\cite{Lipkin:2007yi} has reviewed the latest data and concludes that the `parity selection rule' prediction holds but that a sum rule is violated indicating an additional contribution to the $\eta-\eta'$ system.  Whether this additional contribution is from `gluon hairpin' penguins\cite{Dighe:1995gq,Dighe:1997hm}, an electroweak penguin or admixtures in the $\eta/\eta'$ is still an open question.  For example, an intrinsic charm component of a few percent in the $\eta$ and/or $\eta'$ could explain the sum rule discrepancy\cite{Lipkin:2007yi}.

It should be noted that there are puzzles in the $B \rightarrow \pi \pi$, $\pi K$ and $K K$ processes even without the $\eta$ and $\eta'$.  See, for example, Fleischer\cite{Fleischer:2007hj} for a review of these.

In summary, we find some hadronic decays where the mixing angle can be extracted cleanly with the current experimental data, some where more data will allow this, and some where a more detailed knowledge of the different amplitudes is required.

\subsection{Hadronic Charmed Meson Decays}
\label{sec:HadronicDDecays}

We now apply the same methods to the hadronic charmed meson decays.  In general these are less straightforward: they all couple to both the non-strange and strange component even when we ignore the 'gluon hairpin' diagrams.  In addition, there is a more significant difference in kinematics, and hence form factors, between the $\eta$ and the $\eta'$ modes.  Because of this we will review the different relevant processes but will not attempt to extract mixing parameters.  These decays have been studied in the literature: see, for example, Refs.\ \cite{Lipkin:1980tk,Lipkin:1990us,Lipkin:1992fd,Close:1996ku,Cheng:1998dn,Rosner:1999xd,Cheng:2000fd,Chiang:2001av,Chiang:2002mr}

\begin{itemize}

\item $D^+ \rightarrow K^+ + \eta/\eta'/\pi^0$ (CKM factor $\approx 5\times10^{-2}$):  This is clean if weak-annihilation diagrams ($A2$) can be ignored compared to the colour favoured tree; in this case only the $d\bar{d}$ component contributes.  There are no data on $\eta$ or $\eta'$ modes in the PDG 2007 Update.

\item $D^0 \rightarrow \bar{K}^0 + \eta/\eta'/\pi^0$ (CKM factor $\approx 9\times10^{-1}$) and $D^0 \rightarrow K^0 + \eta/\eta'/\pi^0$ (CKM factor $\approx 5\times10^{-2}$):  These modes are clean if the weak-exchange diagram ($E2$) is negligible compared to the colour suppressed tree ($C$); in this case the transitions are only to $u\bar{u}$.  This approximation's validity is empirically questioned by the fact that the $K_S^0 \rho^0$ is significantly smaller than the $K_S^0 \omega$ mode (they would be equal in this approximation), although the $\bar{K}^{*0} \rho^0$ is compatible with the $\bar{K}^{*0} \omega$ mode\cite{PDG07}.  

Lipkin\cite{Lipkin:1990us} comments that the relative phase of the strange to non-strange weak-exchange contributions will reverse in the $K^{*}$ modes compared with the $K$ modes, analogously to the $B \rightarrow K + \eta/\eta'/\pi$ decays discussed above.  Therefore, comparing with the $D^0 \rightarrow K^* + \eta/\eta'/\pi$ modes could help disentangle the contributions to this decay.

\item $D^+ \rightarrow  \pi^+ + \eta/\eta'/\pi^0$ (leading CKM factor $\approx 2\times10^{-1}$).   These modes would be clean if we could approximate to only the colour favoured tree diagram (the transition would then only go to $d\bar{d}$), but it is not necessarily reasonable to ignore the colour suppressed tree diagram.  There are data in the PDG 2007 Update.

\item $D^0 \rightarrow \eta/\eta'/\pi^0 + \eta/\eta'/\pi^0$ (leading CKM factor $\approx 2\times10^{-1}$):  Extracting the mixing here would require knowledge of the different diagrams' magnitudes and phases.  There is a reasonable amount of data in the PDG 2007 update, but not much on the $\eta$ or $\eta'$ modes.

\item $D_s^+ \rightarrow K^+ + \eta/\eta'/\pi^0$ (leading CKM factor $\approx 2\times10^{-1}$):  These modes would be clean if we could approximate to only the colour favoured tree diagram (the transition would then only go to $s\bar{s}$), but it is not necessarily reasonable to ignore the colour suppressed tree diagram.  There are new measurements of some of these branching ratios data from CLEO\cite{Adams:2007mx}.

\item $D_s^+ \rightarrow \pi^+ + \eta/\eta'/\pi^0$ (CKM factor $\approx 9\times10^{-1}$):  These modes are clean if we can ignore weak-annihilation diagrams compared to the colour favoured tree diagrams; in this case the transition is only to $s\bar{s}$.  There are data on these modes in the PDG 2007 Update.

\end{itemize}

In summary, we find that the charmed meson hadronic decays are less clean than the bottom mesons decays: there are theoretical uncertainties due to interfering diagrams and form factors.  A more detailed analysis of different contributing amplitudes and form factors is needed before drawing any conclusions regarding the $\eta$ and $\eta'$ mesons from these decays.

\section{Conclusions}
\label{sec:Conclusions}

We have determined the $\eta-\eta'$ mixing angle $\phi$ using a number of different processes and give a summary in Table \ref{table:MixingAngleDeterminations}.  We have also considered a possible gluonic component of the $\eta'$ and extracted the mixing parameters.  A summary of these is given in Table \ref{table:GluonicDeterminations}.

The mixing angle $\phi$ is consistent between different determinations and favours an angle close to $\phi=42\degrees$; a precise determination has been possible.  The data are generally consistent with no additional constituents of the $\eta'$ but there is a hint of a small gluonic component.  This is more difficult to pin down precisely: the various determinations are generally consistent but show more model and mode dependence.  From radiative $J/\psi$ decays and $\psi'$ decays we found that the $c\bar{c}$ components of the $\eta$ and $\eta'$ are $\lesssim 5\%$ in amplitude.

We have explained why we, like Escribano and Nadal\cite{Escribano:2007cd}, think that the KLOE analysis\cite{Ambrosino:2006gk} of glue content of the $\eta'$ is inconsistent.  KLOE reach a conclusion on the glue content of the $\eta'$ which we think is too strong because of this inconsistency and the theoretical uncertainties discussed in Section \ref{sec:RadDecays}.

\begin{table}
\begin{center}
\begin{tabular}{|c|c|}
\hline
\textbf{Method} & \textbf{$\phi / \degrees$} \\
\hline
Light Meson Rad. Decays (no form factor) & $41.3 \pm 0.8$ \\
Light Meson Rad. Decays (new KLOE result, no form factor) & $41.7 \pm 0.5$ \\
Light Meson Rad. Decays ($\beta=400\MeV$) & $41.9\pm1.1$ \\
Light Meson Rad. Decays (new KLOE result, $\beta=400\MeV$) & $42.8\pm0.8$ \\
\hline
$\eta/\eta' \rightarrow \gamma \gamma$ & $38.3\pm1.8$ \\
$\eta/\pi \rightarrow \gamma \gamma$ & $41.3\pm2.0$ \\
\hline
$J/\psi \rightarrow \gamma \eta/\eta'$ ($R=0.708 \pm 0.024$) & $41.1\pm 1.4$ \\
\hline
$J/\psi \rightarrow P V$ (no form factor) & $40 \pm 2$ \\
$J/\psi \rightarrow P V$ ($\beta=500\MeV$) & $37 \pm 3$ \\
$\psi' \rightarrow P V$ (no form factor) & $44^{+7}_{-8}$ \\
$\psi' \rightarrow P V$ ($\beta=500\MeV$) & $43^{+5}_{-6}$ \\
\hline
$B^+ \rightarrow \pi^+ + \eta'/\eta$ & $36\pm6$ \\
$B^+ \rightarrow \pi^+ + \eta'/\eta$ (using new BABAR results\cite{Aubert:2007si}) & $42^{+3}_{-4}$   \\
$B^+ \rightarrow \rho^+ + \eta'/\eta$ & $47^{+6}_{-8}$ \\
$B^+ \rightarrow \rho^+ + \eta'/\eta$ (using new Belle result\cite{Wang:2007rzb}) & $57^{+6}_{-9}$ \\
\hline
$B^0 \rightarrow J/\psi + \eta/\pi$ & $44^{+6}_{-7}$ \\
\hline
$B^0 \rightarrow \bar{D}^0 + \eta'/\eta$ & $39^{+3}_{-4}$  \\
$B^0 \rightarrow \bar{D}^{*0} + \eta'/\eta$ & $41^{+6}_{-7}$  \\
\hline
\end{tabular}
\caption{Mixing angle determinations assuming no glue.  Light meson radiative decays are with $\phi_V=3.4\degrees$ and $C_s \neq C_q$.}
\label{table:MixingAngleDeterminations}
\end{center}
\end{table}

\begin{table}
\begin{center}
\begin{tabular}{|c|c|c|}
\hline
\textbf{Method} & \textbf{$\phi / \degrees$} & \textbf{$\cos{\phi_{G2}}$} \\
\hline
Light Meson Rad. Decays (no form factor) & $41.3 \pm 0.9$ & $0.98 \pm 0.03$ \\
Light Meson Rad. Decays (new KLOE result, no form factor) & $41.3 \pm 0.7$ & $0.98 \pm 0.02$ \\
Light Meson Rad. Decays ($\beta=400\MeV$) & $42.0\pm0.9$ & $0.95 \pm 0.03$ \\
Light Meson Rad. Decays (new KLOE result, $\beta=400\MeV$) & $41.9\pm0.7$ & $0.95 \pm 0.02$ \\
\hline
$\eta,\eta'/\pi \rightarrow \gamma \gamma$ & $41.3\pm2.0$ & $0.90 \pm 0.06$ \\
\hline
$J/\psi \rightarrow P V$ (no form factor) & $45 \pm 4$ & $0.84^{+0.10}_{-0.14}$ \\
$J/\psi \rightarrow P V$ ($\beta=500\MeV$) & $46^{+4}_{-5}$ & $0.72^{+0.10}_{-0.12}$ \\
\hline
\end{tabular}
\caption{Mixing angle determinations with glue in the $\eta'$.  Light meson radiative decays are with $\phi_V=3.4\degrees$ and $C_s \neq C_q$.}
\label{table:GluonicDeterminations}
\end{center}
\end{table}

We have surveyed the semileptonic and hadronic decays of bottom and charmed mesons.  We find some modes where the mixing angle can be extracted cleanly with the current experimental data ($B^+ \rightarrow \pi^+ + \eta/\eta'$, $B^0 \rightarrow (c\bar{c}) + \eta/\eta'$, and $B^0 \rightarrow \bar{D}^0 + \eta/\eta'$), some where more data will allow this ($B^0 \rightarrow \eta/\eta'/\pi^0 + \eta/\eta'/\pi^0$ and $B_s^0 \rightarrow (c\bar{c}) + \eta/\eta'$ should be accessible, $B^0 \rightarrow D^0 + \eta/\eta'$ and $B^+ \rightarrow D^+ + \eta/\eta'$ are suppressed by CKM factors, $B_c^+ \rightarrow D^+ + \eta/\eta'$ and $B_c^+ \rightarrow \pi^+ + \eta/\eta'$ require more measurements of $B_c$ decays), and some where a more detailed knowledge of the different amplitudes is required.  Some of the more complicated sets of decays can be simplified by making further approximations which can be tested using other related processes.  We find no definite evidence for a gluonic component in the $\eta$ or $\eta'$ from these processes, but note also that there are open theoretical questions in analysis of some modes.  These decays provide promising avenues of further investigation with more data becoming available from CLEO, the B-factories, LHCb and any future super B-factories.

More experimental data and data with lower uncertainties will help to pin down the mixing and other possible constituents (such as glue, $c\bar{c}$ and radial excitations).  In particular, measuring processes involving $\eta(1295)$ or $\eta(1405/1475)$, such as decays to $\gamma + \rho/\omega/\phi$ in $\psi \rightarrow \gamma \gamma V$, will give another window on the light pseudoscalar mesons\cite{Close:2004ip} and lead to a more robust analysis.  

The $\psi' \rightarrow \psi \eta$ decay has been measured and the first similar kinematically allowed decay involving the $\eta'$ is $\psi(4160) \rightarrow \psi \eta/\eta'$.  The ratio of $\psi \eta$ to $\psi \eta'$ in this or heavier vector charmonium decays could provide a further probe, for example to constrain the charm content\cite{Harari:1975ie}.

Although it is not possible to determine the mixing angle precisely from $\chi_{c0,2}$ decays, they do provide a test of consistency.  Currently there is only a limit on the $\chi_{c0} \rightarrow \eta \eta'$ mode and a measurement of this could be compared to our prediction: $BR(\chi_{c0} \rightarrow \eta \eta') \approx 2\times 10^{-5} - 1\times 10^{-4}$ (for $\phi=40\degrees$).  In addition, measurements of the corresponding $\chi_{c2}$ decays can be used as outlined above.

Measurement of $\psi' \rightarrow \eta' \gamma$ compared to our prediction of $BR(\psi' \rightarrow \eta' \gamma) \approx 1 \times 10^{-5}$ will provide a further check.

Bottonium decays can provide another source of information by probing the gluonic, charmonium (and bottonium) content of the $\eta$ and $\eta'$ mesons in another kinematic region and starting from a different flavour (we assume that any $b\bar{b}$ component of $\eta'$ should be small compared to any possible small $c\bar{c}$ component).  There is preliminary evidence for $\Upsilon(2S) \rightarrow \Upsilon(1S) \eta$ at CLEO\cite{Kreinick:Charm07:Bottonium}.  CLEO has also measured significantly improved upper limits for $\Upsilon(1S) \rightarrow \gamma \eta/\eta'$\cite{Athar:2007hz,Kreinick:Charm07:Bottonium}.  These and other bottonium decays (such as to $\omega/\rho/\phi + \eta/\eta'$) will provide further tests. 

In summary, there are tantilising hints of glue in the $\eta'$.  However, there is a limit to how well the mixing parameters can be extracted from the data because of the lack of precise theoretical control (for example, form factors) and this is particularly significant for gluonic (or $c\bar{c}$, etc.) component.  This is exemplified by the wavefunction overlap factors $C_s$ and $C_q$ used in Section \ref{sec:RadDecays}: with the experimental situation of a few years ago, these were consistent with being equal but they are no longer so.  Prediction of these form factors, for example by lattice QCD or other theoretical approaches, coupled with more experimental data would more strongly constrain the constituents of the enigmatic light pseudoscalar mesons.

\section*{Acknowledgements}

I thank Frank Close for many useful discussions and Qiang Zhao for useful discussion on $\chi_{c0,2}$ decays.  I thank Rafel Escribano for comments on an earlier version of this manuscript.  This work is supported by a studentship from the Science \& Technology Facilities Council (UK).

\bibliography{EtaEtap}

\begin{thebibliography}{100}

\bibitem{Ambrosino:2006gk}
KLOE, F.~Ambrosino {\em et~al.},
\newblock Phys. Lett. {\bf B648}, 267 (2007), hep-ex/0612029.

\bibitem{Li:2007ky}
G.~Li, Q.~Zhao, and C.-H. Chang,
\newblock (2007), hep-ph/0701020.

\bibitem{Escribano:2007cd}
R.~Escribano and J.~Nadal,
\newblock JHEP {\bf 05}, 006 (2007), hep-ph/0703187.

\bibitem{PDG06}
Particle Data Group, W.-M. {Yao} {\em et~al.},
\newblock {Journal of Physics G} {\bf 33}, 1+ (2006).

\bibitem{rosner:1982ey}
J.~L. Rosner,
\newblock Phys. Rev. {\bf D27}, 1101 (1983).

\bibitem{Haber:1985cv}
H.~E. Haber and J.~Perrier,
\newblock Phys. Rev. {\bf D32}, 2961 (1985).

\bibitem{Seiden:1988rr}
A.~Seiden, H.~F.~W. Sadrozinski, and H.~E. Haber,
\newblock Phys. Rev. {\bf D38}, 824 (1988).

\bibitem{Klempt:2004xg}
E.~Klempt,
\newblock `Beijing 2004, ICHEP 2004' {\bf 2}, 1082 (2004), hep-ph/0409148,
\newblock Contributed to 32nd International Conference on High-Energy Physics
  (ICHEP 04), Beijing, China, 16-22 Aug 2004.

\bibitem{Close:1996yc}
F.~E. Close, G.~R. Farrar, and Z.-p. Li,
\newblock Phys. Rev. {\bf D55}, 5749 (1997), hep-ph/9610280.

\bibitem{Close:2004ip}
F.~E. Close,
\newblock Int. J. Mod. Phys. {\bf A20}, 5156 (2005), hep-ph/0411396.

\bibitem{Feldmann:1998vh}
T.~Feldmann, P.~Kroll, and B.~Stech,
\newblock Phys. Rev. {\bf D58}, 114006 (1998), hep-ph/9802409.

\bibitem{Feldmann:1998sh}
T.~Feldmann, P.~Kroll, and B.~Stech,
\newblock Phys. Lett. {\bf B449}, 339 (1999), hep-ph/9812269.

\bibitem{escribano:1999nh}
R.~Escribano and J.~M. Frere,
\newblock Phys. Lett. {\bf B459}, 288 (1999), hep-ph/9901405.

\bibitem{escribano:2005qq}
R.~Escribano and J.-M. Frere,
\newblock JHEP {\bf 06}, 029 (2005), hep-ph/0501072.

\bibitem{escribano:2005ci}
R.~Escribano,
\newblock PoS {\bf HEP2005}, 418 (2006), hep-ph/0512021.

\bibitem{Kaiser:1998ds}
R.~Kaiser and H.~Leutwyler,
\newblock (1998), hep-ph/9806336.

\bibitem{Kaiser:2000gs}
R.~Kaiser and H.~Leutwyler,
\newblock Eur. Phys. J. {\bf C17}, 623 (2000), hep-ph/0007101.

\bibitem{Email:RafelEscribano}
Private communication with {Rafel Escribano}.

\bibitem{kroll:2005sd}
P.~Kroll,
\newblock Mod. Phys. Lett. {\bf A20}, 2667 (2005), hep-ph/0509031.

\bibitem{kroll:2004rs}
P.~Kroll,
\newblock Int. J. Mod. Phys. {\bf A20}, 331 (2005), hep-ph/0409141.

\bibitem{Gilman:1987ax}
F.~J. Gilman and R.~Kauffman,
\newblock Phys. Rev. {\bf D36}, 2761 (1987).

\bibitem{ball:1995zv}
P.~Ball, J.~M. Frere, and M.~Tytgat,
\newblock Phys. Lett. {\bf B365}, 367 (1996), hep-ph/9508359.

\bibitem{Bramon:1997va}
A.~Bramon, R.~Escribano, and M.~D. Scadron,
\newblock Eur. Phys. J. {\bf C7}, 271 (1999), hep-ph/9711229.

\bibitem{Bramon:1997mf}
A.~Bramon, R.~Escribano, and M.~D. Scadron,
\newblock Phys. Lett. {\bf B403}, 339 (1997), hep-ph/9703313.

\bibitem{Cao:1999fs}
F.-G. Cao and A.~I. Signal,
\newblock Phys. Rev. {\bf D60}, 114012 (1999), hep-ph/9908481.

\bibitem{bramon:2000fr}
A.~Bramon, R.~Escribano, and M.~D. Scadron,
\newblock Phys. Lett. {\bf B503}, 271 (2001), hep-ph/0012049.

\bibitem{Gerard:2004gx}
J.~M. Gerard and E.~Kou,
\newblock Phys. Lett. {\bf B616}, 85 (2005), hep-ph/0411292.

\bibitem{Xiao:2005af}
B.-W. Xiao and B.-Q. Ma,
\newblock Phys. Rev. {\bf D71}, 014034 (2005), hep-ph/0501160.

\bibitem{Huang:2006as}
T.~Huang and X.-G. Wu,
\newblock Eur. Phys. J. {\bf C50}, 771 (2007), hep-ph/0612007.

\bibitem{O'Donnell:1981sj}
P.~J. O'Donnell,
\newblock Rev. Mod. Phys. {\bf 53}, 673 (1981).

\bibitem{nasrallah:2004ms}
N.~F. Nasrallah,
\newblock Phys. Rev. {\bf D70}, 116001 (2004), hep-ph/0410240.

\bibitem{Achasov:2006dv}
M.~N. Achasov {\em et~al.},
\newblock Phys. Rev. {\bf D74}, 014016 (2006), hep-ex/0605109.

\bibitem{Karliner:2006fr}
M.~Karliner and H.~J. Lipkin,
\newblock Phys. Lett. {\bf B650}, 185 (2007), hep-ph/0608004.

\bibitem{Godfrey:1985xj}
S.~Godfrey and N.~Isgur,
\newblock Phys. Rev. {\bf D32}, 189 (1985).

\bibitem{JoThesis}
J.~J. Dudek,
\newblock {\em Phenomenology of Exotic Hadrons - Hybrid Mesons and
  Pentaquarks},
\newblock PhD thesis, Rudolf Peierls Centre for Theoretical Physics, Department
  of Physics, Universisty of Oxford, 2004.

\bibitem{Schechter:1992iz}
J.~Schechter, A.~Subbaraman, and H.~Weigel,
\newblock Phys. Rev. {\bf D48}, 339 (1993), hep-ph/9211239.

\bibitem{Agaev:2003kb}
S.~S. Agaev and N.~G. Stefanis,
\newblock Phys. Rev. {\bf D70}, 054020 (2004), hep-ph/0307087.

\bibitem{Kekez:2005ie}
D.~Kekez and D.~Klabucar,
\newblock Phys. Rev. {\bf D73}, 036002 (2006), hep-ph/0512064.

\bibitem{Kawarabayashi:1980dp}
K.~Kawarabayashi and N.~Ohta,
\newblock Nucl. Phys. {\bf B175}, 477 (1980).

\bibitem{Kawarabayashi:1980uh}
K.~Kawarabayashi and N.~Ohta,
\newblock Prog. Theor. Phys. {\bf 66}, 1789 (1981).

\bibitem{Kopke:1988cs}
L.~Kopke and N.~Wermes,
\newblock Phys. Rept. {\bf 174}, 67 (1989).

\bibitem{Zhao:2006gw}
Q.~Zhao, G.~Li, and C.-H. Chang,
\newblock Phys. Lett. {\bf B645}, 173 (2007), hep-ph/0610223.

\bibitem{Harari:1975ie}
H.~Harari,
\newblock Phys. Lett. {\bf B60}, 172 (1976).

\bibitem{Baltrusaitis:1984rz}
MARK-III, R.~M. Baltrusaitis {\em et~al.},
\newblock Phys. Rev. {\bf D32}, 2883 (1985).

\bibitem{Jousset:1988ni}
DM2, J.~Jousset {\em et~al.},
\newblock Phys. Rev. {\bf D41}, 1389 (1990).

\bibitem{Zhao:2005im}
Q.~Zhao,
\newblock Phys. Rev. {\bf D72}, 074001 (2005), hep-ph/0508086.

\bibitem{Adams:2006na}
CLEO, G.~S. Adams,
\newblock Phys. Rev. {\bf D75}, 071101 (2007), hep-ex/0611013.

\bibitem{Kim:2003ka}
C.~S. Kim, S.~Oh, and C.~Yu,
\newblock Phys. Lett. {\bf B590}, 223 (2004), hep-ph/0305032.

\bibitem{Skands:2000ru}
P.~Z. Skands,
\newblock JHEP {\bf 01}, 008 (2001), hep-ph/0010115.

\bibitem{Bashiry:2005nq}
V.~Bashiry,
\newblock Eur. Phys. J. {\bf C47}, 423 (2006), hep-ph/0509143.

\bibitem{Chen:2006qr}
C.-H. Chen and C.-Q. Geng,
\newblock Phys. Lett. {\bf B645}, 197 (2007), hep-ph/0608246.

\bibitem{Akeroyd:2007fy}
A.~G. Akeroyd, C.-H. Chen, and C.-Q. Geng,
\newblock Phys. Rev. {\bf D75}, 054003 (2007), hep-ph/0701012.

\bibitem{Aliev:2002tra}
T.~M. Aliev, I.~Kanik, and A.~Ozpineci,
\newblock Phys. Rev. {\bf D67}, 094009 (2003), hep-ph/0210403.

\bibitem{Aliev:2003ge}
T.~M. Aliev and M.~Savci,
\newblock Eur. Phys. J. {\bf C29}, 515 (2003), hep-ph/0302187.

\bibitem{Ball:2004ye}
P.~Ball and R.~Zwicky,
\newblock Phys. Rev. {\bf D71}, 014015 (2005), hep-ph/0406232.

\bibitem{Ball:2007hb}
P.~Ball and G.~W. Jones,
\newblock (2007), arXiv:0706.3628 [hep-ph].

\bibitem{Charng:2006zj}
Y.-Y. Charng, T.~Kurimoto, and H.-n. Li,
\newblock Phys. Rev. {\bf D74}, 074024 (2006), hep-ph/0609165.

\bibitem{PDG07}
Particle Data Group, W.-M. {Yao} {\em et~al.},
\newblock {Journal of Physics G} {\bf 33}, 1+ (2006),
\newblock 2007 partial update for the 2008 edition available on the PDG WWW
  pages (URL: http://pdg.lbl.gov/).

\bibitem{Aubert:2006gba}
BABAR, B.~Aubert {\em et~al.},
\newblock (2006), hep-ex/0607066.

\bibitem{Lipkin:1992fd}
H.~J. Lipkin,
\newblock Phys. Lett. {\bf B283}, 421 (1992).

\bibitem{Anisovich:1997dz}
V.~V. Anisovich, D.~V. Bugg, D.~I. Melikhov, and V.~A. Nikonov,
\newblock Phys. Lett. {\bf B404}, 166 (1997), hep-ph/9702383.

\bibitem{Datta:2001ir}
A.~Datta, H.~J. Lipkin, and P.~J. O'Donnell,
\newblock Phys. Lett. {\bf B529}, 93 (2002), hep-ph/0111336.

\bibitem{Gronau:1994rj}
M.~Gronau, O.~F. Hernandez, D.~London, and J.~L. Rosner,
\newblock Phys. Rev. {\bf D50}, 4529 (1994), hep-ph/9404283.

\bibitem{Lipkin:2005sn}
H.~J. Lipkin,
\newblock Phys. Lett. {\bf B633}, 540 (2006), hep-ph/0507225.

\bibitem{Aubert:2007si}
BABAR, B.~Aubert {\em et~al.},
\newblock Phys. Rev. {\bf D76}, 031103 (2007), arXiv:0706.3893 [hep-ex].

\bibitem{Wang:2007rzb}
Belle, C.~H. Wang {\em et~al.},
\newblock Phys. Rev. {\bf D75}, 092005 (2007), hep-ex/0701057.

\bibitem{Chang:2006sd}
Belle, M.~C. Chang {\em et~al.},
\newblock Phys. Rev. Lett. {\bf 98}, 131803 (2007), hep-ex/0609047.

\bibitem{Liu:2007yg}
X.~Liu, Z.-J. Xiao, and H.-S. Wang,
\newblock (2007), arXiv:0704.0395 [hep-ph].

\bibitem{Li:2007xf}
J.-W. Li and D.-S. Du,
\newblock (2007), arXiv:0707.2631 [hep-ph].

\bibitem{Lu:2003xc}
C.-D. Lu,
\newblock Phys. Rev. {\bf D68}, 097502 (2003), hep-ph/0307040.

\bibitem{Sun:2002rn}
J.-f. Sun, G.-h. Zhu, and D.-s. Du,
\newblock Phys. Rev. {\bf D68}, 054003 (2003), hep-ph/0211154.

\bibitem{Xiao:2006gf}
Z.-j. Xiao, X.~Liu, and H.-s. Wang,
\newblock Phys. Rev. {\bf D75}, 034017 (2007), hep-ph/0606177.

\bibitem{Chen:2007qm}
X.-f. Chen, D.-q. Guo, and Z.-j. Xiao,
\newblock (2007), hep-ph/0701146.

\bibitem{Liu:2007sh}
X.~Liu, Z.-J. Xiao, and H.-S. Wang,
\newblock (2007), arXiv:0704.1027 [hep-ph].

\bibitem{Du:2001hr}
D.-s. Du, H.-u. Gong, J.-f. Sun, D.-s. Yang, and G.-h. Zhu,
\newblock Phys. Rev. {\bf D65}, 074001 (2002), hep-ph/0108141.

\bibitem{Du:2002up}
D.-s. Du, H.-j. Gong, J.-f. Sun, D.-s. Yang, and G.-h. Zhu,
\newblock Phys. Rev. {\bf D65}, 094025 (2002), hep-ph/0201253.

\bibitem{Beneke:2002jn}
M.~Beneke and M.~Neubert,
\newblock Nucl. Phys. {\bf B651}, 225 (2003), hep-ph/0210085.

\bibitem{Beneke:2003zv}
M.~Beneke and M.~Neubert,
\newblock Nucl. Phys. {\bf B675}, 333 (2003), hep-ph/0308039.

\bibitem{Dutta:2003ha}
B.~Dutta, C.~S. Kim, S.~Oh, and G.-h. Zhu,
\newblock Eur. Phys. J. {\bf C37}, 273 (2004), hep-ph/0312388.

\bibitem{Yang:2000ce}
M.-Z. Yang and Y.-D. Yang,
\newblock Nucl. Phys. {\bf B609}, 469 (2001), hep-ph/0012208.

\bibitem{Liu:2005mm}
X.~Liu, H.-s. Wang, Z.-j. Xiao, L.~Guo, and C.-D. Lu,
\newblock Phys. Rev. {\bf D73}, 074002 (2006), hep-ph/0509362.

\bibitem{Wang:2005bk}
H.-s. Wang, X.~Liu, Z.-j. Xiao, L.-b. Guo, and C.-D. Lu,
\newblock Nucl. Phys. {\bf B738}, 243 (2006), hep-ph/0511161.

\bibitem{Xiao:2006mg}
Z.-j. Xiao, D.-q. Guo, and X.-f. Chen,
\newblock Phys. Rev. {\bf D75}, 014018 (2007), hep-ph/0607219.

\bibitem{Guo:2007vw}
D.-Q. Guo, X.-F. Chen, and Z.-J. Xiao,
\newblock Phys. Rev. {\bf D75}, 054033 (2007), hep-ph/0702110.

\bibitem{Williamson:2006hb}
A.~R. Williamson and J.~Zupan,
\newblock Phys. Rev. {\bf D74}, 014003 (2006), hep-ph/0601214.

\bibitem{Lipkin:1990us}
H.~J. Lipkin,
\newblock Phys. Lett. {\bf B254}, 247 (1991).

\bibitem{Dighe:1995gq}
A.~S. Dighe, M.~Gronau, and J.~L. Rosner,
\newblock Phys. Lett. {\bf B367}, 357 (1996), hep-ph/9509428.

\bibitem{Dighe:1997hm}
A.~S. Dighe, M.~Gronau, and J.~L. Rosner,
\newblock Phys. Rev. Lett. {\bf 79}, 4333 (1997), hep-ph/9707521.

\bibitem{Gronau:1999hq}
M.~Gronau and J.~L. Rosner,
\newblock Phys. Rev. {\bf D61}, 073008 (2000), hep-ph/9909478.

\bibitem{Lipkin:2000sf}
H.~J. Lipkin,
\newblock Phys. Lett. {\bf B494}, 248 (2000), hep-ph/0009241.

\bibitem{Datta:2002nm}
A.~Datta, H.~J. Lipkin, and P.~J. O'Donnell,
\newblock Phys. Lett. {\bf B540}, 97 (2002), hep-ph/0202235.

\bibitem{Gronau:2006eb}
M.~Gronau, Y.~Grossman, G.~Raz, and J.~L. Rosner,
\newblock Phys. Lett. {\bf B635}, 207 (2006), hep-ph/0601129.

\bibitem{Lipkin:2007yi}
H.~J. Lipkin,
\newblock (2007), arXiv:0705.2557 [hep-ph].

\bibitem{Chiang:2001ir}
C.-W. Chiang and J.~L. Rosner,
\newblock Phys. Rev. {\bf D65}, 074035 (2002), hep-ph/0112285.

\bibitem{Chiang:2003rb}
C.-W. Chiang, M.~Gronau, and J.~L. Rosner,
\newblock Phys. Rev. {\bf D68}, 074012 (2003), hep-ph/0306021.

\bibitem{Gronau:2006qh}
M.~Gronau, J.~L. Rosner, and J.~Zupan,
\newblock Phys. Rev. {\bf D74}, 093003 (2006), hep-ph/0608085.

\bibitem{Escribano:2007mq}
R.~Escribano, J.~Matias, and J.~Virto,
\newblock (2007), arXiv:0708.0119 [hep-ph].

\bibitem{Atwood:1997bn}
D.~Atwood and A.~Soni,
\newblock Phys. Lett. {\bf B405}, 150 (1997), hep-ph/9704357.

\bibitem{Hou:1997wy}
W.-S. Hou and B.~Tseng,
\newblock Phys. Rev. Lett. {\bf 80}, 434 (1998), hep-ph/9705304.

\bibitem{Kagan:1997qn}
A.~L. Kagan and A.~A. Petrov,
\newblock (1997), hep-ph/9707354.

\bibitem{Ahmady:1997fa}
M.~R. Ahmady, E.~Kou, and A.~Sugamoto,
\newblock Phys. Rev. {\bf D58}, 014015 (1998), hep-ph/9710509.

\bibitem{Fritzsch:1997ps}
H.~Fritzsch,
\newblock Phys. Lett. {\bf B415}, 83 (1997), hep-ph/9708348.

\bibitem{Du:1997hs}
D.-s. Du, C.~S. Kim, and Y.-d. Yang,
\newblock Phys. Lett. {\bf B426}, 133 (1998), hep-ph/9711428.

\bibitem{He:1997xk}
X.-G. He, W.-S. Hou, and C.~S. Huang,
\newblock Phys. Lett. {\bf B429}, 99 (1998), hep-ph/9712478.

\bibitem{Ali:1997ex}
A.~Ali, J.~Chay, C.~Greub, and P.~Ko,
\newblock Phys. Lett. {\bf B424}, 161 (1998), hep-ph/9712372.

\bibitem{Du:1998rp}
D.-s. Du, Y.-d. Yang, and G.-h. Zhu,
\newblock Phys. Rev. {\bf D59}, 014007 (1999), hep-ph/9805451.

\bibitem{Cheng:1997if}
H.-Y. Cheng and B.~Tseng,
\newblock Phys. Lett. {\bf B415}, 263 (1997), hep-ph/9707316.

\bibitem{Kou:2001pm}
E.~Kou and A.~I. Sanda,
\newblock Phys. Lett. {\bf B525}, 240 (2002), hep-ph/0106159.

\bibitem{Gerard:2006ch}
J.~M. Gerard and E.~Kou,
\newblock Phys. Rev. lett. {\bf 97}, 261804 (2006), hep-ph/0609300.

\bibitem{Halperin:1997as}
I.~E. Halperin and A.~Zhitnitsky,
\newblock Phys. Rev. {\bf D56}, 7247 (1997), hep-ph/9704412.

\bibitem{Yuan:1997ts}
F.~Yuan and K.-T. Chao,
\newblock Phys. Rev. {\bf D56}, 2495 (1997), hep-ph/9706294.

\bibitem{Datta:2002pk}
A.~Datta, H.~J. Lipkin, and P.~J. O'Donnell,
\newblock Phys. Lett. {\bf B544}, 145 (2002), hep-ph/0206155.

\bibitem{Chang:2006dh}
Q.~Chang, X.-Q. Li, and Y.-D. Yang,
\newblock JHEP {\bf 06}, 038 (2007), hep-ph/0610280.

\bibitem{Fleischer:2007hj}
R.~Fleischer,
\newblock (2007), arXiv:0705.1121 [hep-ph].

\bibitem{Lipkin:1980tk}
H.~J. Lipkin,
\newblock Phys. Rev. Lett. {\bf 46}, 1307 (1981).

\bibitem{Close:1996ku}
F.~E. Close and H.~J. Lipkin,
\newblock Phys. Lett. {\bf B405}, 157 (1997), hep-ph/9607349.

\bibitem{Cheng:1998dn}
H.-Y. Cheng and B.~Tseng,
\newblock Phys. Rev. {\bf D59}, 014034 (1999), hep-ph/9806209.

\bibitem{Rosner:1999xd}
J.~L. Rosner,
\newblock Phys. Rev. {\bf D60}, 114026 (1999), hep-ph/9905366.

\bibitem{Cheng:2000fd}
H.-Y. Cheng and B.~Tseng,
\newblock Chin. J. Phys. {\bf 39}, 28 (2001), hep-ph/0006081.

\bibitem{Chiang:2001av}
C.-W. Chiang and J.~L. Rosner,
\newblock Phys. Rev. {\bf D65}, 054007 (2002), hep-ph/0110394.

\bibitem{Chiang:2002mr}
C.-W. Chiang, Z.~Luo, and J.~L. Rosner,
\newblock Phys. Rev. {\bf D67}, 014001 (2003), hep-ph/0209272.

\bibitem{Adams:2007mx}
CLEO, G.~S. Adams {\em et~al.},
\newblock (2007), arXiv:0708.0139 [hep-ex].

\bibitem{Kreinick:Charm07:Bottonium}
CLEO, D.~Kreinick,
\newblock Recent results in bottonium ties to charmonium,
\newblock talk presented at Charm 2007, Cornell, August 5-8, 2007.

\bibitem{Athar:2007hz}
CLEO, S.~B. Athar {\em et~al.},
\newblock (2007), arXiv:0704.3063 [hep-ex].

\end{thebibliography}

\appendix

\section{Electroweak Decay Amplitudes}
\label{sec:EWDecayAmps}

We consider the following types of diagram: colour favoured tree ($T$), colour suppressed tree ($C$), penguin ($P2_{u,d,s}$), `gluon hairpin' penguin ($P1_{u,d,s}$), weak-exchange ($E2_{u,d,s}$), `gluon hairpin' weak-exchange ($E1_{u,d,s}$), weak-annihilation ($A2_{u,d,s}$), and `gluon hairpin' weak-annihilation ($A1_{u,d,s}$).  The subscript on the penguin, weak-exchange and weak-annihilation amplitudes refers to the quark-antiquark pair produced at the gluon vertex.  By default the second meson (i.e. $Y$ in $B \rightarrow X + Y$) contains the spectator quark and we denote with a $'$ those amplitudes where the other meson contains the spectator quark.  We refer to the lowest order diagrams but these are just an example of the topologies: any topologically equivalent correction can be incorporated in to these amplitudes.

\begin{figure}[htb]
\begin{center}
\subfigure[Colour favoured tree, $T$]{\includegraphics[width=7cm]{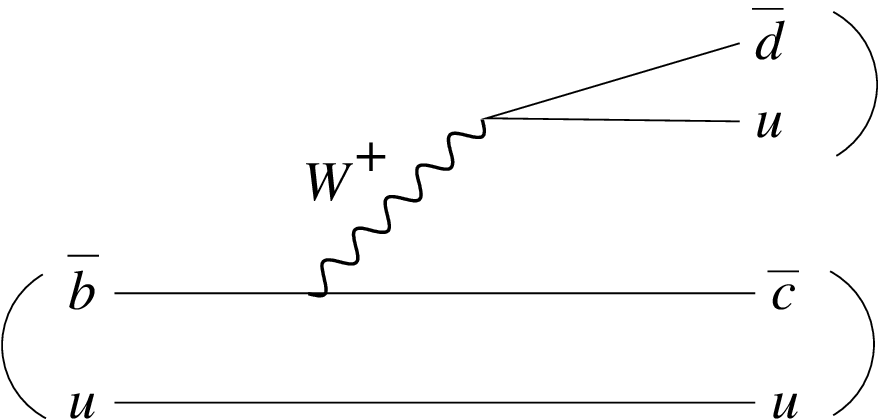}}
\hfill
\subfigure[Colour suppressed tree, $C$]{\includegraphics[width=7cm]{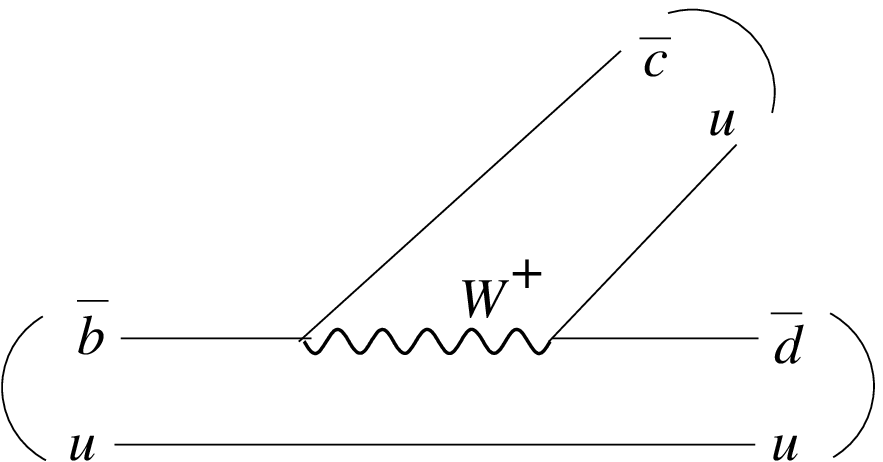}}
\subfigure[Penguin, $P2_q$]{\includegraphics[width=7cm]{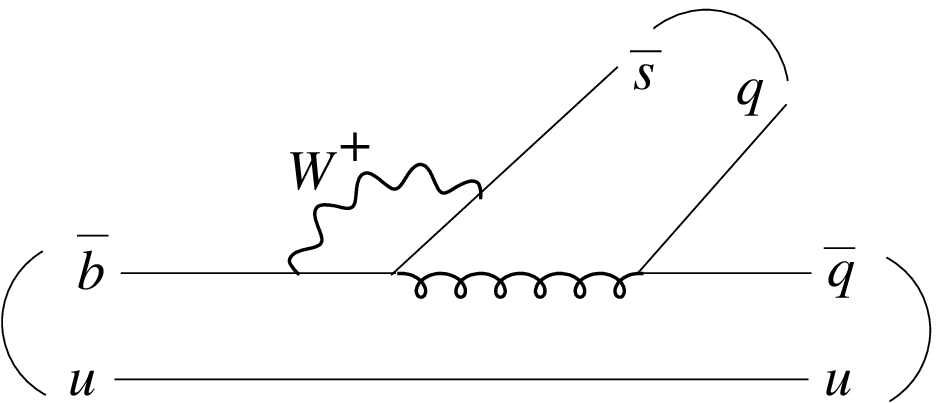}}
\hfill
\subfigure[`Gluon hairpin' penguin, $P1_q$]{\includegraphics[width=7cm]{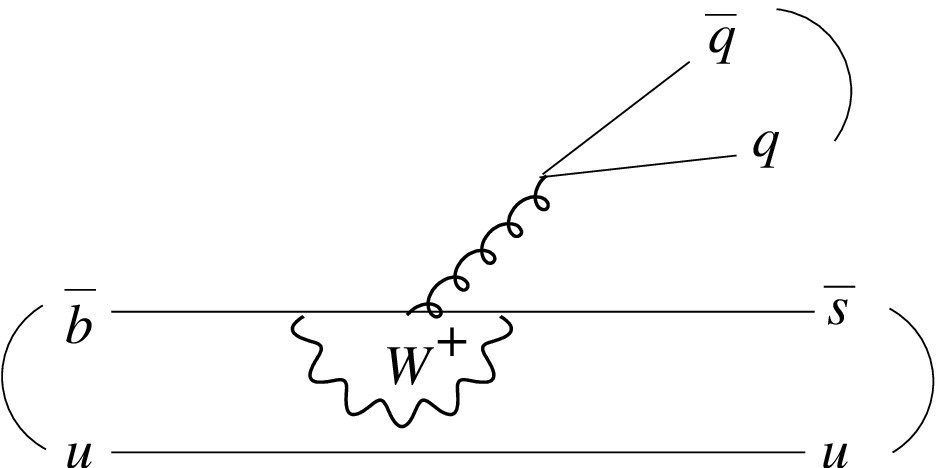}}
\subfigure[Weak-exchange, $E2_q$]{\includegraphics[width=7cm]{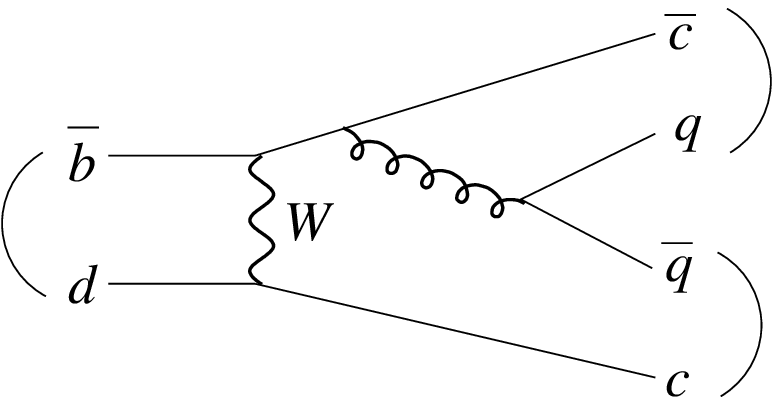}}
\hfill
\subfigure[`Gluon hairpin' weak-exchange, $E1_q$]{\includegraphics[width=7cm]{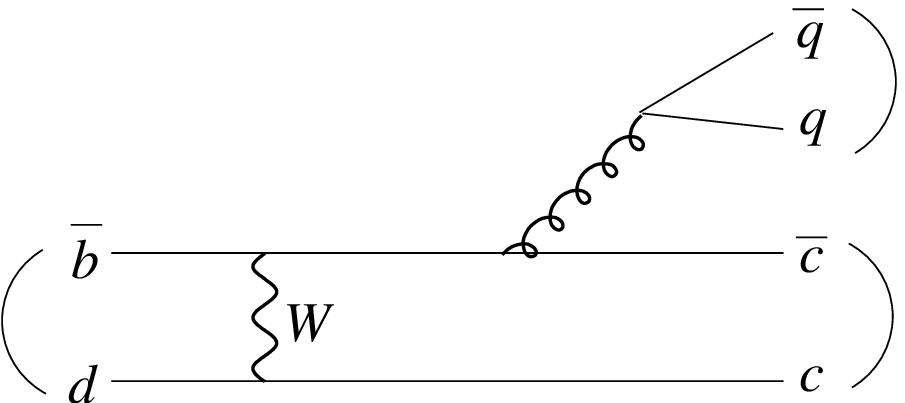}}
\subfigure[Weak-annihilation, $A2_q$]{\includegraphics[width=7cm]{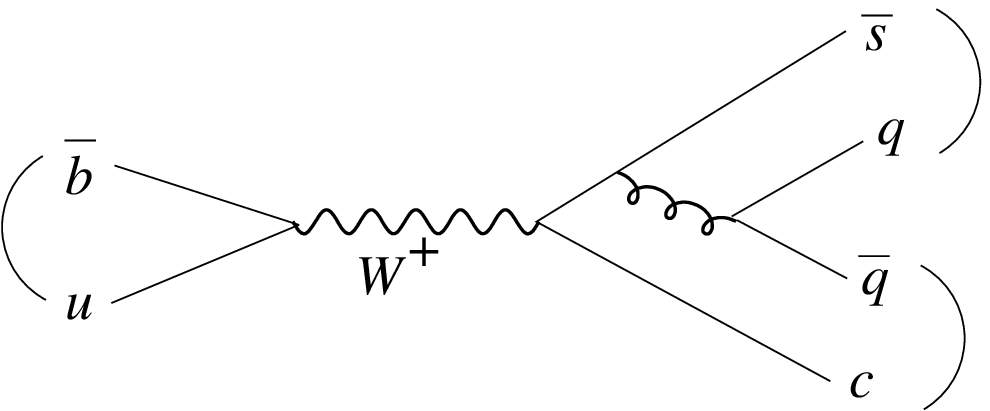}}
\hfill
\subfigure[`Gluon hairpin' weak-annihilation, $A1_q$]{\includegraphics[width=7cm]{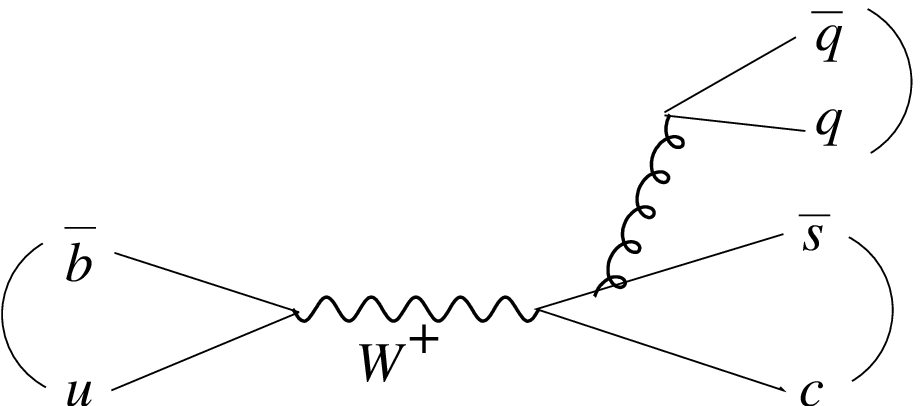}}
\caption{Examples of the types of diagram considered}
\label{fig:WeakDecayDiagrams}
\end{center}
\end{figure}

Examples of the different types of diagrams are given in Fig.\ \ref{fig:WeakDecayDiagrams}.  Approximate magnitudes of relevant CKM factors are given in Table \ref{table:CKMFactors}.  The amplitudes and CKM factors for the relevant decays are shown in Tables \ref{table:BDecayAmps} - \ref{table:DDsDecayAmps}.

\begin{table}[htb]
\begin{center}
\begin{tabular}{|c|c|}
\hline
\textbf{CKM Factor} & \textbf{Approximate Magnitude}  \\
\hline
$|V_{ud}V_{bc}|$ & $4\times10^{-2}$  \\
$|V_{us}V_{bc}|$ & $1\times10^{-2}$  \\
$|V_{cd}V_{bu}|$ & $9\times10^{-4}$  \\
$|V_{cs}V_{bu}|$ & $4\times10^{-3}$  \\
\hline
$|V_{us}V_{bu}|$ & $9\times10^{-4}$  \\
$|V_{cs}V_{bc}|$ & $4\times10^{-2}$  \\
$|V_{ts}V_{bt}|$ & $4\times10^{-2}$  \\
\hline
$|V_{ud}V_{bu}|$ & $4\times10^{-3}$  \\
$|V_{cd}V_{bc}|$ & $1\times10^{-2}$  \\
$|V_{td}V_{bt}|$ & $8\times10^{-3}$  \\
\hline
$|V_{su}V_{cd}|$ & $5\times10^{-2}$  \\
$|V_{du}V_{cs}|$ & $9\times10^{-1}$  \\
\hline
$|V_{du}V_{cd}|$ & $2\times10^{-1}$  \\
$|V_{su}V_{cs}|$ & $2\times10^{-1}$  \\
$|V_{bu}V_{cb}|$ & $2\times10^{-4}$  \\
\hline
\end{tabular}
\caption{Approximate magnitudes of relevant CKM factors from the global fit in the PDG Review 2006\cite{PDG06}.}
\label{table:CKMFactors}
\end{center}
\end{table}

\begin{table}[htb]
\begin{center}
\begin{tabular}{|c|l|l|}
\hline
\textbf{Mode} & \textbf{Amplitudes} & \textbf{CKM Factor(s)} \\
\hline
$B^+ \rightarrow D^+ + (u\bar{u})$ & $T, A1_u$ & $V_{cd}V_{bu}$ \\
$B^+ \rightarrow D^+ + (d\bar{d})$ & $A2_d, A1_d$ &  \\
$B^+ \rightarrow D^+ + (s\bar{s})$ & $A1_s$ &  \\
$B^+ \rightarrow D_s^+ + \bar{K}^0$ & $A2_s$ &  \\
$B^+ \rightarrow D^0 + \pi^+$ & $C, A2_u$ &  \\
\hline
$B^0 \rightarrow D^0 + (u\bar{u})$ & $E2'_u, E1'_u$ & $V_{cd}V_{bu}$ \\
$B^0 \rightarrow D^0 + (d\bar{d})$ & $C, E1'_d$ &  \\
$B^0 \rightarrow D^0 + (s\bar{s})$ & $E1'_s$ &  \\
$B^0 \rightarrow D_s^+ + K^-$ & $E2'_s$ &  \\
\hline
$B^0 \rightarrow \bar{D}^0 + (u\bar{u})$ & $E2_u, E1'_u$ & $V_{ud}V_{bc}$ \\
$B^0 \rightarrow \bar{D}^0 + (d\bar{d})$ & $C, E1'_d$ &  \\
$B^0 \rightarrow \bar{D}^0 + (s\bar{s})$ & $E1'_s$ &  \\
$B^0 \rightarrow D_s^- + K^+$ & $E2_s$ & \\
\hline
$B^+ \rightarrow D_s^+ + (u\bar{u})$ & $T, A1_u$ & $V_{cs}V_{bu}$  \\
$B^+ \rightarrow D_s^+ + (d\bar{d})$ & $A1_d$ &  \\
$B^+ \rightarrow D_s^+ + (s\bar{s})$ & $A2_s, A1_s$ &  \\

$B^0 \rightarrow D_s^+ + \pi^-$ & $T$ &  \\
$B^+ \rightarrow D^+ + K^0$ & $A2_d$ &  \\
$B^+ \rightarrow D^0 + K^+$ & $C, A2_u$ &  \\
$B^0 \rightarrow D^0 + K^0$ & $C$ &  \\
\hline
$B^0 \rightarrow (c\bar{c}) + (u\bar{u})$ & $E1_c, E1'_u$ & $P1: V_{ud}V_{bu}, V_{cd}V_{bc}, V_{td}V_{bt}$ \\
$B^0 \rightarrow (c\bar{c}) + (d\bar{d})$ & $C, P1_c, E1'_d$ & $C, E1'_{u,d,s}: V_{cd}V_{bc}$ \\
$B^0 \rightarrow (c\bar{c}) + (s\bar{s})$ & $E1'_s$ & $E1_c: V_{ud}V_{bu}$ \\
$B^+ \rightarrow (c\bar{c}) + \pi^+$ & $C, P1_c, A1_c$ &  \\
\hline
$B^0 \rightarrow (c\bar{c}) + K^0$ & $C, P1_c$ & $C: V_{cs}V_{bc}$ \\
$B^+ \rightarrow (c\bar{c}) + K^+$ & $C, P1_c, A1_c$ & $P1: V_{us}V_{bu}, V_{cs}V_{bc}, V_{ts}V_{bt}$ \\
 & & $A1: V_{us}V_{bu}$ \\
\hline
$B^+ \rightarrow K^+ + (u\bar{u})$ & $T, C', P2_u, P1'_u, A2_u, A1_u$ & $P1,P2: V_{us}V_{bu}, V_{cs}V_{bc}, V_{ts}V_{bt}$ \\
$B^+ \rightarrow K^+ + (d\bar{d})$ & $P1'_d, A1_d$ & $T, C, A2, A1: V_{us}V_{bu}$ \\
$B^+ \rightarrow K^+ + (s\bar{s})$ & $P2'_s, P1'_s, A2_s, A1_s$ &  \\
$B^+ \rightarrow K^0 + \pi^+$ & $P2_d, A2_d$ &  \\
\hline
$B^0 \rightarrow K^0 + (u\bar{u})$ & $C', P1'_u$ & $P1,P2: V_{us}V_{bu}, V_{cs}V_{bc}, V_{ts}V_{bt}$ \\
$B^0 \rightarrow K^0 + (d\bar{d})$ & $P2_d, P1'_d$ & $C: V_{us}V_{bu}$ \\
$B^0 \rightarrow K^0 + (s\bar{s})$ & $P2'_s, P1'_s$ &  \\
$B^0 \rightarrow K^+ + \pi^-$ & $T, P2_u$ &  \\
\hline
$B^+ \rightarrow \pi^+ + (u\bar{u})$ & $T, C', P2_u, P1'_u, A2_u, A1_u$ & $P1, P2: V_{ud}V_{bu}, V_{cd}V_{bc}, V_{td}V_{bt}$ \\
$B^+ \rightarrow \pi^+ + (d\bar{d})$ & $P2'_d, P1'_d, A2_d, A1_d$ & $T, C, A2, A1: V_{ud}V_{bu}$ \\
$B^+ \rightarrow \pi^+ + (s\bar{s})$ & $P1'_s, A1_s$ &  \\
$B^+ \rightarrow \bar{K}^0 + K^+$ & $P2_s, A2_s$ &  \\
\hline
$B^0 \rightarrow (u\bar{u}) + (u\bar{u})$ & $E2_u, E1_u$ & $P1, P2: V_{ud}V_{bu}, V_{cd}V_{bc}, V_{td}V_{bt}$ \\ 
$B^0 \rightarrow (u\bar{u}) + (d\bar{d})$ & $C, P1_u, E1'_d$ & $T, C, E2, E1: V_{ud}V_{bu}$ \\ 
$B^0 \rightarrow (d\bar{d}) + (d\bar{d})$ & $P2_d, P1_d$ &  \\ 
$B^0 \rightarrow (u\bar{u}) + (s\bar{s})$ & $E1'_s$ &  \\ 
$B^0 \rightarrow (d\bar{d}) + (s\bar{s})$ & $P1'_s$ &  \\ 
$B^0 \rightarrow \pi^+ + \pi^-$ & $T, P2_u, E2'_d$ &  \\
$B^0 \rightarrow \bar{K}^0 + K^0$ & $P2_s$ &  \\
$B^0 \rightarrow K^- + K^+$ & $E2_s$ &  \\
\hline
\end{tabular}
\caption{Electroweak decays of $B^+$ and $B^0$ mesons to final states including $\eta/\eta'$ and related modes.  See the text for an explanation of the notation used.}
\label{table:BDecayAmps}
\end{center}
\end{table}

\begin{table}[htb]
\begin{center}
\begin{tabular}{|c|l|l|}
\hline
\textbf{Mode} & \textbf{Amplitudes} & \textbf{CKM Factor(s)} \\
\hline
$B_s^0 \rightarrow D^0 + (u\bar{u})$ & $E2'_u, E1'_u$ & $V_{cs}V_{bu}$ \\
$B_s^0 \rightarrow D^0 + (d\bar{d})$ & $E1'_d$ &  \\
$B_s^0 \rightarrow D^0 + (s\bar{s})$ & $C, E1'_s$ &  \\
$B_s^0 \rightarrow D^+ + \pi^-$ & $E2'_d$ &  \\
\hline
$B_s^0 \rightarrow \bar{D}^0 + (u\bar{u})$ & $E2_u, E1'_u$ & $V_{us}V_{bc}$ \\
$B_s^0 \rightarrow \bar{D}^0 + (d\bar{d})$ & $E1'_d$ &  \\
$B_s^0 \rightarrow \bar{D}^0 + (s\bar{s})$ & $C, E1'_s$ &  \\
$B_s^0 \rightarrow D^- + \pi^+$ & $E2_d$ &  \\
\hline
$B_s^0 \rightarrow (c\bar{c}) + (u\bar{u})$ & $E1_c, E1'_u$ & $P1: V_{us}V_{bu}, V_{cs}V_{bc}, V_{ts}V_{bt}$ \\
$B_s^0 \rightarrow (c\bar{c}) + (d\bar{d})$ & $E1'_d$ & $C, E1_{u,d,s}: V_{cs}V_{bc}$ \\
$B_s^0 \rightarrow (c\bar{c}) + (s\bar{s})$ & $C, P1_c, E1'_s$ & $E1_c: V_{us}V_{bu}$ \\
\hline
$B_s^0 \rightarrow \bar{K}^0 + (u\bar{u})$ & $C', P1'_u$ & $P2, P1: V_{ud}V_{bu}, V_{cd}V_{bc}, V_{td}V_{bt}$ \\
$B_s^0 \rightarrow \bar{K}^0 + (d\bar{d})$ & $P2'_d, P1'_d$ & $C: V_{ud}V_{bu}$ \\
$B_s^0 \rightarrow \bar{K}^0 + (s\bar{s})$ & $P2_s, P1'_s$ &  \\
\hline
$B_s^0 \rightarrow (u\bar{u}) + (u\bar{u})$ & $E2_u, E1_u$ & $P2, P1: V_{us}V_{bu}, V_{cs}V_{bc}, V_{ts}V_{bt}$ \\
$B_s^0 \rightarrow (u\bar{u}) + (d\bar{d})$ & $E1'_d$ & $T, C, E2, E1: V_{us}V_{bu}$ \\
$B_s^0 \rightarrow (d\bar{d}) + (d\bar{d})$ & $-$ &  \\
$B_s^0 \rightarrow (u\bar{u}) + (s\bar{s})$ & $C, P1_u, E1'_s$ &  \\
$B_s^0 \rightarrow (d\bar{d}) + (s\bar{s})$ & $P1_d$ &  \\
$B_s^0 \rightarrow (s\bar{s}) + (s\bar{s})$ & $P2_s, P1_s$ &  \\
$B_s^0 \rightarrow \pi^- + \pi^+$ & $E2_d$ &  \\
$B_s^0 \rightarrow K^+ + K^-$ & $T, P2_u, E2'_s$ &  \\
$B_s^0 \rightarrow K^0 + \bar{K}^0$ & $P2_d$ &  \\
\hline
\hline
$B_c^+ \rightarrow D^+ + (u\bar{u})$ & $C', P1'_u, A1_u$ & $P2, P1: V_{ud}V_{bu}, V_{cd}V_{bc}, V_{td}V_{bt}$ \\
$B_c^+ \rightarrow D^+ + (d\bar{d})$ & $P2'_d, P1'_d, A2_d, A1_d$ & $T, C: V_{ud}V_{bu}$ \\
$B_c^+ \rightarrow D^+ + (s\bar{s})$ & $P1'_s, A1_s$ & $A2, A1: V_{cd}V_{bc}$ \\
$B_c^+ \rightarrow D^0 + \pi^+$ & $T', P2'_u, A2_u$ &  \\
$B_c^+ \rightarrow D_s^+ + \bar{K}^0$ & $P2'_s, A2_s$ &  \\
\hline
$B_c^+ \rightarrow D_s^+ + (u\bar{u})$ & $C', P1'_u, A1_u$ & $P2, P1: V_{us}V_{bu}, V_{cs}V_{bc}, V_{ts}V_{bt}$ \\
$B_c^+ \rightarrow D_s^+ + (d\bar{d})$ & $P1'_d, A1_d$ & $T, C: V_{us}V_{bu}$ \\
$B_c^+ \rightarrow D_s^+ + (s\bar{s})$ & $P2'_s, P1'_s, A2_s, A1_s$ & $A2, A1: V_{cs}V_{bc}$ \\
$B_c^+ \rightarrow D^0 + K^+$ & $T', P2'_u, A2_u$ &  \\
$B_c^+ \rightarrow D^+ + K^0$ & $P2'_d, A2_d$ &  \\
\hline
$B_c^+ \rightarrow K^+ + (u\bar{u})$ & $A2_u, A1_u$ & $V_{us}V_{bc}$ \\
$B_c^+ \rightarrow K^+ + (d\bar{d})$ & $A1_d$ &  \\
$B_c^+ \rightarrow K^+ + (s\bar{s})$ & $A2_s, A1_s$ &  \\
$B_c^+ \rightarrow K^0 + \pi^+$ & $A2_d$ &  \\
\hline
$B_c^+ \rightarrow \pi^+ + (u\bar{u})$ & $A2_u, A1_u$ & $V_{ud}V_{bc}$ \\
$B_c^+ \rightarrow \pi^+ + (d\bar{d})$ & $A2_d, A1_d$ &  \\
$B_c^+ \rightarrow \pi^+ + (s\bar{s})$ & $A1_s$ &  \\
$B_c^+ \rightarrow K^+ + \bar{K}^0$ & $A2_s$ &  \\
\hline
\end{tabular}
\caption{Electroweak decays of $B_s$ and $B_c$ mesons to final states including $\eta/\eta'$ and related modes.  See the text for an explanation of the notation used.}
\label{table:BsBcDecayAmps}
\end{center}
\end{table}

\begin{table}[htb]
\begin{center}
\begin{tabular}{|c|l|l|}
\hline
\textbf{Mode} & \textbf{Amplitudes} & \textbf{CKM Factor(s)} \\
\hline
 & & (All penguins $P2, P1$: $V_{du}V_{cd}, V_{su}V_{cs}, V_{bu}V_{cb}$) \\
\hline
$D^0 \rightarrow K^0 + (u\bar{u})$ & $C, E1'_u$ & $V_{su}V_{cd}$ \\
$D^0 \rightarrow K^0 + (d\bar{d})$ & $E2'_d, E1'_d$ &  \\
$D^0 \rightarrow K^0 + (s\bar{s})$ & $E2_s, E1'_s$ &  \\
$D^0 \rightarrow K^+ + \pi^-$ & $T, E2'_u$ &  \\
$D^+ \rightarrow K^0 + \pi^+$ & $C, A2_d$ &  \\
\hline
$D^0 \rightarrow \bar{K}^0 + (u\bar{u})$ & $C, E1'_u$ & $V_{du}V_{cs}$ \\
$D^0 \rightarrow \bar{K}^0 + (d\bar{d})$ & $E2_d, E1'_d$ &  \\
$D^0 \rightarrow \bar{K}^0 + (s\bar{s})$ & $E2'_s, E1'_s$ &  \\
$D^0 \rightarrow K^- + \pi^+$ & $T', E2_u$ &  \\
$D^+ \rightarrow \bar{K}^0 +\pi^+$ & $T', C$ &  \\
\hline
$D^+ \rightarrow K^+ + (u\bar{u})$ & $A2_u, A1_u$ & $V_{su}V_{cd}$ \\
$D^+ \rightarrow K^+ + (d\bar{d})$ & $T, A1_d$ &  \\
$D^+ \rightarrow K^+ + (s\bar{s})$ & $A2_s, A1_s$ &  \\
\hline
$D^0 \rightarrow (u\bar{u}) + (u\bar{u})$ & $P2_u, P1_u$ & $C': V_{du}V_{cd}$ \\
$D^0 \rightarrow (u\bar{u}) + (d\bar{d})$ & $C', P1'_d, E1_u$ & $C': V_{su}V_{cs}$  \\
$D^0 \rightarrow (d\bar{d}) + (d\bar{d})$ & $E2_d, E1_d$ & $E2_d: V_{du}V_{cd}$ \\
$D^0 \rightarrow (u\bar{u}) + (s\bar{s})$ & $C', P1'_s, E1_u$ & $E2_s: V_{su}V_{cs}$ \\
$D^0 \rightarrow (d\bar{d}) + (s\bar{s})$ & $E1_d, E1'_s$ &  $E1_u((u\bar{u})(d\bar{d})), E1_d((d\bar{d})(d\bar{d})), E1'_s((d\bar{d})(s\bar{s})): V_{du}V_{cd}$ \\
$D^0 \rightarrow (s\bar{s}) + (s\bar{s})$ & $E2_s, E1_s$ &  $E1_u((u\bar{u})(s\bar{s})),E1_d((d\bar{d})(s\bar{s})), E1_s((s\bar{s})(s\bar{s})): V_{su}V_{cs}$ \\
\hline
$D^+ \rightarrow \pi^+ + (u\bar{u})$ & $P2'_u, P1'_u, A2_u, A1_u$ & $T, C, A2, A1, E2: V_{du}V_{cd}$ \\
$D^+ \rightarrow \pi^+ + (d\bar{d})$ & $T, C', P2_d, P1'_d, A2_d, A1_d$ &  \\
$D^+ \rightarrow \pi^+ + (s\bar{s})$ & $C', P1'_s, A1_s$ &  $C'(\pi^+ (s\bar{s})): V_{su}V_{cs}$ \\
$D^0 \rightarrow \pi^+ + \pi^-$ & $T, P2_d, E2'_u$ &  \\
\hline
$D^0 \rightarrow K^+ + K^-$ & $T, P2_s, E2'_u$ & $V_{su}V_{cs}$ \\
\hline
$D^0 \rightarrow K^0 + \bar{K}^0$ & $E2_s, E2'_d$ & $E2_s: V_{du}V_{cd};\ E2'_d: V_{su}V_{cs}$ \\
\hline
$D^+ \rightarrow K^+ + \bar{K}^0$ & $T, P2_s, A2_s$ & $T: V_{su}V_{cs};\ A2_s: V_{du}V_{cd}$ \\
\hline
\hline
$D_s^+ \rightarrow K^+ + (u\bar{u})$ & $P2'_u, P1'_u, A2_u, A1_u$ & $V_{su}V_{cs}$ \\
$D_s^+ \rightarrow K^+ + (d\bar{d})$ & $C', P1'_d, A1_d$ &  $C'(K^+(d\bar{d})), T'(K^0 \pi^+): V_{du}V_{cd}$ \\
$D_s^+ \rightarrow K^+ + (s\bar{s})$ & $T, C', P2_s, P1'_s, A2_s, A1_s$ &  \\
$D_s^+ \rightarrow K^0 + \pi^+$ & $T',P2'_d, A2_d$ &  \\
\hline
$D_s^+ \rightarrow \pi^+ + (u\bar{u})$ & $A2_u, A1_u$ & $V_{du}V_{cs}$ \\
$D_s^+ \rightarrow \pi^+ + (d\bar{d})$ & $A2_d, A1_d$ &  \\
$D_s^+ \rightarrow \pi^+ + (s\bar{s})$ & $T, A1_s$ &  \\
$D_s^+ \rightarrow K^+ + \bar{K}^0$ & $C', A2_s$ &  \\
\hline
$D_s^+ \rightarrow K^+ + K^0$ & $T, C'$ & $V_{su}V_{cd}$ \\
\hline
\end{tabular}
\caption{Electroweak decays of $D$ and $D_s$ mesons to final states including $\eta/\eta'$ and related modes.  See the text for an explanation of the notation used.}
\label{table:DDsDecayAmps}
\end{center}
\end{table}

\end{document}